%% file: article.tex
\documentclass[aps,                
               prc,                
               showpacs,           
               twocolumn,          
               10pt,               
               twoside,            
               floatfix,           
               superscriptaddress] 
               {revtex4-1}

\input{packages.tex}

\begin{document}

  \input{title.tex}

  \input{section_1_intro.tex}
  \input{section_2_reldyn.tex}
  \input{section_3_njl.tex}
  \input{section_4_integral.tex}
  \input{section_5_results.tex}
  \input{section_6_summary.tex}

  \input{acknowledgment.tex}

  \input{appendix.tex}

  \bibliography{biblio}

\end{document}

%% file: packages.tex
\usepackage{mathptmx}
\usepackage[T1]{fontenc}
\usepackage{geometry}         
\usepackage{fancyhdr}         
\usepackage{graphicx}         
\usepackage[svgnames]{xcolor} 
\usepackage{hyperref}         
\usepackage{amsmath}          
\usepackage{amssymb}          
\usepackage{amsfonts}         
\usepackage{amstext}          
\usepackage{textcomp}         

\hypersetup{colorlinks=true,linkcolor=Blue,citecolor=Blue,urlcolor=Blue}

\DeclareMathAlphabet{\mathcal}{OMS}{cmsy}{m}{n}

\newcommand{\feyn}[1]{{#1}\!\!\!{\slash}}

\fancypagestyle{firststyle}
{%
  \fancyhf{}
  \fancyhead[C]{PHYSICAL REVIEW C {\bf 87}, 034912 (2013)} 
  \fancyfoot[L]{\small 0556-2813/2013/87(1)/034912(25)} 
  \fancyfoot[C]{\small 034912-\thepage} 
  \fancyfoot[R]{\small \textcopyright 2013 American Physical Society} 
}

%% file: title.tex
\title{\texorpdfstring{\begin{minipage}[c]{\textwidth}\centering Molecular dynamics description of an expanding $\text{\bf\emph{q}}$/$\bar{\text{\bf\emph{q}}}$ plasma with the Nambu--Jona-Lasinio model and applications to heavy ion collisions at energies available at the BNL Relativistic Heavy Ion Collider and the CERN Large Hadron Collider\end{minipage}}{Molecular dynamics description of an expanding q/qb plasma with the Nambu--Jona-Lasinio model and applications to heavy ion collisions at energies available at the BNL Relativistic Heavy Ion Collider and the CERN Large Hadron Collider}}

\author{R.~Marty}
\email{marty@fias.uni-frankfurt.de}
\affiliation{\begin{minipage}[c]{0.98\textwidth}Subatech, UMR 6457, IN2P3/CNRS, Universit\'e de Nantes, \'Ecole des Mines de Nantes, 4 rue Alfred Kastler, 44307 Nantes cedex 3, France\end{minipage}}
\affiliation{\begin{minipage}[c]{\textwidth}Frankfurt Institute for Advanced Studies and Institute for Theoretical Physics, Johann Wolfgang Goethe Universit\"at, Ruth-Moufang-Strasse 1,\end{minipage}\\
60438 Frankfurt am Main, Germany\\
\normalfont{(Received 15 October 2012; revised 5 February 2013; published 28 March 2013)}\vspace{2.5mm}}

\author{J.~Aichelin}
\email{aichelin@subatech.in2p3.fr}
\affiliation{\begin{minipage}[c]{0.98\textwidth}Subatech, UMR 6457, IN2P3/CNRS, Universit\'e de Nantes, \'Ecole des Mines de Nantes, 4 rue Alfred Kastler, 44307 Nantes cedex 3, France\end{minipage}}

\pacs{24.10.Jv, 02.70.Ns, 12.38.Mh, 24.85.+p%
\hfill DOI:\href{http://link.aps.org/doi/10.1103/PhysRevC.87.034912}{10.1103/PhysRevC.87.034912}}

\begin{abstract}
We present a relativistic molecular dynamics approach based on the Nambu--Jona-Lasinio Lagrangian. We derive the relativistic time evolution equations for an expanding plasma, discuss the hadronization cross section, and explain how they act in such a scenario. We present in detail how one can transform the time evolution equation to a simulation program and apply this program to study the expansion of a plasma created in experiments at the Relativistic Heavy Ion Collider and the Large Hadron Collider. We present first results on  the centrality dependence of $v_2$ and of the transverse momentum spectra of pions and kaons and discuss in detail the hadronization mechanism.
\end{abstract}

\maketitle

%% file: section_1_intro.tex
\section{Introduction}
\thispagestyle{firststyle}
\vskip -2mm
The interpretation of the results of ultrarelativistic heavy ion collisions is presently one of the most challenging problems in theoretical nuclear physics. In these collisions, investigated at the Relativistic Heavy Ion Collider  (RHIC) at the Brookhaven National Laboratory and at the Large Hadron Collider (LHC) at CERN, more than a thousand particles are observed in central collisions. Although the multiplicity and the single particle transverse momentum spectra at midrapidity of the different particle species are of interest in their own right, the purpose of the experiments is to find out whether during the reaction the matter has made a transition towards a new state of matter, a quark-gluon plasma (QGP). This information is not directly visible in the measured hadron spectra and therefore theoretical approaches have to be employed to verify whether the measured observables are compatible with the existence of such a QGP or, even more desired, whether they can even lead to the conclusion that such a state is necessary to explain the measured quantities.

State-of-the-art theoretical approaches aim at a complete description of the heavy ion reaction, from the initial separation  of projectile and target up to the momenta of the finally observed particles \cite{Hirano2011,Werner2010,Luzum2008,Gale2013,Song2011,Petersen2008,Teaney2012,Drescher2007}. Almost all of the presently developed models assume that the reaction can be subdivided into four different phases, each of them described by a different model. The transition between these phases is local in time.  The first phase is a fast local equilibration of the system.  In view of the magnitude of perturbative quantum chromo dynamics (pQCD) cross sections it is hard to understand how this can occur. No detailed theory is available yet for this initial phase. The second phase is the expansion of the plasma described by hydrodynamical equations employing the equation of state, calculated by lattice gauge calculations. Initial geometrical fluctuations  are taken into account. This phase is followed by a phase transition. The transition toward on-shell hadrons which carry their free mass is described by the Cooper-Frye formalism. Finally, the hadrons expand and interact among each other by the measured free cross sections. Despite the quite severe and differing assumptions of each of the models for all of these phases these models have been quite successful in describing many of the observables.

From data alone it is not possible to judge which of the assumptions are justified. As an example of the ambiguity of the theoretical interpretation of experimental results we just mention the centrality dependence of the elliptical flow, one of the key observables, which is equally well reproduced in three quite different approaches. In approaches which use \emph{viscous} hydrodynamics \cite{Luzum2008} this centrality dependence serves to determine the viscosity of the QGP and therefore of the interaction among the constituents of the plasma, the quarks, and the gluons. In \emph{ideal} hydrodynamical approaches with fluctuating initial conditions \cite{Werner2010, Aichelin2010}, in which only regions of a high energy density form a QGP, this centrality dependence is due to the impact parameter dependence of the relative contributions of high energy density and of low energy density regions. Finally, the centrality dependence is also well described in the core-corona model \cite{Aichelin2010}, in which it is assumed that nucleons which suffer from only one hard initial scattering fragment like a proton in $p$-$p$ collisions whereas the rest forms a QGP whose properties are impact parameter independent.  

In order to check the assumptions and to calculate the transport coefficients used in this these multiphase models one needs models in which one does not assume right from the beginning that a local equilibrium is established. The microscopic color quark dynamics model \cite{Scherer2001} and the parton hadron string dynamics (PHSD) model \cite{Cassing2009,Cassing2009a,Bratkovskaya2011} are such models which allow one to study the plasma evolution by solving a Boltzmann-type equation. The former is a nonrelativisitc approach with effective potentials. It allowed for the first time to the study of the expansion of a colored plasma. In the latter the potentials among the plasma constituents are chosen in such a way that the equation of state from lattice calculations is respected. Cross sections can be derived from the spacelike part of the interaction and are employed for the scattering interactions among the plasma constituents. In this model gluons as well as quarks acquire a large mass when approaching the phase transition. Therefore the prehadrons which are created at the phase transition are rather heavy. Pions and other light hadrons are produced by the decay of these prehadrons. Another model which allows for these studies is that of a gluonic cascade realized in the Boltzmann approach to multiparton scattering (BAMPS) \cite{Uphoff2010}. The gluon emissions and interactions during the expansion of the QGP move the system toward equilibrium.

A while ago a third approach has been advanced \cite{Rehberg1999a} which is based on the Nambu--Jona-Lasinio (NJL) Lagrangian \cite{Klimt1990,Nebauer2002}. This Lagrangian is an approximation to the QCD Lagrangian which respects all its symmetries. In the version which includes a Polyakov loop (PNJL) this approach also describes the equation of state of the lattice QCD data \cite{Ratti2007}. It has the advantage that all free parameters of the Lagrangian can be determined by static meson properties, such as meson masses and decay constants. It contains no explicit gluons and the in-medium mass of the gluons is assumed to be large as compared to the transferred momentum and therefore the interaction of the quarks is effectively a contact interaction. The quarks interact by scalar fields and by cross sections which can be as well derived from the Lagrangian \cite{Rehberg1996a,Gastineau2005}. Mesons can be produced, even in the deconfined phase, but they are unstable there and may decay. Only when the system approaches the crossover does the finite width of the meson mass disappear and can stable mesons can emerge from the system [but for low chemical potential the (P)NJL Lagragian shows a cross over and not a phase transition]. The light mesons are directly produced by $q \bar{q}$ scattering. Similar to the PHSD approach this Lagrangian offers therefore the opportunity to study the evolution of the system from the creation of a plasma up to the finally observed mesons. By using an $N$-body molecular dynamics approach it is possible to study correlations and fluctuations which are built up during the expansion phase and to investigate whether observables can be identified which are sensitive to them. 

The cross sections calculated in this approach are quite small deep inside the plasma phase but, due to $s$-channel resonances, they are quite large close to the crossover \cite{Gastineau2005} where the system behaves like a liquid. Deep in the plasma phase the particles have only their bare mass and move with a velocity which is close to the velocity of light. In order to study the time evolution of the $N$-body system with the NJL Lagrangian we have therefore to develop a molecular dynamics approach for interacting particles which move relativistically. Such an approach was advanced in the original paper on the relativistic quantum molecular dynamics (RQMD) \cite{Sorge1989} approach but has never been used in practice because of conceptual and numerical problems. Some of the conceptual problems are related to the choice of constraints which one has to impose to construct such a relativistic molecular dynamics. 

The papers which contain the mathematical tools to develop a relativistic molecular dynamics approach are widely scattered. Therefore, and in order to present a comprehensive approach, we will start out in the next section with a presentation of the formalism and its derivation. We will describe how a relativistic molecular dynamics can be developed, how one can avoid the no interaction theorem (NIT), and how the Dirac approach for a Hamiltonian system with constraints is of importance for the development of relativistic dynamics. Finally, we present the formalism which is used. We discuss the constraints and their consequences for the dynamics. The third section presents the NJL model as far as it is necessary to understand our approach; in particular we discuss how masses and cross sections can be calculated. The fourth section is devoted to the details of the numerical realization of the approach. In the fifth section we present how we validated the program and some results. Finally, in the sixth section, we draw our conclusions.

%% file: section_2_reldyn.tex
\section{Relativistic quantum molecular dynamics}
\vskip -2mm
\subsection{Molecular dynamics}
\vskip -1mm
\subsubsection{\bf\emph{One-body classical molecular dynamics}}
\vskip -3mm
In the classical molecular dynamics approach particles are moving under the mutual influence of forces. The goal is to describe the trajectories of these particles in phase-space (${\bf q}_i(t),{\bf p}_i(t) $). Knowing the phase space point for a given initial condition  (${\bf q}_i(0),{\bf p}_i(0) $) and the Hamiltonian $\mathcal{H}$ we can predict the phase-space points at any given moment $t$ and can calculate the value of each observable which is defined on the classical phase space. The trajectory may depend in a very sensible way on the initial condition and may therefore become chaotic. Such systems are, however, not of interest here.

We start the discussion of the nonrelativistic approach by providing the formalism. We employ the Hamilton-Jacobi approach to formulate the motion of one particle in phase space. The equation of motion for an observable $A$, defined on the classical phase space, $A({\bf q},{\bf p},t) $, where  ${\bf q},{\bf p},t $ are the independent variables, is given by
\begin{equation}
  \frac{d}{d t} A({\bf q},{\bf p},t) = \frac{\partial A}{\partial t}
  + \frac{\partial A}{\partial {\bf q}} \frac{\partial {\bf q}}{\partial t}
  + \frac{\partial A}{\partial {\bf p}} \frac{\partial {\bf p}}{\partial t}.
\label{at}
\end{equation}
The Hamilton-Jacobi equations, which present the equations of motion of the phase-space coordinates ${\bf q}$ and ${\bf p}$  in time can be obtained by a variational principle
\begin{equation}
  \frac{d \bf q}{d t} = \frac{\partial \mathcal{H}}{\partial \bf p},
  \quad
  \frac{d \bf p}{d t} =-\frac{\partial \mathcal{H}}{\partial \bf q},
  \label{ham}
\end{equation}
where $\mathcal{H} ({\bf q},{\bf p})$ is the Hamiltonian of the system. We can bring Eq. \eqref{at} into the form
\begin{equation}
  \frac{d A}{d t} = \frac{\partial A}{\partial t} + \{ A,\mathcal{H} \}.
\label{phsfct}
\end{equation}
 $\{ A,B \}$ is the Poisson's bracket of $A$ and $B$ defined for $N$ particles as
\begin{equation}
 \{ A,B \} = \sum_k^N   \frac{\partial A}{\partial {\bf q}_k}
                        \frac{\partial B}{\partial {\bf p}_k}
		              - \frac{\partial A}{\partial {\bf p}_k}
		                \frac{\partial B}{\partial {\bf q}_k}. 
\end{equation}
In the special case where $A$ does not explicitly depend on time we find
\begin{equation}
  \frac{d A}{d t} = \{ A,\mathcal{H} \} . 
\end{equation}
If we replace $A$ by either $\bf q$ or $ \bf p$ we recover the Hamilton-Jacobi equations, Eq. \eqref{ham},
\begin{equation}
  \frac{d \bf q}{d t} = \{ \bf q , \mathcal{H} \}
  = \frac{\partial \mathcal{H}}{\partial \bf p},
  \quad
  \frac{d \bf p}{d t} = \{ \bf p , \mathcal{H} \}
  =-\frac{\partial \mathcal{H}}{\partial \bf q}.
  \label{hameq}
\end{equation}
For a given initial condition (${\bf q}_0,{\bf p}_0 $) these equations can be solved, analytically or at least numerically, and we obtain the desired trajectory of the particle in phase space. For the later discussion it is important to note that Eqs. \eqref{hameq} are the differential equations for the trajectory on which the energy $\mathcal{H}$ is conserved.

\subsubsection{\bf\emph{N-body (quantum) molecular qynamics}}
\vskip -3mm
This approach can be easily extended toward several mutually interacting particles and also toward quantum mechanics. The starting point for finding the time evolution of a classical $N$-body system is the $N$-body Hamiltonian 
\begin{equation}
 \mathcal{H} = \sum_i^N\frac{{\bf p}_i^2}{2m} + \sum_{i \ne j}^N V( {\bf q} _i, {\bf p} _i,{\bf q} _j, {\bf p} _j),
\end{equation}
where  $V( {\bf q}_i, {\bf p}_i,{\bf q}_j, {\bf p}_j)  $ is the two-body potential between particles $i$ and $j$. For a given initial conditions $({\bf q} _i (t=0)$ , ${\bf p} _i (t=0))$ the Hamilton-Jacobi equations \eqref{ham} can be solved analytically or numerically. This approach has been extended in the 1980s toward the quantum molecular dynamics (QMD) approach, a theory which has been successfully applied to simulate heavy ion reactions in the energy range of 50 $A$MeV $\le E_{kin} \le 2$ $A$GeV \cite{Aichelin1991}. This approach allowed clarification of the origin of multifragmentation \cite{LeFevre2009}, the production of mesons close to threshold \cite{Hartnack2012}, and the equation of state of hadronic matter \cite{Hartnack2006} well above normal nuclear matter density. It is based on a time-dependent version of the Ritz variational principle and starts out from a trial wave function of Gaussian form. The Wigner density of this trial wave function is defined on the phase space and has the form
\begin{equation}
  \begin{aligned}
    f({\bf q}_i, {\bf p}_i, t) \propto
          &\exp \left( -\frac{\left[{\bf q}_i - {\bf q}_i^0(t)\right]^2}{L} \right) \\
    \times&\exp \left( -      \left[{\bf p}_i - {\bf p}_i^0(t)\right]^2  L  \right).
  \end{aligned}
  \label{wigner}
\end{equation}
By assuming now that the wave function of the $N$-body system is a product of the single-particle wave functions and that the centroids ${\bf q}_i^0(t)$ and ${\bf p}_i^0(t)$ of the Gaussians depend on time whereas the width is constant, the variational principle gives the following equations of motion:
\begin{equation}
  \frac{d \bf q_i^0}{d t} = \frac{\partial \langle \mathcal{H} \rangle}{\partial \bf p_i^0},
  \quad
  \frac{d \bf p_i^0}{d t} =-\frac{\partial \langle \mathcal{H} \rangle}{\partial \bf q_i^0},
  \label{ham1}
\end{equation}
with $\langle \mathcal{H} \rangle$ being the expectation value of the Hamiltonian with respect to the trial wave function. For details we refer to \cite{Aichelin1991}.

\subsection{\texorpdfstring{Relativistic phase space and transformations\\between inertial systems}{Relativistic phase space and transformations between inertial systems}}
\vskip -1mm
\subsubsection{\bf\emph{Minkowski phase space}}
\vskip -3mm
One may have the idea that the equations of motion for relativistic particles can be obtained by replacing in Eq. \eqref{hameq} the three-dimensional vectors ${\bf q}$ and ${\bf p}$ by four-dimensional vectors $q^\mu$ and $p^\mu$. This is, however, not true for the following reasons:
\begin{itemize}
  \itemsep0em
  \item[(i)] Replacing in Eq. \eqref{ham} ${\bf q}$ by $q^\mu$ and  ${\bf p}$ by $p^\mu$ one finds an equation which is not covariant because $\mathcal{H}$ is the zero component of the energy-momentum four-vector.
  \item[(ii)] Equation \eqref{ham} contains a derivative with respect to the time $t$. In a relativistic theory the time is just the zero component of the space-time four-vector. For $N$ particles we have furthermore $N$ different times and it is not evident how these times are related to the variable $t$ of Eq. \eqref{ham}.
  \item[(iii)] These equations describe the motion of particles in an eight-dimensional phase space in which neither is the energy conserved nor are the times of the different particles synchronized. In a molecular dynamics approach we are interested in obtaining physical trajectories in a $(6N+1)$-dimensional phase space (${\bf q}_i(\tau),{\bf p}_i (\tau)$), i.e., world lines of the particles, and we want to know at which position in coordinate and momentum space the particle is located for a given value of the time evolution parameter $\tau$ (whose nature will be discussed later).
\end{itemize}

Thus a Hamiltonian in a nonrelativistic sense (the total energy of the system) has no place in a relativistic approach. If we talk later of a ``Hamiltonian" in relativistic dynamics and of time evolution equations which have a form similar to Eq. \eqref{ham} the meaning of the different terms in this equation will be completely different compared to that in a non-relativistic theory. 

The starting point is the definition of the four-position and four-momentum coordinates ($q^\mu_k,p^\mu_k$) as \emph{canonical} variables which obey
\begin{equation}
  \{ q^\mu_a,q^\nu_b \} = \{ p^\mu_a,p^\nu_b \} = 0,
  \quad
  \{ q^\mu_a,p^\nu_b \} = \delta_{ab} \ g^{\mu\nu},
  \label{pq}
\end{equation}
with $g^{\mu\nu}$ being the Minkowski metric with the diagonal $\{ 1 , -1 , -1 , -1 \}$ and zero otherwise. Here we have introduced the Poisson brackets for four-vectors:
\begin{equation}
  \{ A,B \} = \sum_{k=1}^N
  \frac{\partial A}{\partial q_k^{\mu}}
  \frac{\partial B}{\partial p_{k \mu}} -
  \frac{\partial A}{\partial p_k^{\mu}}
  \frac{\partial B}{\partial q_{k \mu}}.
\end{equation}
Because in a dynamical system $q^\mu$ and $p^\mu$ depend on the time evolution parameters $\tau$, these quantities have to be taken at equal $\tau$.

\subsubsection{\bf\emph{Poincar\'e Group and Algebra}}
\vskip -3mm
Relativistic theories have to be invariant under Lorentz transformations $\Lambda$ and space-time translations $a$. Both transformations form the Poincar\'e group with the group element $R(\Lambda,a)$.  It consists of all transformations of the form 
\begin{equation}
  R(\Lambda ,a):  q^\mu \to {q'}^\mu = R(\Lambda ,a) q^\mu;
  \quad
  {q'}^\mu = \Lambda_\nu^\mu q^\nu + a^\mu,
\end{equation}
which leave the scalar product between two four-vectors unchanged:
\begin{equation}
 {q'}_\mu {q'}^{\mu} = q_\mu q^\mu,
\label{norm}
\end{equation}
with $q^\mu =(t,{\bf q})$ and $q_\mu = g_{\mu\nu} q^\nu$. 

\setlength{\belowdisplayskip}{5pt} \setlength{\belowdisplayshortskip}{5pt}
\setlength{\abovedisplayskip}{5pt} \setlength{\abovedisplayshortskip}{5pt}   

The algebra associated  with the continuous symmetry group is given by the algebra of the generators of infinitesimal transformations. Finite transformations can be built with help of the infinitesimal ones. To determine the algebra of the Poincar\'e group we start from a Lorentz transformation which differs only infinitesimally from the neutral element $R(\openone,0)$:
\begin{equation}
  \Lambda_\nu^\mu = \delta^\mu_\nu + \Delta \omega^\mu_\nu 
  \label{lt}
\end{equation}
with $\Delta \omega$ being small. The invariance of the scalar product of four-vectors under a Lorentz transformation can be expressed as
\begin{equation}
\begin{aligned}
  {q'}_{\mu} {q'}^{\mu}&= {q'}^\mu g_{\mu\nu} {q'}^\nu = \Lambda^\mu_\sigma q^\sigma g_{\mu\nu} \Lambda^\nu_\rho q^{\rho}\\
                       &=    q^\mu g_{\mu\nu}    q^\nu = q_\mu q^\mu
\end{aligned}
\end{equation}
and hence
\begin{equation}
\begin{aligned}
g_{\sigma\rho}
&= \Lambda^\mu_\sigma g_{\mu\nu} \Lambda^\nu_\rho\\
&= g_{\mu\nu} (\delta^\mu_\sigma +\Delta \omega^\mu_\sigma)(\delta^\nu_\rho +\Delta \omega^\nu_\rho)\\
&= g_{\sigma\rho} + \Delta \omega_{\sigma \rho} + \Delta \omega_{\rho \sigma} + {\cal O}(\Delta \omega^2).
\end{aligned}
\label{gmn1}
\end{equation}
Consequently $\Delta \omega_{\mu \nu}$ has to be  antisymmetric. There are six independent elements which satisfy 
\begin{equation}
\Delta \omega_{\sigma \rho} = - \Delta \omega_{\rho \sigma}.
\end{equation}
In matrix form we can write the infinitesimal Lorentz transformation as
\begin{equation}
\begin{aligned}
\Lambda (\Delta \omega_{\mu\nu})
&= \openone - \frac{i}{2}\Delta \omega_{\mu\nu} \hat M^{\mu\nu}\\
&= \openone + \frac{1}{2}\Delta \omega_{\mu\nu} (q^\mu \partial^\nu - q^\nu \partial^\mu)
\end{aligned}
\label{lg}
\end{equation}
(where the factor $\frac{1}{2}$ is convention in order to obtain the standard definition of the angular momentum $J_i=\frac{1}{2}\epsilon_{ijk}M_{jk}$), where $\hat M_{\mu\nu} = - \hat M_{\nu\mu}$ are the generators of the Lorentz group, and we find [compare Eq. \eqref{lt}]
\begin{equation}
\begin{aligned}
\Lambda (\Delta \omega_{\mu\nu} ) q^\sigma
&= q^\sigma + \frac{1}{2}(\Delta \omega_{\mu\sigma} - \Delta \omega_{\sigma\mu}) q_\mu \\
&= q^\sigma + \Delta \omega^\sigma_\mu q^\mu.
\end{aligned}
\end{equation}
Similarly for the infinitesimal translation
\begin{equation}
 T(\Delta a) = \openone + i\Delta a^\mu P_\mu = \openone +  \Delta a^\mu \partial_\mu
\label{tg}
\end{equation}
we find
\begin{equation}
\begin{aligned}
  {q'}^\nu = T(\Delta a) q^\nu = q^\nu + \Delta a^\nu .
\end{aligned}
\end{equation}
If the system is composed of several particles we find for the generators for the translation group 
\begin{equation}
  P^\mu = \sum_k^N p_k^\mu,
  \label{bigP}
\end{equation}
and for the Lorentz group [$SL(n=2,\mathbb{C}) \rightarrow \textrm{dim} = 2 (n^2 - 1) = 6$]
\begin{equation}
  M^{\mu\nu} = \sum_{k=1}^N q_k^\mu p_k^\nu - q_k^\nu p_k^\mu.
\end{equation}
These 10 generators respect the algebra of the group, which is called Poincar\'e algebra:
\begin{equation}
  \begin{split}
  [P_\mu      , P_\nu         ]&= 0, \quad
  [M_{\mu \nu}, P_\rho        ] = g_{\mu \rho} P_\nu  - g_{\nu\rho} P_\mu,\\
  [M_{\mu\nu} , M_{\rho\sigma}]&= g_{\mu\rho} M_{\nu\sigma} - g_{\mu\sigma} M_{\nu\rho}
                                - g_{\nu\rho} M_{\mu\sigma} + g_{\nu\sigma} M_{\mu\rho}.
  \end{split}
  \label{poinal}
\end{equation}
This can be directly verified by going back to the definition, Eqs. \eqref{lg} and \eqref{tg}, and calculating the brackets. The generators $M_{\mu\nu}$ and $P_\mu$ do not commute. Physically, this comes from the fact that there is a length contraction in the Lorentz boost (and a time dilatation). 

The generator of a Poincar\'e transformation is given by the combination of that of the Lorentz transformation and of a translation:
\begin{equation} 
  G = \frac{1}{2} \omega^{\mu\nu} M_{\mu\nu} - a^\mu P_\mu.
  \label{gmn}
\end{equation}
If two inertial frames ${\cal O}$ and ${\cal O}'$ are connected by an infinitesimal element of the Poincar\'e group, $R(\Lambda, a)$, then the space-time coordinates of the same event in ${\cal O}$ and ${\cal O}'$ are related by
\begin{equation}
  {q'}^\mu = q^\mu + \{q,G\}
           = q^\mu + \omega^\mu_\nu q^\nu + a^\mu
           =        \Lambda^\mu_\nu q^\nu + a^\mu.
  \label{poibra}
\end{equation}

\subsubsection{\bf\emph{Reduction of the dimension of the phase space}}
\vskip -3mm
Relativistic theories are based on four-vectors whose transformation between two inertial systems is given by elements of the Poincar\'e group. As a consequence the phase space of an $N$-particle system no longer has $6N$ dimensions as in nonrelativistic dynamics but $8N$. World lines are given by (${\bf q}_i(\tau),{\bf p}_i (\tau)$) and therefore physical trajectories (position and momentum of the particles as a function of the time $\tau$) have $6N+1$ dimensions. Thus we need \emph{constraints} to reduce the number of degrees of freedom in the relativistic phase space. After an introduction to the $8N$ dimensional phase space and to the Poincar\'e group and algebra we will illustrate the reduction of the degrees of freedom first for the example of one free particle and then we extend systematically the approach to $N$ interacting particles.

\subsection{From 1 to N-body relativistic system}
\vskip -1mm
\subsubsection{\bf\emph{The case of 1 free particle}}
\vskip -3mm
We start with the most simple case of one free particle \cite{Sudarshan1981}. The Hamilton equations for the time evolution of a nonrelativistic particle determine the trajectory in phase space for which energy is conserved. This suggests defining a constraint, the mass shell constraint, which for a noninteracting particle is
\begin{equation}
  K = p^\mu p_\mu - m^2 = 0.
  \label{const3}
\end{equation}
This constraint reduces the phase space from eight to seven dimensions by relating the energy of the particle with its three-momentum (and also with its position if we include a potential). It defines therefore the seven-dimensional subspace $\Sigma$ of the eight-dimensional phase space on which this condition is fulfilled. Because $K$ is a Poincar\'e invariant quantity, we find
\begin{equation}
\{K, M_{\mu\nu}\} = 0 , \quad \{K,P_\mu\}=0.
\label{const4}
\end{equation}
Of course the seven-dimensional phase space region $\Sigma$ is also Poincar\'e invariant:
\begin{equation}
R(\Lambda,a) \Sigma = \Sigma.
\end{equation}
The trajectory in phase space on which this constraint is satisfied is given by the solution of
\begin{equation} 
  \begin{aligned}
    &\frac{d q^\mu(\tau)}{d \tau} = \lambda \{ q^\mu(\tau),K \},\\
    &\frac{d p^\mu(\tau)}{d \tau} = \lambda \{ p^mu(\tau),K \},
  \end{aligned}
  \label{ham00}
\end{equation}
with the initial condition $q(0)=q_0$ and $p(0)=p_0$. $\lambda$ is a free parameter. In order to associate to each value of $\tau$ {\it one} point in phase space $( q(\tau), p(\tau))$ or, in other words, in order to create a worldline a second constraint, $\chi(q^\mu,p^\mu,\tau)= 0$, has to be employed to fix $\lambda$. It relates the time  $q^0$ of the particle with a Lorentz-invariant system time $\tau$.  This time constraint $\chi$ has been chosen quite differently in the literature, giving quite different time evolution equations. The subspace we are interested in is determined by a conserved $\chi$ and $K$ constraint. This is expressed by
\begin{equation} 
  \frac{d \chi}{d \tau} = \frac{\partial \chi}{\partial \tau}
  + \lambda \{ \chi(\tau),K \}=0.
  \label{ham0}
\end{equation}
This equation determines $\lambda$ as
\begin{equation}
  \lambda = - \frac{\partial \chi}{\partial \tau} \{\chi,K \}^{-1}.
\label{vcon}
\end{equation} 
$\lambda$ depends therefore on the choice of the constraint $\chi$. Formally, we can define 
\begin{equation}
  \mathcal{Z} = \lambda K = - \frac{\partial \chi}{\partial \tau} \{\chi,K \}^{-1} K
\label{hamrel}
\end{equation}
and obtain a time evolution equation for a phase space function $f$,
\begin{equation}
  \frac{d f }{d \tau}
  = \frac{\partial f}{\partial \tau} + \lambda \{ f,K \}
  = \frac{\partial f}{\partial \tau} + \{ f,\mathcal{Z} \},
  \label{ham11}
\end{equation}
an equation which is formally identical with the nonrelativistic evolution equation \eqref{phsfct} but $\mathcal{Z} $ is not the classical Hamiltonian but given by Eq. \eqref{hamrel}.

How to treat a Hamilton system with constraints has been developed by Dirac \cite{Dirac1950}. To determine the time evolution for any function of the phase-space variables along the trajectory determined by the two constraints $\phi_1=K$ and $\phi_2=\chi$ is given by the Dirac bracket, which is defined as
\begin{equation}
  \{ A,B \}_D = \{ A,B \} - \{ A,\phi_i \} C_{ij} \{ \phi_j,B \},
  \label{dirac_bracket}
\end{equation}
with the matrix of these constraints
\begin{equation}
  C_{ij}^{-1} = \{ \phi_i,\phi_j \}.
\end{equation}
\vskip 2mm
On the hypersurface, where the constraints are fulfilled, Dirac brackets and Poisson brackets agree. Dirac introduced the symbol $\approx$ to describe that two functions are identical at the subspace defined by the constraints: $\{A,B\}_D \approx \{A,B\}$. For our example we find
\begin{equation}
  \{A,B\}_D = \{A,B\} - \frac{\{A,K\} \{\chi,B\}}{\{K,\chi\}}
                      - \frac{\{A,\chi\} \{K,B\}}{\{\chi,K\}}.
  \label{db}
\end{equation}
The Dirac brackets of the 10 generators of the Poincar\'e group yield the same result as the Poisson brackets, Eq. \eqref{poinal}, because $K$ commutes with them. Therefore we can use also the Dirac brackets to construct a transformation between the two inertial systems $\mathcal{O}$ and $\mathcal{O}'$. This transformation we call $R^*(\Lambda,a)$. Both transformations, $R(\Lambda,a)$ as well as $R^*(\Lambda,a)$, therefore map $\Sigma$ to $\Sigma$ but $R^*(\Lambda,a)$ and $R(\Lambda,a)$ map the same point in  $\mathcal{O}$ to different points in  $\mathcal{O}'$ . Because $\{G,\chi\}_D=0 $, $\chi$ is unchanged under a transformation $R^*(\Lambda,a)$ and the Dirac brackets transform a phase-space point on $\mathcal{O}$ to a phase-space point on $\mathcal{O}'$ which has the same value of $\tau$. For the transformation using Poisson brackets this is generally not the case. Therefore the Dirac brackets are the proper tool to determine world lines in the two inertial frames \cite{Kihlberg1981}. Using Eq. \eqref{poibra} and replacing the Poisson bracket $\{ \cdot,\cdot \}$ by the Dirac bracket $\{ \cdot,\cdot \}_D$ we find the \emph{canonical} transformation between two inertial systems:
\begin{equation}
  {q'}^\mu(\tau) = q^\mu(\tau) + \{q^\mu(\tau),G \}_D,
  \label{ddieom}
\end{equation}
where $G$ is given by Eq. \eqref{gmn}. If we use the Poisson brackets for the transformation between the two inertial systems we obtain the \emph{geometrical} transformation
\begin{equation}
  \begin{aligned}
    {q'}^\mu(\tau)&= q^\mu(\tau + \Delta \tau) + \{q^\mu(\tau+\Delta \tau),G\} \\
            &\approx q^\mu(\tau) + \frac{d q^\mu}{d \tau} \Delta \tau
            + \{q^\mu(\tau),G\}.
  \end{aligned}
  \label{poieom}
\end{equation}
Applying the general equation \eqref{db} we can relate $\{q^\mu,G\}_D$ and $\{q^\mu,G\}$ ($\{K,G\}=0$):
\begin{equation}
  \{q^\mu,G\}_D = \{q^\mu,G\} - \frac{\{q^\mu,K\}\{\chi,G\}}{\{K,\chi\}}.
  \label{dipoi}
\end{equation}
Using furthermore the time evolution equation \eqref{ham11} we find 
\begin{equation}
  \frac{d q^\mu}{d \tau} =
  - \frac{\partial \chi}{\partial \tau} \frac{\{ q^\mu,K \}}{\{\chi,K \}}
  \label{ham17}
\end{equation}
and therefore Eq. \eqref{dipoi} can be rewritten in the form
\begin{equation}
  \{q^\mu,G\}_D = \{q^\mu,G\} - \{\chi,G\} \left(\frac{\partial \chi}{\partial \tau}\right)^{-1} \frac{d q^\mu}{d \tau} .
  \label{dipoii}
\end{equation}
Consequently, if we can ensure that 
\begin{equation}
  \{\chi,G\} \left(\frac{\partial \chi}{\partial \tau}\right)^{-1} = \Delta \tau
  \label{wlc}
\end{equation}
the transformation between two inertial systems using Dirac brackets [the canonical transformation, Eq. \eqref{ddieom}] becomes identical to that using Poisson brackets ]the geometrical transformation, Eq. \eqref{poieom}]. If the condition \eqref{wlc} is fulfilled the world lines of particles remain the same under the two transformations. They are therefore frame independent. This requirement of frame independence of the trajectories is called the world line condition (WLC). 

The constraint $\chi$ determines the time evolution of the system and therefore whether the world line condition is fulfilled. If we impose the constraint $\chi = q^0 - \tau = 0$ \cite{Sudarshan1981} we find for the time evolution of $q^\mu$ Eq. \eqref{ham17}
\begin{equation}
  \frac{d q^\mu}{d \tau} = \lambda \{ q^\mu,K \} = \frac{p^\mu}{p^0},
  \label{ham112}
\end{equation}
whereas for the condition $\chi = x_\mu p^\mu - m \tau = 0$ \cite{Rohrlich1979} we obtain
\begin{equation}
  \frac{d q^\mu}{d \tau} = \lambda \{ q^\mu,K \} = \frac{p^\mu}{m}.
  \label{ham113}
\end{equation}
In both cases we have $d p^\mu / d \tau =0$ compatible with the fact that we have a single free particle. The different time evolution equations remind us that $\tau$ is a parameter introduced by the constraint $\chi$ and not an independently defined time. Thus the time evolution of a relativistic system is only determined after the constraint $\chi$ is imposed. Different choices of the constraint yield different time evolutions of the system.

\vspace{-1cm}
\subsubsection{\bf\emph{Extension to 2 interacting particles}}
\vskip -3.5mm
The above discussed construction of world lines on which a particle moves independent of the chosen reference system has been extended to a larger number of particles in \cite{Rohrlich1979,Sudarshan1981,Samuel1982}. For a system with two interacting particles \cite{Kihlberg1981} the mass shell constraints [Eq. \eqref{const3}] have the form
\begin{equation}
  \begin{aligned}
    K_1&= p_1^\mu p_{1\mu} - m^2 + V = 0, \\
    K_2&= p_2^\mu p_{2\mu} - m^2 + V = 0
  \end{aligned}
  \label{con2}
\end{equation}
in order to have reference-frame-independent world lines. In addition, they have to be \emph{first-class} constraints in the notation of Dirac \cite{Dirac1950} :
\begin{equation}
\{K_1,K_2\} = 2 \left(  p_1^\mu \frac{\partial}{\partial q_1^\mu}
                      - p_2^\mu \frac{\partial}{\partial q_2^\mu} \right) V = 0,
\label{kt}
\end{equation}
a condition which can be fulfilled if the potential $V$ depends on $q_T^\mu$ \cite{Sudarshan1981}, which is the part of $q^\mu = q_1^\mu - q_2^\mu$ which is transverse with respect to the center-of-mass motion $P^\mu = p_1^\mu + p_2^\mu$, and which is defined as
\begin{equation}
  q_T^\mu = q^\mu - \frac{q_\nu P^\nu}{P^2} P^\mu.
\end{equation}
Poincar\'e transformations map the $7N$ dimensional phase space, on which the constraints [Eqs. \eqref{con2}] are fulfilled, on itself. The evolution equations can be extended to
\begin{equation}
  \begin{aligned}
    \frac{d q_i^\mu}{d \tau} &= v_1 \{ q_i^\mu,K_1 \} + v_2 \{ q_i^\mu,K_2 \}, \\
    \frac{d p_i^\mu}{d \tau} &= v_1 \{ p_i^\mu,K_1 \} + v_2 \{ p_i^\mu,K_2 \}  
  \end{aligned}
  \label{ceq1}
\end{equation}
with arbitrary parameters $v_1$ and $v_2$. For an interacting system the time components of particles $q_i^0$ become connected  by the potential term and consequently the spatial position of each particle $q_i^k$ depends on both times $q_1^0$ and  $q_2^0$. This does not correspond to a world line but to a sheet. To obtain a world line we have to synchronize first the times of both particles by a constraint without any parameter,
  \begin{equation}
  \chi_{1}(q_1,q_2,p_1,p_2) = 0,
  \label{chi2}
\end{equation}
and finally to connect the synchronized times to a clock time $\tau$,
\begin{equation}
  \chi_{2}(q_1,q_2,p_1,p_2,\tau) = 0,
\label{con1}
\end{equation}
with the property $\det \{K_i,\chi_j\} \neq 0$. If $\{K_i,\chi_j\}=0$ we cannot assign to each point on the trajectory uniquely a value of the parameter $\tau$. These two additional constraints reduce the $7N$-dimensional phase space to a $6N$-dimensional phase space with a parameter $\tau$  so effectively to a $(6N+1)$-dimensional phase space \cite{Mukunda1981}). Condition \eqref{con1} allows for fixing the free parameters $v_i$ of Eq. \eqref{ceq1} [see Eq. \eqref{ham11}]:
\begin{equation}
  \frac{d \chi_2(\tau) }{d \tau}
  = \frac{\partial \chi_2(\tau)}{\partial \tau} + \{\chi_2(\tau),K_i \} v_i =0
\end{equation}
yields
\begin{equation}
  v_i = - \{\chi_2(\tau),K_i \}^{-1} \frac{\partial \chi_2(\tau)}{\partial \tau}.
  \label{vfct}
\end{equation}
Consequently,  the general evolution equation for a phase-space function $f$ is 
\begin{equation}
  \frac{d f}{d \tau} = \frac{\partial f}{\partial \tau} - S_{i2} \frac{\partial \chi_2}{\partial \tau} \{f,K_i\},
\end{equation}
with
\begin{equation}
  S_{ij} = \{\chi_j,K_i\}^{-1}.
  \label{inv}
\end{equation}

As in the one particle case the WLC requires that the two transformations between the inertial systems, the one expressed by Dirac brackets [Eq. \eqref{db} (canonical transformation)],
\begin{equation}
  {q'}_i^\mu(\tau)= q_i^\mu(\tau) + \{q_i^\mu(\tau),G\}_D,
\end{equation}
and the one expressed by Poisson brackets (geometrical transformation),
\begin{equation}
  \begin{aligned}
    {q'}_i^\mu(\tau)&= q_i^\mu(\tau + \Delta \tau_i) + \{q^\mu(\tau+\Delta \tau_i),G\} \\
              &\approx q_i^\mu(\tau) + \frac{d q_i^\mu}{d \tau} \Delta \tau_i + \{q_i^\mu(\tau),G\},
  \end{aligned}
\end{equation}
lead to points on the same world lines. Employing the Dirac brackets with Eq. \eqref{inv} and taking advantage of $\{K_i,G\} = 0$ we find
\begin{equation}
  \begin{aligned}
    \frac{d q_i^\mu(\tau)}{d \tau} \Delta \tau_i
    &= \{q_i^\mu(\tau),K_j\} S_{lj} \{\chi_l,G\}\\
    &= \{q_i^\mu(\tau),K_j\} S_{lj} \frac{d \chi_l}{d \tau} \Delta \tau_i,
  \end{aligned}
  \label{twobodycond}
\end{equation}
or, with help of Eqs. \eqref{ceq1} and \eqref{vfct} if $\{q_i^\mu(\tau),K_j\}\ne 0$ and $S_{lj} \ne 0$,
\begin{equation}
\{\chi_l,G\} = \frac{d \chi_l}{d \tau} \Delta \tau_i.
\end{equation}
In reality the last equation poses two conditions: $\Delta \tau_1=\Delta \tau_2$ and that $\chi_1$, which does not depend on $\tau$, is Poincar\'e invariant to fulfill  ${\{\chi_1,G\}} =0$. These conditions cannot be fulfilled by every choice of $\chi_i$. Indeed, if we relate the $\tau$ to the fourth component of $q^\mu$, the instant form of Dirac \cite{Dirac1950},
\begin{equation}
  \chi_1 = \tfrac{1}{2} (q_1^0 - q_2^0) = 0,
  \quad
  \chi_2 = \tfrac{1}{2} (q_1^0 + q_2^0) - \tau = 0,
\end{equation}
we recover ${\{\chi_1,G\}} \neq 0$ and hence the \emph{no-interaction theorem} stating that relativistic molecular dynamics can only be formulated for non-interacting particles \cite{Sudarshan1981}. If, on the other hand, the $\chi_i$ are defined kinematically as
\begin{equation}
  \chi_1 = \tfrac{1}{2}    q^\mu            U_\mu        = 0,
  \quad
  \chi_2 = \tfrac{1}{2} (q_1^\mu + q_2^\mu) U_\mu - \tau = 0,
  \label{twopartchi}
\end{equation}
with $U_\mu = P_\mu / \sqrt{P^2}$, which gives $U_\mu = (1,\vec 0)$ in the center of mass of two particles, the world line condition can be fulfilled \cite{Sudarshan1981} by setting $\Delta \tau_i = -\{\chi_2,G\}$. With the latter time constraints, Eqs. \eqref{twopartchi}, we can compute the $v_i$ [Eq. \eqref{vfct}]. We start out from the matrix of constraints :
\begin{equation}
  \begin{aligned}
  S_{ij}^{-1}&=
  \begin{pmatrix}
    \{ \chi_1 , K_1 \} & \{ \chi_1 , K_2 \} \\[1.5mm]
    \{ \chi_2 , K_1 \} & \{ \chi_2 , K_2 \}
  \end{pmatrix}\\
  &=
  \begin{pmatrix}
    p_1^\mu U_\mu & - p_2^\mu U_\mu \\[1.5mm]
    p_1^\mu U_\mu &   p_2^\mu U_\mu
  \end{pmatrix},
   \end{aligned} 
\end{equation}
which can be inverted :
\begin{equation}
  S_{ij}=
  \begin{pmatrix}
    ( 2 \ p_1^\mu U_\mu )^{-1} & ( 2 \ p_1^\mu U_\mu )^{-1} \\[1.5mm]
   -( 2 \ p_2^\mu U_\mu )^{-1} & ( 2 \ p_2^\mu U_\mu )^{-1}
  \end{pmatrix},
\end{equation}
and the parameters $v_i$ become
\begin{equation}
  \begin{split}
    v_1&= ( 2 \ p_1^\mu U_\mu )^{-1} \stackrel{\text{cms}}{=} \frac{1}{2 E_1}, \\
    v_2&= ( 2 \ p_2^\mu U_\mu )^{-1} \stackrel{\text{cms}}{=} \frac{1}{2 E_2}.
  \end{split}
\end{equation}
We obtain then the equations of motion for 2 interacting particles in their center of mass:
\begin{equation}
  \frac{d q_i^\mu}{d \tau} = \frac{p_i^\mu}{E_i},
  \quad
  \frac{d p_i^\mu}{d \tau} = - \sum_{k=1}^2 \frac{1}{2 E_k} \frac{\partial V (q_T)}{\partial {q_i}_\mu}.
  \label{twoparteom}
\end{equation}
We can easily see that the classical non-relativistic limit of these equations gives the same result as QMD by taking  ${\bf p} \ll m$.

In this example of two interacting particles we can also address another problem: the \emph{separability} of clusters. In contradiction to nonrelativistic dynamics this separability is not trivially fulfilled by taking a potential which vanishes for large distances because the potential enters the constraint matrix which determines the time evolution. Cluster separability means that we have the equations of motion of two free particles if the distance between them is large. 

\subsubsection{\bf\emph{Extention of  the formalism}}
\vskip -3mm

For the two body case, for which $\{K_i,K_j\}=0$, the formalism has been developed in the last section. Here we extend the formalism to $N>2$, where  $\{K_i,K_j\}$ may be different from 0. In this case the time evolution equations for
the $2N$ constraints are
\begin{equation}
  \begin{aligned}
  \frac{d \phi_i}{d \tau}
  &= \frac{\partial \phi_i}{\partial \tau} + \sum_k^{2N} \lambda_k \{ \phi_i,\phi_k \} = 0\\
  &= \frac{\partial \phi_i}{\partial \tau} + \sum_k^{2N} C_{ik}^{-1} \lambda_k = 0.
  \end{aligned}
  \label{const_conv}
\end{equation}
with
\begin{equation}
  \phi_k =
  \begin{cases}
   &   K_k(q^\mu,p^\mu)      = 0 \text{ for } 1 < k < N, \\
   &\chi_k(q^\mu,p^\mu)      = 0 \text{ for } N+1 < k < 2N-1, \\
   &\chi_N(q^\mu,p^\mu,\tau) = 0.
  \end{cases}
\end{equation}
Only the constraint $i=2N$ depends on $\tau$. Rewriting the last line of Eq. (\ref{const_conv}) as
\begin{equation}
\sum_k^{2N} C_{ik}^{-1} \lambda_k = - a_i
\end{equation}
with $a_i$ being $\frac{\partial \phi_i}{\partial \tau}$ and hence a vector in which only the $2N^{th}$ component is different from zero we find
\begin{equation}
  \lambda_k = - \sum_i^{2N} C_{ki} a_i = - C_{k2N} \frac{\partial \phi_{N}}{\partial \tau}.
  \label{lambda}
\end{equation}
Also this equation shows that different choices of constraints will yield different values of $\lambda_k$ and different $\lambda_k$ will give a different time evolution. Therefore the relativistic kinematics is only defined after the constraints are defined. By defining
\begin{equation}
  \mathcal{Z} = \sum_k^{2N} \lambda_k \phi_k
\end{equation}
the trajectory in phase space for which the constraints are fulfilled is given by 
\begin{equation}
  \frac{d q^\mu_i }{d \tau } = \{ q^\mu_i(\tau),\mathcal{Z} \},
  \quad
  \frac{d p^\mu_i }{d \tau } = \{ p^\mu_i(\tau),\mathcal{Z} \}.
  \label{eom}
\end{equation}
and therefore have a form which reminds us of the Hamilton-Jacobi equations. This similarity can be easily understood by recalling that this is the genuine form of trajectories under the conditions that constraints are conserved. The non relativistic equations of motion lead to trajectories on which the total energy of the system is conserved whereas Eqs. \eqref{eom} lead to trajectories on which the constraints $\phi_k$ are conserved.

\subsubsection{\bf\emph{N-particle system}}
\vskip -3mm
The N-body system is actually a trivial generalization of the 3-body problem. Therefore we will discuss here the three-body case, which was studied in detail in \cite{Sudarshan1981,Samuel1982,Sorge1989}. As compared to the two-body case we are confronted here with the new feature
that the commutation of the on-shell mass constraints $\{ K_i,K_j \}$ which is easily fulfilled in the two-particle case [Eq. \eqref{kt}], and which avoids that $\chi$ constraints that appear explicitly in the equations of motion, is not necessarily fulfilled for more than two particles.

We start out with the definition of the projector to the two-body center-of-mass system (called the frame projector in \cite{Todorov1982}),
\begin{equation}
  u_{ij}^\mu = \frac{p_{ij}^\mu}{\sqrt{p_{ij}^2}}  \stackrel{\text{cms}}{=} (1,0,0,0),
\end{equation}
where $p_{ij}^\mu = p_i^\mu + p_j^\mu$ and to the overall center-of-mass system,
\begin{equation}
  U^\mu = \frac{P^\mu}{\sqrt{P^2}}  \stackrel{\text{lab}}{=} (1,0,0,0),
\end{equation}
where $P^\mu$ is the Poincar\'e generator \eqref{bigP}. We call the latter system the laboratory system (to avoid confusion with the two-body center-of-mass system) because in collider physics the total center-of-mass system is identical to the laboratory system. Then
\begin{equation}
  \frac{\partial u_{ij}^\mu}{\partial q_k^\nu} = 0,
  \quad
  \frac{\partial u_{ij}^\mu}{\partial p_k^\nu} =
    \frac{1}{\sqrt{p_{ij}^2}} \left( g^{\mu\nu} - u_{ij}^\mu u_{ij}^\nu \right) (\delta_{ik} + \delta_{jk})
\end{equation}
and
\begin{equation}
  \frac{\partial U^\mu}{\partial q_k^\nu} = 0,
  \quad
  \frac{\partial U^\mu}{\partial p_k^\nu} =
  \frac{1}{\sqrt{P^2}} \left( g^{\mu\nu} - U^\mu U^\nu \right).
\end{equation}
We define as well the projector to the transverse distances as
\begin{equation}
  \theta^{\mu\nu} = \left( g^{\mu\nu} - u_{ij}^\mu u_{ij}^\nu \right) \stackrel{\text{cms}}{=}
  \begin{pmatrix}
    0 & 0 & 0 & 0\\[1mm]
    0 &-1 & 0 & 0\\[1mm]
    0 & 0 &-1 & 0\\[1mm]
    0 & 0 & 0 &-1
  \end{pmatrix}
\end{equation}
and
\begin{equation}
  \Theta^{\mu\nu} = \left( g^{\mu\nu} - U^\mu U^\nu \right) \stackrel{\text{lab}}{=}
  \begin{pmatrix}
    0 & 0 & 0 & 0\\[1mm]
    0 &-1 & 0 & 0\\[1mm]
    0 & 0 &-1 & 0\\[1mm]
    0 & 0 & 0 &-1
  \end{pmatrix},
\end{equation}
with
\begin{equation}
  \begin{aligned}
    {q_T}_{ij}^\mu =&q_{ij}^\sigma \theta_{\sigma\mu}
                   = q_{ij}^\mu - [(q_{ij})_\sigma u_{ij}^\sigma] u_{ij}^\mu,\\
    {q_T}_{ij}^2   =&q_{ij}^2   - [(q_{ij})_\sigma u_{ij}^\sigma]^2
  \end{aligned}
\end{equation}
and
\begin{equation}
  \begin{aligned}
    {q_T'}_{ij}^\mu =&q_{ij}^\sigma \Theta_{\sigma\mu}
                    = q_{ij}^\mu - [(q_{ij})_\sigma U^\sigma] U^\mu\\
    {q_T'}_{ij}^2   =&q_{ij}^2   - [(q_{ij})_\sigma U^\sigma]^2.
  \end{aligned},
\end{equation}
We notice the following properties:
\begin{equation}
  ({q_T}_{ij})_\mu u_{ij}^\mu = 0,
  \quad
  ({q_T}_{ij})_\mu \theta^{\mu\nu} = {q_T}_{ij}^\nu
\end{equation}
and
\begin{equation}
  ({q_T'}_{ij})_\mu U^\mu = 0,
  \quad
  ({q_T'}_{ij})_\mu \Theta^{\mu\nu} = {q_T'}_{ij}^\nu.
  \label{propqt}
\end{equation}
The derivatives of these transverse distances are relegated to the appendix.

We start the discussion of three-particle dynamics with the constraints used in the RQMD approach of ref. \cite{Sorge1989}. In this paper the authors extend the previously discussed two-body system to an $N$-body system and employ as mass shell constraints for an interacting system
\begin{equation}
  K_i = p_i^\nu p_{i \nu} - m_i^2 + V_i({q_T}_{ij}^2) = 0
\end{equation}
whereas the time constraints are defined by 
\begin{equation}
  \chi_i = \frac{\sum_{j \ne i} q_{ij}^\nu}{N} (u_{ij})_\nu = 0, \quad  1 \leq i \leq N-1.
  \label{sorgechi}
\end{equation}
and
\begin{equation}
  \chi_N = \frac{\sum_j q_j^\nu}{N} U_\nu - \tau = 0.
  \label{sorgechi1}
\end{equation}
The last constraint ensures that all times $q_j^0$ are related to the time evolution parameter $\tau$.

All these constraints fulfill the WLC \cite{Sorge1989}. In order to ensure the separability of clusters two-particle distances are weighted in RQMD with the weight function
\begin{equation}
  g_{ij} = \frac{L}{{q_T}_{ij}^2} \exp \left( \frac{{q_T}_{ij}^2}{L} \right).
  \label{weightfct}
\end{equation}

There are several problems with this approach. The first is that the Komar-Todorov (KT) \cite{Todorov1982} condition, $ \{ K_i , K_j \}=0$, is not fulfilled. This means that the mass shell constraints of different particles are not independent:
\begin{equation}
  \begin{aligned}
    \{ K_i , K_j \}&= 2    p_j^\mu \frac{\partial V_i}{\partial q_j^\mu}
                    - 2    p_i^\mu \frac{\partial V_j}{\partial q_i^\mu} + \{ V_i,V_j \} \\
                   &= 2 p_{ij}^\mu \frac{\partial V_i}{\partial q_j^\mu} + \{ V_i,V_j \} \ne 0
  \end{aligned}
  \label{ktcond}
\end{equation}
using the fact that $\partial V_i / \partial q_j^\mu = - \partial V_j / \partial q_i^\mu$. We notice that neither $V(q_T)$ nor $V(q_T')$ can fulfill this condition. In the RQMD paper \cite{Sorge1989} it is assumed that $\{ K_i , K_j \}$ remains negligible  and consequently the time constraints do not appear in the full equations of motion based on Eq. \eqref{eom}:
\begin{equation}
  \begin{aligned}
    \frac{d q_i^\mu}{d \tau}
    &= \ \ \ \sum_{k=1}^N      \lambda_k \ \frac{\partial    K_k}{\partial {p_i}_\mu}
          +  \sum_{k=N+1}^{2N} \lambda_k \ \frac{\partial \chi_k}{\partial {p_i}_\mu},\\
    \frac{d p_i^\mu}{d \tau}
    &=    -  \sum_{k=1}^N      \lambda_k \ \frac{\partial    K_k}{\partial {q_i}_\mu}
          -  \sum_{k=N+1}^{2N} \lambda_k \ \frac{\partial \chi_k}{\partial {q_i}_\mu}
  \end{aligned}
\end{equation}
because if we assume $\{ K_i , K_j \} = 0$, then $\lambda_k = 0$ for $N+1 < k < 2N$ [see Eq. \eqref{lambda}]. With this assumption and the time constraint equations \eqref{sorgechi} and \eqref{sorgechi1} the parameter $\lambda$ becomes
\begin{equation}
  \lambda_k = S_{kN},
\end{equation}
where $S_{Nk}$ is defined in Eq. \eqref{inv}. The equations of motion of \cite{Sorge1989} are then given by
\begin{equation}
  \frac{d q_i^\mu}{d \tau}
  = 2 {p_i}_\mu S_{iN},
  \quad
  \frac{d p_i^\mu}{d \tau}
  =    -  \sum_{k=1}^N S_{kN} \frac{\partial V(q_T)}{\partial {q_i}_\mu}.
\end{equation}
Even if we deal with three free particles and therefore the KT condition is trivially fulfilled, the RQMD approach of ref. \cite{Sorge1989} poses problems. The mass shell
\begin{equation}
  \begin{split}
    K_1&= {p_1}^2 - {m_1}^2 = 0, \\
    K_2&= {p_2}^2 - {m_2}^2 = 0, \\
    K_3&= {p_3}^2 - {m_3}^2 = 0
  \end{split}
\end{equation}
and time constraints
\begin{equation}
  \begin{split}
    \chi_1&= (q_{12}^\mu u_{12 \mu} + q_{13}^\mu u_{13 \mu}) / 3 = 0, \\
    \chi_2&= (q_{21}^\mu u_{21 \mu} + q_{23}^\mu u_{23 \mu}) / 3 = 0, \\
    \chi_3&= (q_1 + q_2 + q_3)^\mu U_\mu / 3 - \tau = 0
  \end{split}
\end{equation}
\begin{widetext}
give the matrix of constraints,
\vskip 4mm
\begin{equation}
  S_{ij}^{-1} =
  \begin{pmatrix}
      4/3 \ p_1^\mu (u_{12} + u_{13})_\mu &
    - 2/3 \ p_2^\mu  u_{12 \mu}           &
    - 2/3 \ p_3^\mu  u_{13 \mu}           \\[1.5mm]
    - 2/3 \ p_1^\mu  u_{12 \mu}           &
      4/3 \ p_2^\mu (u_{21} + u_{23})_\mu &
    - 2/3 \ p_3^\mu  u_{23 \mu}           \\[1.5mm]
      2/3 \ p_1^\mu  U_\mu                &
      2/3 \ p_2^\mu  U_\mu                &
      2/3 \ p_3^\mu  U_\mu
  \end{pmatrix},
\end{equation}
\vskip 4mm
\end{widetext}
whose inverse is highly non-trivial. The numerical calculation of $\lambda_k$ gives non physical trajectories with velocities above the speed of light. Moreover, if we include the weight function, Eq. \eqref{weightfct}, for the separability of clusters we encounter another numerical problem: the matrix inversion fails because the numerical values of the matrix elements cover many orders of magnitude \cite{Maruyama1991}.

Last but not least the energy is not conserved locally in time (but is conserved on average over a long time). This is because $q_T$ is used as a variable of the potential. This choice forces us to compute the forces in each two-body center-of-masses system. Subsequently, an inverse Lorentz boost has to be  applied to transform all forces into the same common frame. This transformation is ill defined because the Lorentz transformation is not valid when we study accelerated particles and indeed this transformations creates fluctuations of the energy of the system. This has been discussed in \cite{Maruyama1991} without a solution being offered.

To avoid these problems we made a different choice of constraints. Instead of formulating the constraints in the two-body systems we define them in the common center-of-mass system. These means that we have to replace the $u_{ij}^\mu$ by $U^\mu$ in the constraint formulas, which then read as
\begin{equation}
  \begin{aligned}
    &K_i = p_i^\nu p_{i \nu} - m_i^2 + V_i({q_T'}^2) = 0,\\
    &\chi_i = \frac{\sum_{j \ne i} q_{ij}^\nu}{N} U_\nu = 0,\\
    &\chi_N = \frac{\sum_j q_j^\nu}{N} U_\nu - \tau = 0.
  \end{aligned}
  \label{finalconst}
\end{equation}
We cannot fulfill the KT condition using $q_T'$ (see the appendix), but the weaker condition $\{ K_i , \sum_{i\neq j}K_j \}=0$ holds. This means that the mass shell constraint of particle $i$ commutes with the sum of those of all other particles, which is not the case in the approach of ref. \cite{Sorge1989}. As in \cite{Sorge1989} we assume that $\{ K_i , K_j \}$ is negligible. 


Our choice of constraints avoids all the other problems of the approach of \cite{Sorge1989}. We can check that by using the previous example of 3 free particles. Changing the time constraints to
\begin{equation}
  \begin{split}
    \chi_1&= (q_{12} + q_{13})^\mu U_\mu / 3 = 0, \\
    \chi_2&= (q_{21} + q_{23})^\mu U_\mu / 3 = 0
  \end{split}
\end{equation}
we obtain
\begin{equation}
  S_{ij}^{-1} =
  \begin{pmatrix}
      4/3 \ p_1^\mu U_\mu & - 2/3 \ p_2^\mu U_\mu & - 2/3 \ p_3^\mu U_\mu \\[1.5mm]
    - 2/3 \ p_1^\mu U_\mu &   4/3 \ p_2^\mu U_\mu & - 2/3 \ p_3^\mu U_\mu \\[1.5mm]
      2/3 \ p_1^\mu U_\mu &   2/3 \ p_2^\mu U_\mu &   2/3 \ p_3^\mu U_\mu
  \end{pmatrix},
\end{equation}
with the solution
\begin{equation}
  S_{ij}=
  \begin{pmatrix}
      ( 2 \ p_1^\mu U_\mu )^{-1} &
                     0           &
      ( 2 \ p_1^\mu U_\mu )^{-1} \\[1.5mm]
                     0           &
      ( 2 \ p_2^\mu U_\mu )^{-1} &
      ( 2 \ p_2^\mu U_\mu )^{-1} \\[1.5mm]
    - ( 2 \ p_3^\mu U_\mu )^{-1} &
    - ( 2 \ p_3^\mu U_\mu )^{-1} &
      ( 2 \ p_3^\mu U_\mu )^{-1}
  \end{pmatrix}.
\end{equation}
The last column is the $\lambda$ parameter which has an analytical and trivial solution
\begin{equation}
  \lambda_k = ( 2 \ p_k^\mu U_\mu )^{-1} \stackrel{\text{lab}}{=} \frac{1}{2 E_k}
\end{equation}
which is in perfect agreement with the solution we found for the 2-particle case. The equations of motion in the global (laboratory) frame, where $\sum_i^N {\bf p}_i = 0$, are then
\begin{equation}
  \begin{aligned}
    \frac{d q_i^\mu}{d \tau}&= \frac{p_i^\mu}{E_i}, \\
    \frac{d p_i^\mu}{d \tau}&= - \sum_{k=1}^N \frac{1}{E_i}
                              \frac{\partial V_k (q_T')}{\partial {q_i}_\mu}.
  \end{aligned}
  \label{finaleom}
\end{equation}
These equations conserve energy and ensure physical trajectories with velocities below the speed of light. Moreover, the analytical solution for the $\lambda_k$ is useful to avoid the numerical inversion of the matrix of constraints at each time step of the evolution, which is not possible with presently available computers. The equations of motion of Eqs. \eqref{finaleom} are finally  identical to the relativistic equations which are currently used in other approaches \cite{Bass1998}.

Our approach also avoids the problem of cluster separability. This can easily be seen by dividing the system into two subsystems $a$ and $b$ with 
\begin{equation}
  P^\mu = P_a^\mu + P_b^\mu.
\end{equation}
If we calculate the time evolution equations for the partons in each subsystem separately we obtain the  same result as if we calculate them for the full system. This means that one cluster does not influence the motion of the other. 

%% file: section_3_njl.tex
\section{Nambu-Jona-Lasinio Model}
\vskip -2mm
In this paper we study the expansion of a $q / \bar q$ plasma employing the Nambu--Jona-Lasinio (NJL) model. The NJL model is the simplest low-energy approximation of QCD. It describes the interaction between two quark currents as a pointlike exchange of a perturbative gluon \cite{Klevansky1992}. By assuming that the mass of the gluon is large compared to its momentum the interaction reduces to an effective four point interaction and is given by
\begin{equation}
  {\cal L}^{\text{int}} = \kappa \sum ^{N_c^2-1}_{c=1} \sum_{i,j}^3
  (\bar{q}_{i,\alpha} [\gamma_{\mu}\lambda^c]_{\alpha \delta} q_{i,\delta})
  (\bar{q}_{j,\gamma} [\gamma^{\mu}\lambda^c]_{\gamma  \beta} q_{j,\beta})
\end{equation}
where we have explicitly shown the color Dirac $\alpha,\beta,\gamma,\delta$ and flavor $i,j$ indices. We normalize $\sum_{i=0}^8 \lambda^i_{\alpha \beta} \lambda^i_{\beta \alpha} = 2$. By applying a Fierz transformation in color space to this interaction the Lagrangian separates into two pieces \cite{Ebert1994}: an attractive color singlet interaction between a quark and an antiquark $({\cal L}_{(q\bar{q})})$ and a repulsive color anti-triplet interaction between two quarks $({\cal L}_{(qq)})$ which disappears in the large-$N_c$ limit. Usually a six-point interaction in the form of the 't Hooft determinant is added  $({\cal L}_A)$ to break the unwanted $\text{U}_{A}(1)$ symmetry of the Lagrangian. For this study we are only interested in the color singlet channel:
\begin{equation}
  {\cal L} = {\cal L}_0 + {\cal L}_{(q\bar{q})} + {\cal L}_A.
\end{equation}
(The color octet channel gives diquarks and can be used to study baryons \cite{Gastineau2002,Gastineau2002a}.) ${\cal L}_0$ is the Lagrangian for a particle without interaction. Concentrating on the dominant scalar and pseudo scalar part in Dirac space we find the following explicit form of the Lagrangian:
\begin{equation}
  \begin{aligned}
  {\cal L} =&\sum_{f=\{u,d,s\}} \Bigg[ \bar{q}_f (i\feyn{\partial} - m^0_{f}) q_f + G_S \sum^8_{a=0} \big[   (\bar{q}_f \lambda_F^a q_f)^2\\
      &+ (\bar{q}_f i\gamma_5 \lambda_F^a q_f)^2 \big] \Bigg] - G_D \{ \det[\bar{q}_f(1-\gamma_5)q_f]\\
      &+ \det [\bar{q}_f(1+\gamma_5)q_f] \}.
  \end{aligned}
  \label{Lagr_allgemein}
\end{equation}
The first term is the free kinetic part, including the flavor-dependent current quark masses $m^0_f$, which break explicitly the chiral symmetry of the Lagrangian. The second part is the scalar-pseudoscalar interaction in the mesonic channel, invariant under $\text{SU}_A(3) \otimes \text{U}_A(1)$. It is diagonal in color as is the third part, the 't Hooft determinant. The $\det$ runs over the flavor degrees of freedom. Consequently, the flavors become connected. $G_S$ is the $q{\bar q}$ coupling constant and $G_D$ is the coupling constant of the 't Hooft term. The quarks in the NJL Lagrangian have four-point interactions (with a coupling constant $G_S$) and six-point interactions (with a coupling constant $G_D$). 

The thermodynamic properties of this Lagragian are summarized in ref. \cite{Nebauer2002}. The NJL Lagrangian has been discussed in many review articles \cite{Klevansky1992,Lutz1992,Buballa2005}, where all details of this model can be found. We concentrate here on those quantities which enter directly in our calculation, the quark and meson masses as well as the cross sections.

\subsection{Quark Masses}
\vskip -4mm
\begin{figure} [b]
  \begin{center}
    \includegraphics[width=7.6cm]{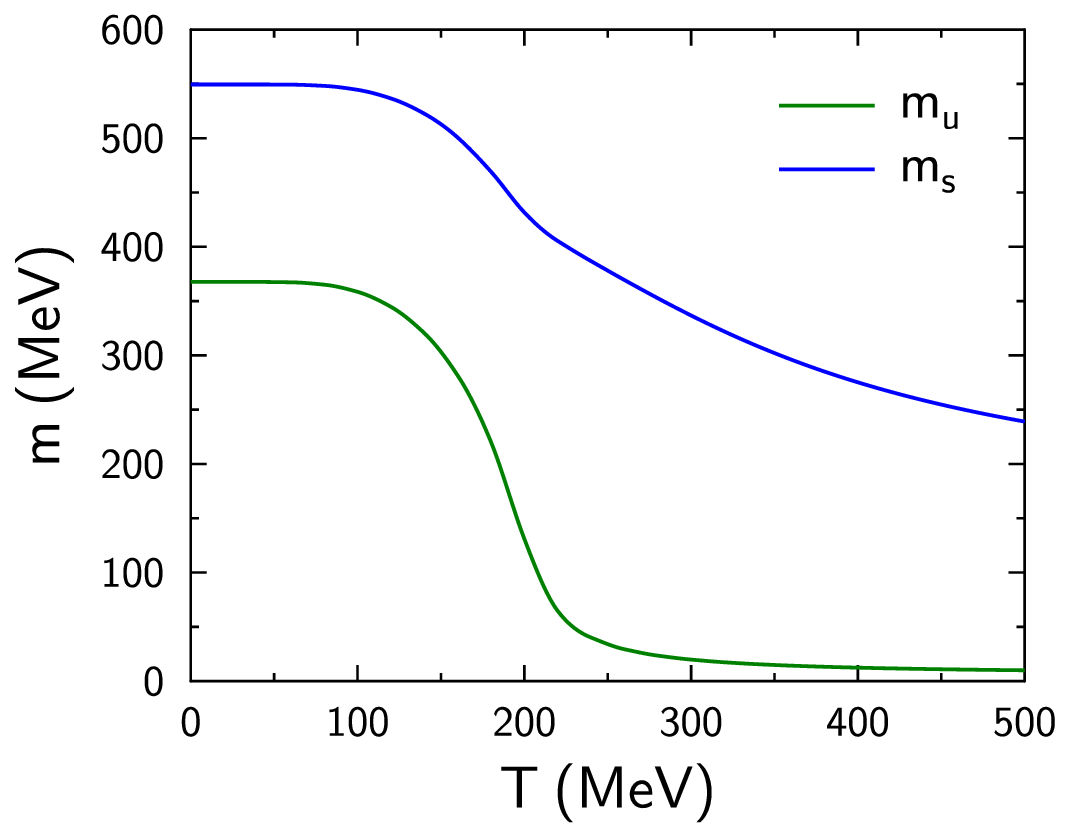}
  \end{center}
  \vskip -5mm
  \caption{(Color online )Masses of $u$ and $s$ quarks as a function of $T$.\label{qmasses}}
\end{figure}

\begin{figure*}
  \begin{center}
    \includegraphics[width=13.5cm]{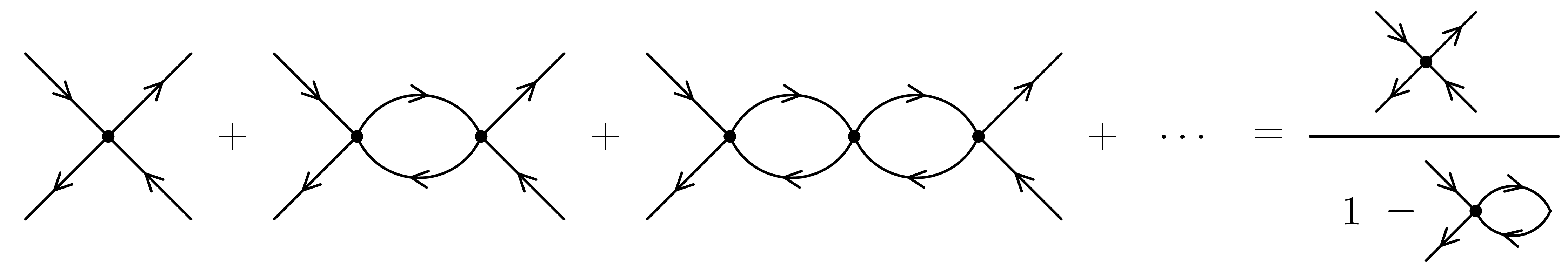}
  \end{center}
  \vskip -5mm
  \caption{Effective interaction between two quarks in the Random Phase Approximation (RPA).\label{col1}}
\end{figure*}

In the NJL model the mass of a free quark of flavor $i$ is given by \cite{Klevansky1992}
\begin{equation}
  \begin{aligned}
    m_i =& m^0_i + G_S (4N_c)(i \text{Tr } S_i) \\
         & - G_D (2N_c^2+3N_c+1)(i\text{Tr } S_j)(i\text{Tr } S_k),
  \end{aligned}
  \label{mass}
\end{equation}
where $m^0_i$ is the bare quark mass, $N_c$ is the number of colors, with $i \neq j \neq k$, and 
\begin{equation}
  \begin{aligned}
    \text{Tr } S_k&= \text{Tr } S_k(x=0) = \int^\Lambda \frac{d^4p}{(2\pi)^4} \text{Tr } S_k(p) \\
    &= \int^\Lambda \frac{d^4p}{(2\pi)^4} \text{Tr } \frac{1}{\feyn p - m_k+ i\epsilon}
     = \frac{2}{i} \int^\Lambda \frac{d^3p}{(2\pi)^3} \frac{m_k}{E_p} 
  \end{aligned}
\end{equation}
with $E_p = \sqrt{{\bf p}^2 + m_k^2}$. We use here a 3-momentum cut-off to regularize the integrals. If the quark is brought into matter with a finite baryon density $\mu$ and a finite temperature $T$, thermal field theory has to be employed and we have to replace 
\begin{equation}
  \begin{aligned}
    p = (p_0, \bf p)&\to p_n = (i\omega_n \pm \mu,\bf p), \\
    i \int^\Lambda \frac{d^4p}{(2\pi)^4}&\to - T \sum_n \int^\Lambda \frac{d^3p}{(2\pi)^3}
  \end{aligned}
\end{equation}
where $\omega_n = (2n+1)\pi T$ with $n=1,2,\dots$ are the Matsubara frequencies for fermions. Hence we find for the propagator
\begin{equation}
  \begin{aligned}
    S(p) \to {\cal S}(\omega_n ,{\bf  p})
    =&\frac{1}{\feyn p_n - m + \gamma^0 \mu} \\
    =&\frac{\feyn p  + m}{2E_p} \frac{1}{i\omega_n - (E_p - \mu)} \\
    &+ \frac{\feyn p' - m}{2E_p} \frac{1}{i\omega_n + (E_p + \mu)}
  \end{aligned}
\end{equation}
with
\begin{equation}
  \feyn p  = \gamma^0 E_p - \gamma\!\!\!\!\!\gamma{\bf p}
  \quad \text{and} \quad
  \feyn p' = \gamma^0 E_p + \gamma\!\!\!\!\!\gamma{\bf p}.
\end{equation}
This yields \cite{Klevansky1992}
\begin{equation}
  \text{Tr } S_k = \frac{2}{i} \int^\Lambda \frac{d^3p}{(2\pi)^3}
  \frac{m}{E_p} [1 - f(E_p-\mu) - f(E_p+\mu)]
\end{equation}
with the Fermi-Dirac distribution
\begin{equation}
  f(E_p \pm \mu) = \{1 + \exp((E_p \pm \mu)/T)\}^{-1}.
  \label{Fermi0}
\end{equation}
Equation \eqref{mass} allows us to calculate the quark masses which are displayed on Fig. \ref{qmasses}.

\begin{figure} [b]
  \begin{center}
    \includegraphics[width=5.4cm]{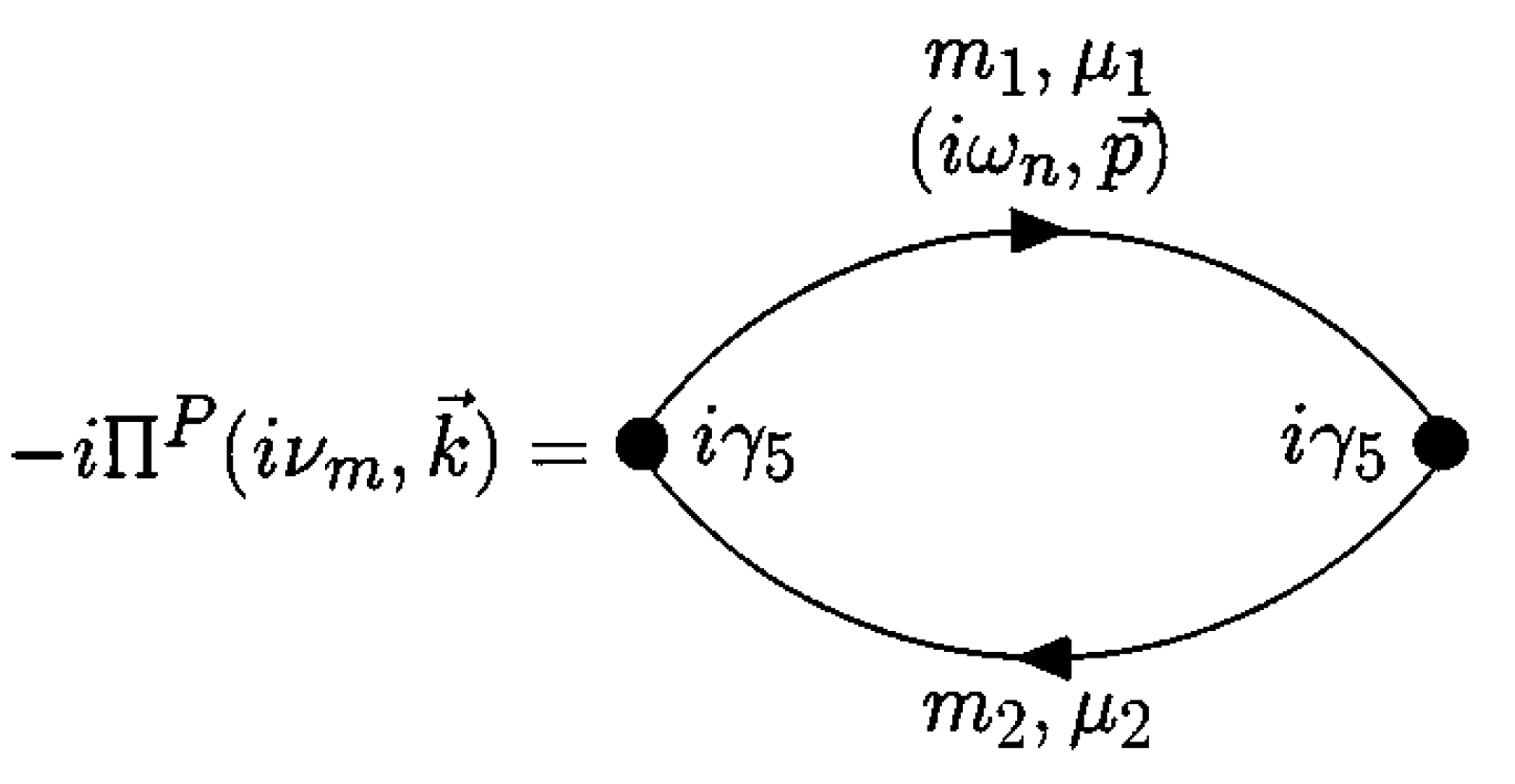}
  \end{center}
  \vskip -6mm
  \caption{The quark-antiquark polarization propagator for pseudoscalar coupling \cite{Rehberg1996b}.\label{col2}}
\end{figure}

\subsection{Meson masses and coupling constants}
\vskip -3mm
How do mesons appear in a theory whose Lagrangian has only quarks as degrees of freedom ? This is shown in Fig. \ref{col1}, which displays the scattering of a quark and an antiquark in our theory with four-point interactions. The left-hand side displays the series of exchange terms which appear in the random phase approximation. This series can be summed. The sum is formally displayed on the right-hand side of this figure. This sum corresponds in leading order of $N_c$ to the propagator of a meson with the proper quantum numbers.

The central building block for the random phase approximation is the quark-antiquark polarization propagator (Fig. \ref{col2})
\vspace{-2mm}
\begin{equation}
  \begin{aligned}
    \frac{1}{i} [\Pi&^{P/S}(q^2,m_1,m_2 )]_{ij}\\
    =&- N_c \sum_{f,f'} \int \frac{d^4p}{(2\pi)^4} \text{Tr } \gamma_5 (T_i)_{ff'} S^f\\
     &\times \left(p+\frac{1}{2}q\right) \gamma_5 (T_j)_{f'f} S^{f'}\left(p-\frac{1}{2}q\right),
  \end{aligned}
  \label{pol}
\end{equation}
where $f$ and $f'$ are the explicit flavor indices and Tr refers therefore to the spinor trace only. $T_i$ and $T_j$ select the appropriate flavor channel:
\begin{equation}
  T_i =
  \begin{cases} 
    \lambda_3                                   &\text{for  } \pi^0,\\
    \frac{1}{\sqrt{2}}(\lambda_1\pm i \lambda_2)&\text{for  } \pi^+, \pi^-,\\
    \frac{1}{\sqrt{2}}(\lambda_6\pm i \lambda_7)&\text{for  } K^0, \bar K^0,\\
    \frac{1}{\sqrt{2}}(\lambda_4\pm i \lambda_5)&\text{for  } K^+, K^-.
  \end{cases}
\end{equation}
For the more complicated $\eta$ and $\eta'$, where $\Pi^{P/S}$ is not diagonal, we refer to \cite{Klevansky1992} where this is treated in detail. After the traces and sums of the polarization propagator [Eq. \eqref{pol}] are carried out one arrives at 
\begin{equation}
  \begin{aligned}
    \frac{1}{i} \Pi^{P/S}(q^2,m_1,m_2 ) =&4 N_c I_1(m_1) + 4 N_c I_1(m_2) \\
                                         &- 4 N_c q^2 I_2(q^2,m_1,m_2)
  \end{aligned}
\end{equation}
with
\begin{equation}
  \begin{aligned}
    I_1(m)          &= \int \frac{d^4p}{(2\pi)^4} \frac{1}{p^2-m^2}, \\
    I_2(q^2,m_1,m_2)&= \int \frac{d^4p}{(2\pi)^4} \frac{1}{p^2-m^2} \frac{1}{(p+q)^2-m^2}.
  \end{aligned}
  \label{poll}
\end{equation}

\begin{figure} [b]
  \begin{center}
    \includegraphics[width=6.2cm]{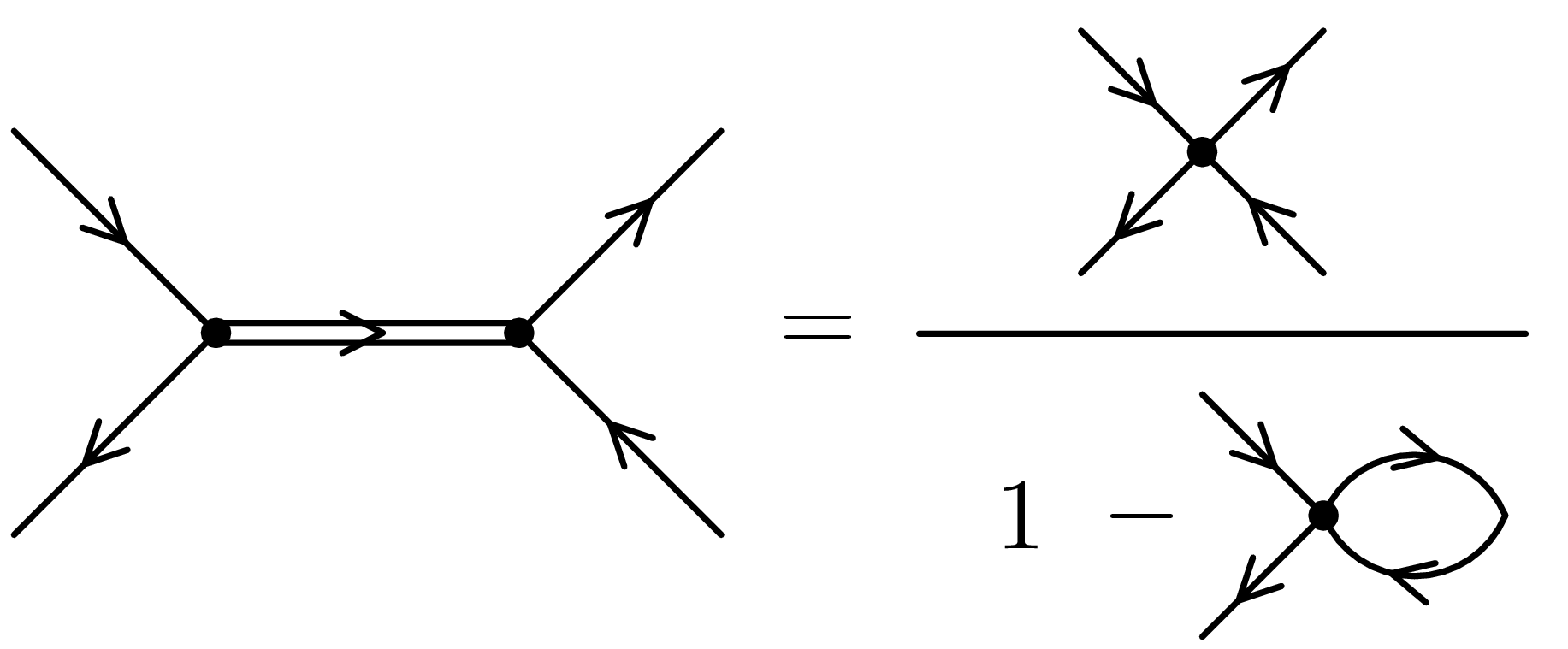}
  \end{center}
  \vskip -5mm
  \caption{The meson propagator corresponding to the RPA sum.\label{col3}}
\end{figure}

\begin{figure*}
  \begin{center}
    \includegraphics[width=7.4cm]{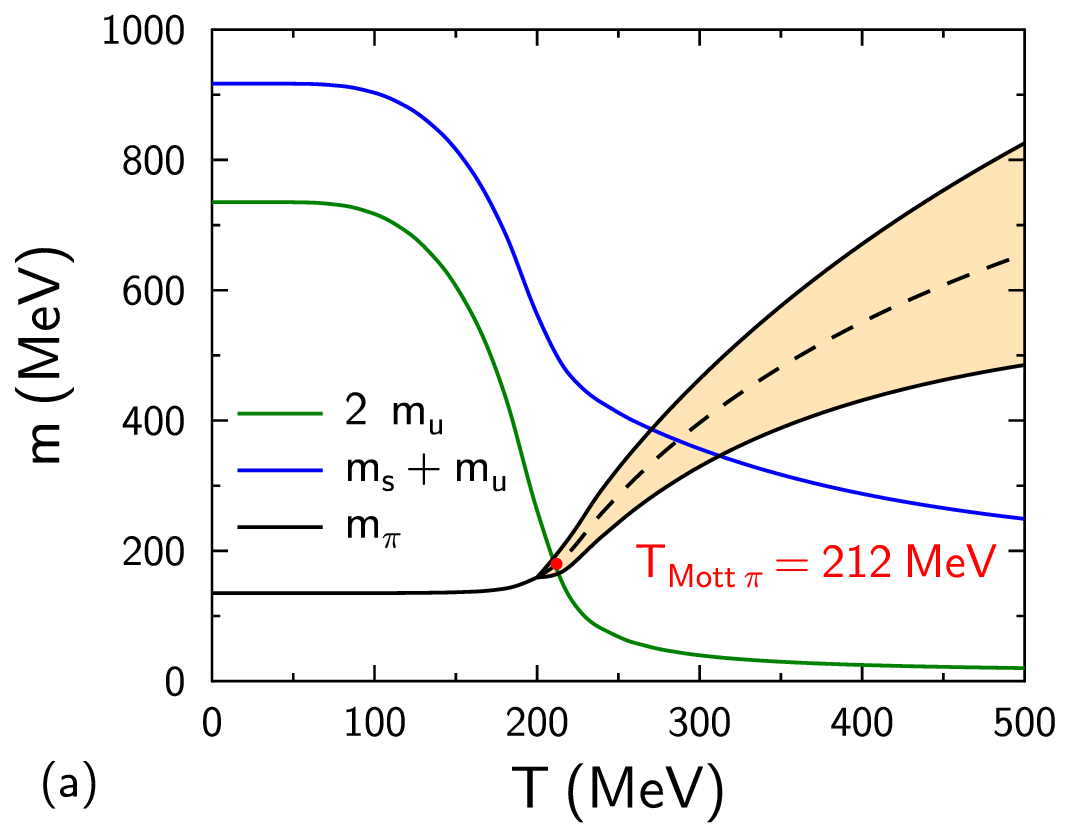} \ \ \
    \includegraphics[width=7.4cm]{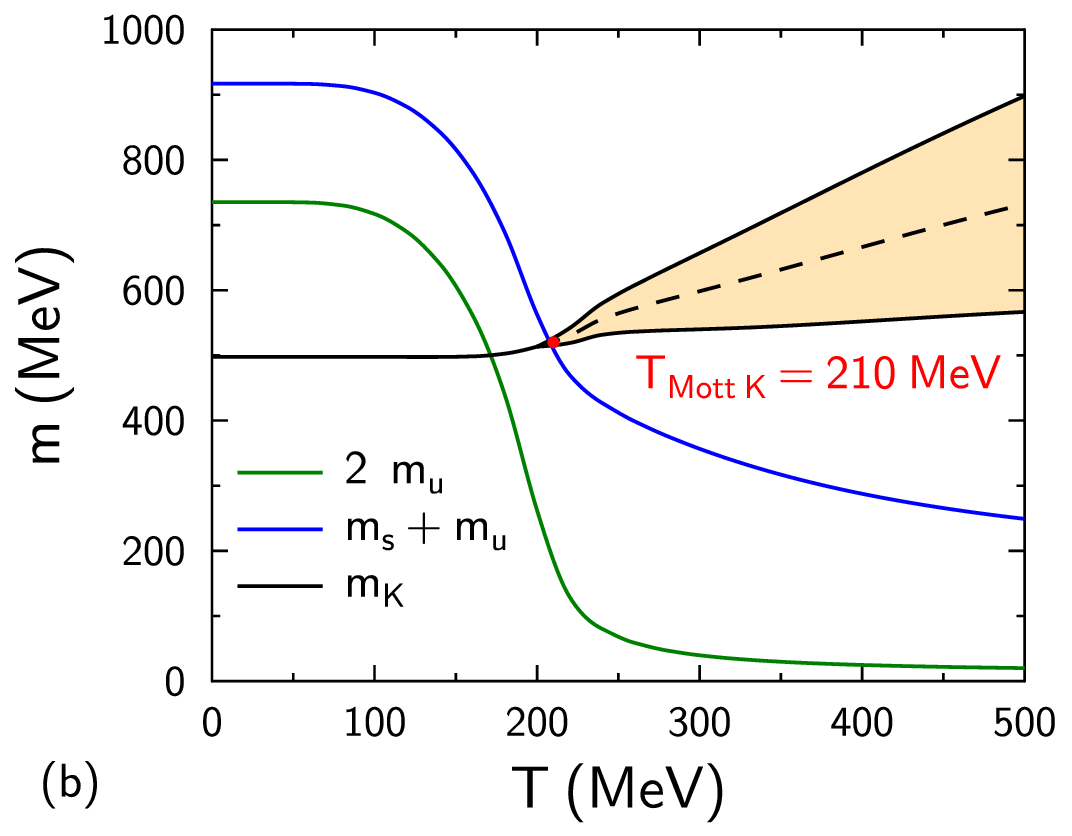}
  \end{center}
  \vskip -5mm
  \caption{Masses of $\pi$ (a) and $K$ (b) mesons as a function of $T$ with the NJL model.\label{masses}}
\end{figure*}

Having the polarization propagator we can sum up the terms of Fig. \ref{col1}. The interactions among a quark and an antiquark with pseudoscalar coupling in the random phase approximation can be written as
\begin{equation}
  \begin{aligned}
      i\gamma_5 T_k&\left[2iG_S + 2iG_S \frac{1}{i} \Pi^{P/S} 2iG_S + \dots\right] i\gamma_5 T_l \\
    = i&\gamma_5 T_k \       \frac{2iG_S}{1 - 2G_S \Pi^{P/S}}                     \ i\gamma_5 T_l .
  \end{aligned}
  \label{pia}
\end{equation}
If a pseudoscalar meson with a mass $M$ is exchanged between the quarks, Fig. \ref{col3}, we find for the interaction
\begin{equation}
  i\gamma_5 T_k \frac{-i g^2_{\pi q \bar q}(M)}{k^2-M^2} i\gamma_5 T_l .
  \label{pib}
\end{equation}
Eqs. \eqref{pia} and \eqref{pib} have the same structure and therefore we can identify the exchange of a pseudoscalar meson with the RPA summation of $q\bar q$ exchanges:
\begin{equation}
  \frac{2 i G_S}{1 - 2 G_S \Pi^{P/S}} = \frac{- i g^2_{\pi q \bar q}}{k^2 - M^2}.
\end{equation}
The mass of the meson can be obtained by solving the equation
\begin{equation}
  1 - 2G_S \Pi^{P/S}\Big|_{k^2=M^2} = 0
  \label{polmass}
\end{equation}
while the coupling constant $g_{\pi q \bar q}$ can be related to the residue of the pole. Expanding Eq. \eqref{pia} around its pole $k^2 = M^2$ we find
\begin{equation}
  \frac{2iG_S}{1-2G_S \Pi^{P/S}} = 
  \frac{-i\Big(\frac{\partial \Pi^{P/S}}{\partial k^2}\Big)^{-1}\Big|_{k^2=M^2}}{k^2-M^2},
  \label{piaa}
\end{equation}
and therefore we can identify
\begin{equation}
  \Bigg(\frac{\partial \Pi^{P/S}}{\partial k^2}\Bigg)^{-1}\Bigg|_{k^2=M^2} = g^2_{\pi q \bar q}.
\end{equation}
For finite temperature and finite chemical potential we have to replace in Eq. \eqref{pol} the propagators $S$ by imaginary time propagators ${\cal S}$:
\begin{equation}
  \begin{aligned}
    \Pi^{P/S}({i\nu_n,\bf q} ) =&N_c T \sum_\omega \sum_{f,f'} \int \frac{d^3p}{(2\pi)^3} \text{Tr }
                                 \gamma_5 (T_i)_{ff'}{\cal S}^f(\omega_l,{\bf p}) \\
                        &\times  \gamma_5 (T_j)_{f'f}{\cal S}^{f'}(\omega_l+\nu_n,{\bf p+q}).
  \end{aligned}
  \label{polt}
\end{equation}
\vskip 5cm
The boson frequencies $\nu_n$ are even: $\nu_n = \pm 2n\ \pi\ T$, $n = 0,1,2,3,\dots$, while the fermion frequencies $\omega_l$ can take odd values only: $\omega_m = \pm (2m+1)\ \pi\ T$, $m = 0,1,2,3,\dots$. So in order to find the pole mass of the pseudoscalar mesons one has to calculate Eq. \eqref{polt} and then solve Eq. \eqref{polmass}. The mass obtained by this procedure does not have to be real. Indeed, when the mass of the meson is larger than that of its constituents the meson can decay into its constituents

After carrying out the frequency sum $\Pi^{P/S}({i\nu_n,\bf q} )$ can be brought as well into the form of Eq. \eqref{poll} with $I_1$ and $I_2$ given by
\begin{widetext}
\begin{equation}
  \begin{aligned}
  I_1(m)      &= -i \int^\Lambda\frac{d^3p}{(2\pi)^3} \frac{1}{2E_p} [1 - f(E_p-\mu) - f(E_p+\mu)],\\
  I_2(m_1,m_2)&=  i \int^\Lambda\frac{d^3p}{(2\pi)^3} \frac{1}{2E_p2E_{p+q}}
  \frac{f(E_p+\mu) + f(E_p-\mu) - f(E_{p+q}+\mu) - f(E_{p+q}-\mu)}{\omega + E_p - E_{p+q}+i\epsilon}\\
  &+ i \int^\Lambda \frac{d^3p}{(2\pi)^3} \frac{1 - f(E_{p}-\mu) - f(E_{p+q}+\mu)}{2E_p2E_{p+q}}
  \left[\frac{1}{\omega +E_p+E_{p+q}+i\epsilon} - \frac{1}{\omega -E_p-E_{p+q}+i\epsilon} \right],
  \end{aligned}
\end{equation}
\end{widetext}
with $E_p = \sqrt{m_1^2+\bf p^2}$ and $ E_{p+q} = \sqrt{m_2^2+(\bf p + \bf q) ^2}$. In the present approach we limit our mesons to the pseudoscalar mesons.

The model contains five parameters: the current mass of the light and strange quarks, the coupling constants $G_D$ and $G_S$, and the momentum cutoff $\Lambda$. These are fixed by physical observables : the pion and kaon masses, the pion decay constant, the scalar quark condensate $\langle \bar qq\rangle$, and the mass difference between $\eta$ and $\eta'$. We will employ the parameters set : $m^0_q = 5.5$ MeV, $m^0_s = 140.7$ MeV , $G_S / \Lambda^2 = 1.835$, $G_D / \Lambda^5 = 12.36$, and $\Lambda = 602.3$ MeV. The masses for up and strange quarks, as well as for $\pi$ and $K$ for this parameter set \cite{Nebauer2002, Gastineau2002a}, are displayed in Fig. \ref{masses}. We see that for small $\mu$ and $T$ the meson masses are smaller than the masses of their constituents. For large $\mu$ and $T$ the opposite is true. If the mass of the constituents become smaller than the meson mass the meson mass becomes complex and the mesons become quasiparticles. They exist in the plasma but with a lifetime which decreases with increasing $\mu$ and/or $T$ (where the width $\Gamma = 2G_S \Im \Pi^{P/S}$ is displayed by the yellow band in Fig. \ref{masses}).

\subsection{Cross sections}
\vskip -4mm
If created in heavy ion collisions the QGP will expand rapidly. Therefore, the cross sections between constituents become dominant over the static properties of the theory. In the NJL model these cross sections can be calculated via a $1 /N_c$  expansion \cite{Quack1994}. All the details can be found in \cite{Rehberg1996a,Rehberg1996b,Quack1994}. Therefore we mention here only the essential facts. 

\subsubsection{\bf\emph{Elastic Collisions}}
\vskip -3mm
\begin{figure}
  \begin{center}
    \includegraphics[width=5.4cm]{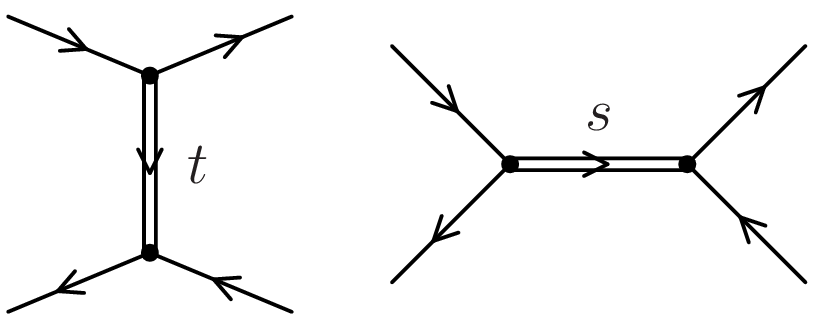}
  \end{center}
  \vskip -5mm
  \caption{Feynman diagrams for elastic $q \bar q$ scattering.\label{colelast}}
\end{figure}

\begin{figure} [b]
  \begin{center}
    \includegraphics[width=6.8cm]{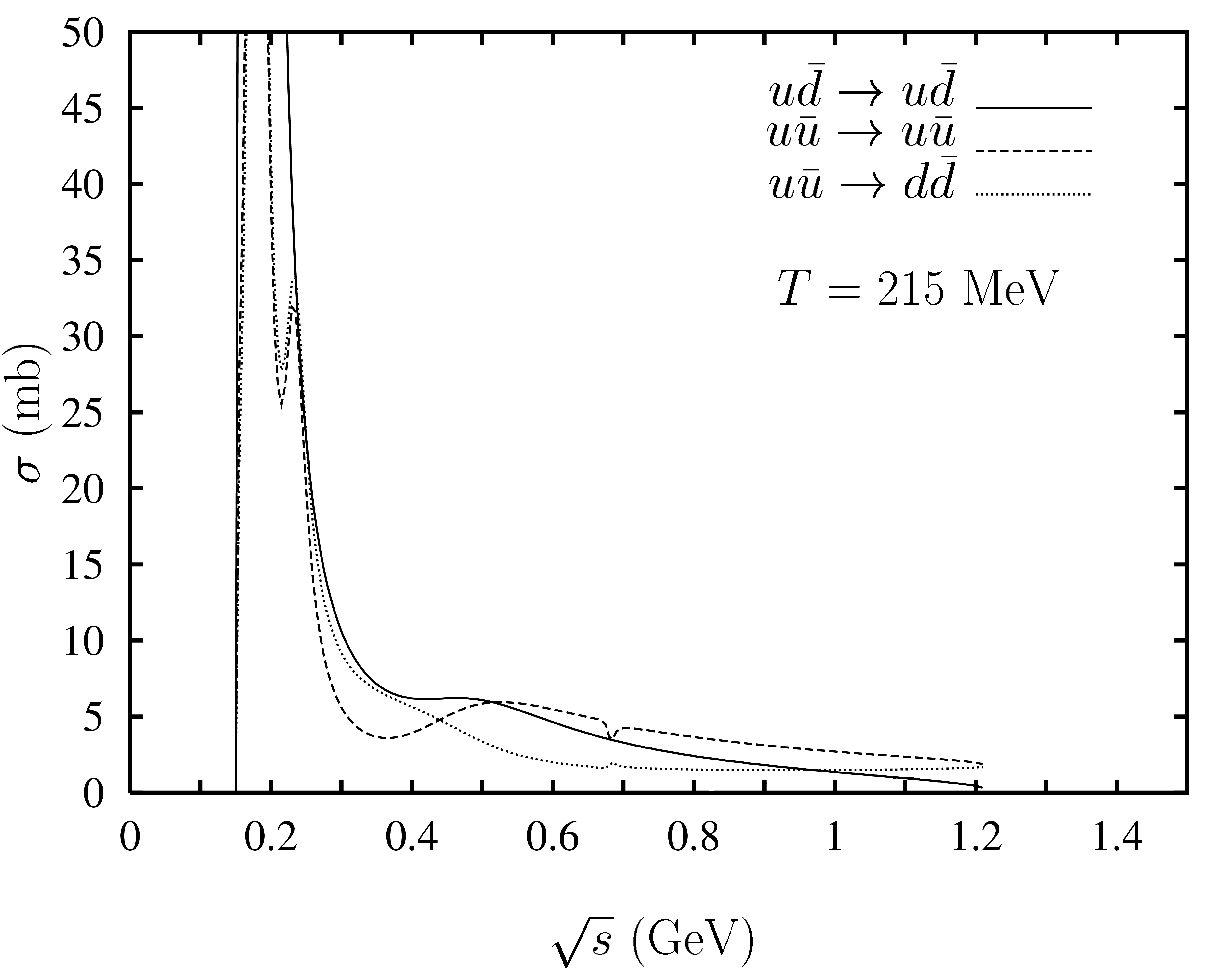} \ \
  \end{center}
  \vskip -5mm
  \caption{Elastic cross section for the different channels as a function of $\sqrt{s}$ for a temperature close to $T_{\text{Mott}}$ and at $\mu=0$ \cite{Rehberg1996b}.\label{sela}}
\end{figure}

The Feynman diagrams for the $q\bar q \rightarrow q\bar q$ cross sections are displayed in Fig. \ref{colelast}. We see contributions from the $s$-channel and from the t-channel. The matrix elements are given by \cite{Rehberg1996a}
\begin{equation}
  \begin{aligned}
    -i{\mathcal{M}}_t =&\delta_{c_1,c_3} \delta_{c_2,c_4} \bar{u}(p_3) T u(p_1) \\
                       &\times[i{\mathcal{D}}^S_t(p_1-p_3)] v(p_4) T \bar{v}(p_2) \\
                       &+ \delta_{c_1,c_3} \delta_{c_2,c_4} \bar{u}(p_3)(i\gamma_5T) u(p_1) \\
                       &\times[i{\mathcal{D}}^P_t(p_1-p_3)] v(p_4) (i\gamma_5T) \bar{v}(p_2)
  \end{aligned}
\end{equation}
and
\begin{equation}
  \begin{aligned}
    -i{\mathcal{M}}_s =&\delta_{c_1,c_2} \delta_{c_3,c_4}  \bar{v}(p_2) T u(p_1) \\
                       &\times[i{\mathcal{D}}^S_s(p_1+p_2)] v(p_4) T \bar{u}(p_3) \\
                       &+ \delta_{c_1,c2} \delta_{c_3,c_4} \bar{v}(p_2) (i\gamma_5T) u(p_1) \\
                       &\times[i{\mathcal{D}}^P_u(p_1+p_2)] v(p_4) (i\gamma_5T) \bar{u}(p_3),
  \end{aligned}
\end{equation}
where $p_1$ $(p_2)$ is the momentum of the incoming $q$ $(\bar q)$ and $p_3$ $(p_4)$ that from the outgoing $q$ $(\bar q)$. The $c_i$ are the color indices and $T$ are the isospin projections on the mesons. ${\mathcal{D}}^S$ and ${\mathcal{D}}^P$ are the meson propagators of the form 
\begin{equation}
  {\mathcal{D}}^{S/P} = \frac{2 G_S}{ 1 - 2 G_S \Pi^{P/S}}
  \label{res}
\end{equation}
with $\Pi^{P/S}$ being the polarization tensor in the pseudoscalar-scalar channel. This cross section is displayed in Fig. \ref{sela}. We see that for most of the center-of-mass energies this cross section is of the order of several millibarns. Close to the Mott transition the cross section increases dramatically  to more than a hundred millibarns, because the incoming quarks in the $s$ channel become resonant with the intermediate meson \cite{Gastineau2005}. This increase is observed for all reactions which have a $s$ channel (see Fig. \ref{sela}). This means that at the end of the expansion of the plasma, shortly before the Mott temperature $T_{\text{Mott}}$ is reached, the system comes almost certainly to a local equilibrium. Whether at temperatures much higher than $T_{\text{Mott}}$ local equilibrium can be established or maintained is, in view of the size of the cross section, not evident. The elastic $qq$ and $\bar q \bar q$ cross sections are of the order of a couple of millibrans. Because they do not have an $s$ channel they do not increase close to $T_{\text{Mott}}$.
\vspace{-2mm}
\subsubsection{\bf\emph{Hadronization Cross Section}}
\vskip -3mm
\begin{figure}
  \begin{center}
    \includegraphics[width=6.7cm]{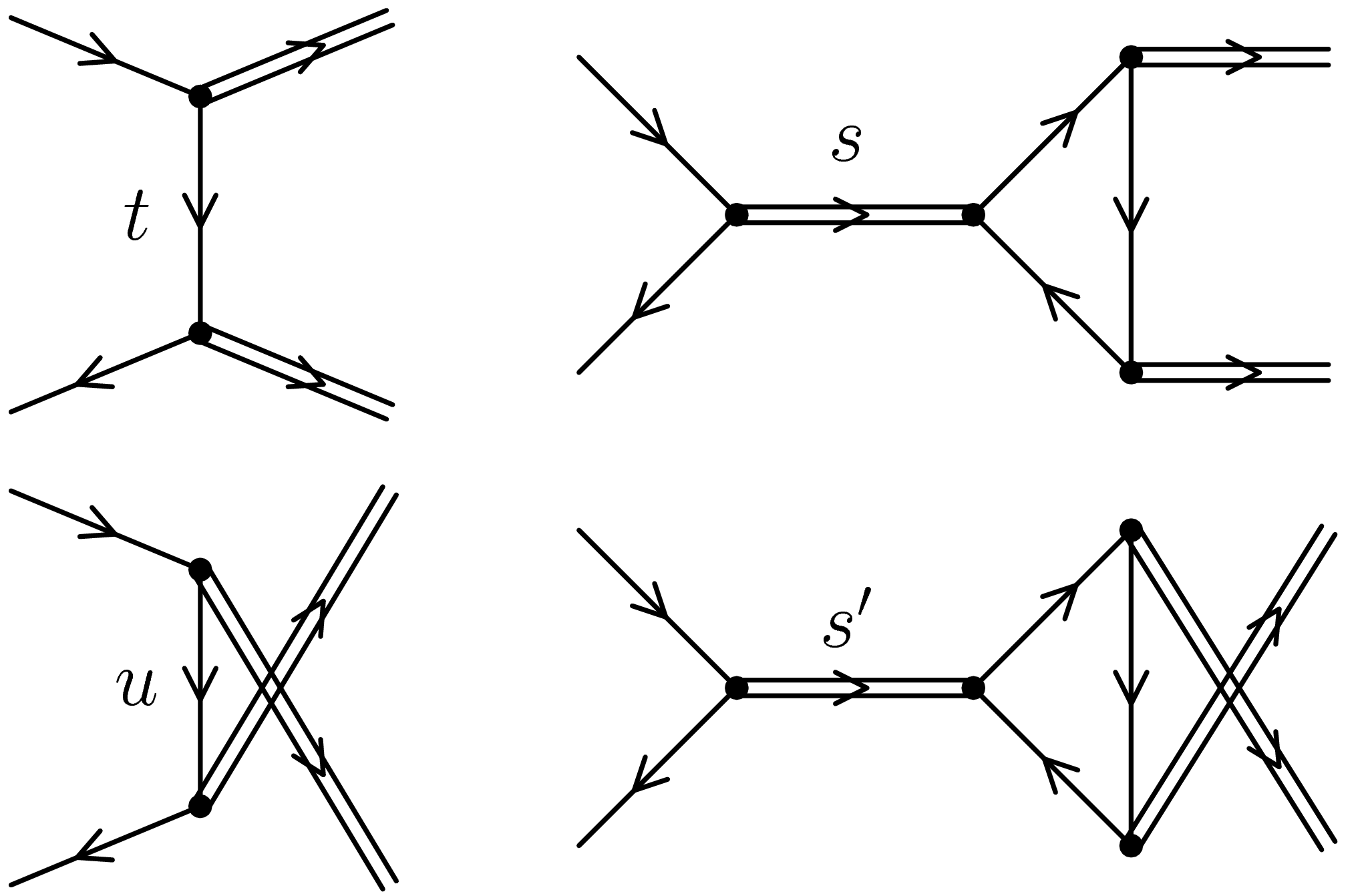}
  \end{center}
  \vskip -5mm
  \caption{Feynman diagrams for hadronisation.\label{colhad}}
  \vspace{-2mm}
\end{figure}

\begin{figure} [b]
  \begin{center}
    \includegraphics[width=6.8cm]{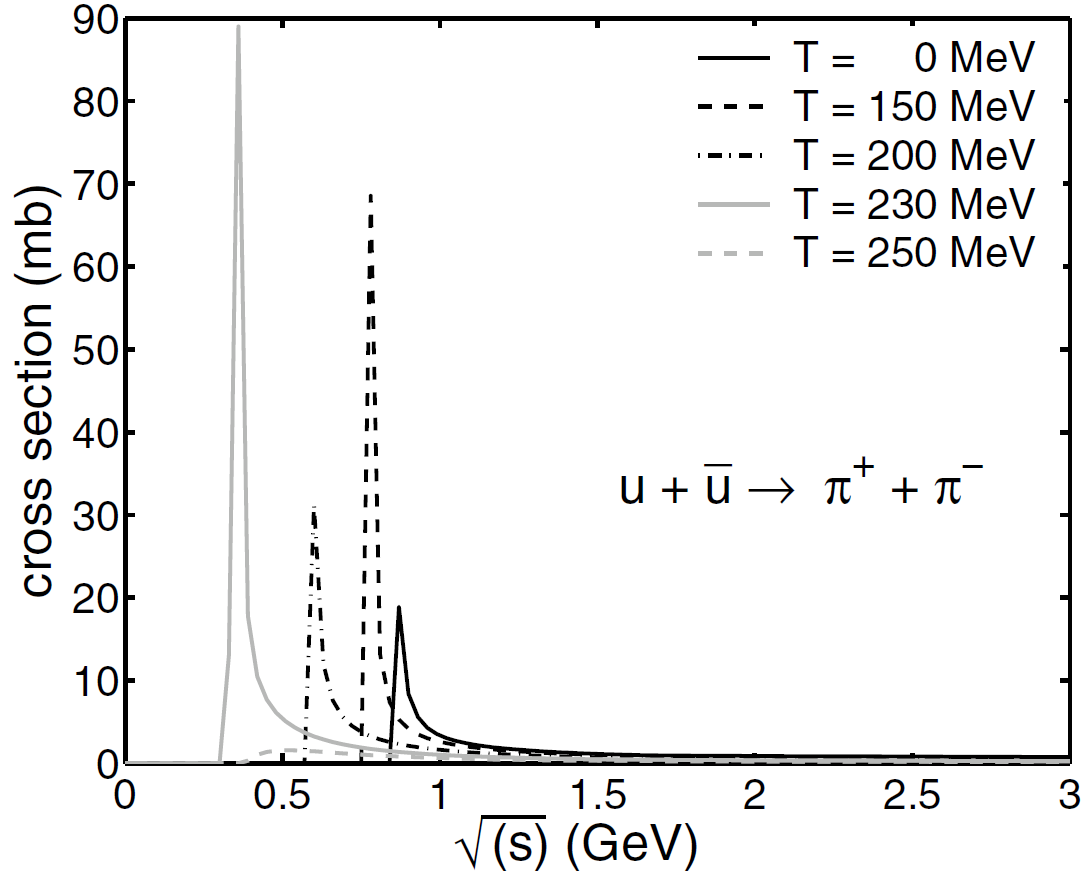}
  \end{center}
  \vskip -5mm
  \caption{Example of an inelastic $q\bar{q}$ cross section \cite{Gastineau2005} as a function of $\sqrt{s}$.\label{NJL_cross_section_1}}
\end{figure}

The $1/N_c$ expansion provides as well the hadronization cross sections in which a $q\bar q$ pair creates two pseudoscalar mesons. The Feynman diagrams are displayed in Fig. \ref{colhad}. For the details of the calculation we refer to  \cite{Rehberg1996a,Hufner1994}. Figure \ref{NJL_cross_section_1} displays the cross section $u\bar u \to \pi^+ \pi^-$ as a function of $\sqrt{s}$ for different temperatures. We observe that the cross section increases close to the kinematical threshold. Close the Mott transition the cross section can reach 100 mb. Although the NJL model has no confinement this large cross section means that close to the crossover $q \bar q$ pairs create mesons very effectively (the cross section for the backward reaction being kinematically suppressed) and therefore most of the quarks are converted into mesons when the system reaches the Mott transition.

We created tables of all elastic and inelastic cross sections for our simulations.

%% file: section_4_integral.tex
\begin{figure} [b]
  \begin{center}
  \includegraphics[width=5cm]{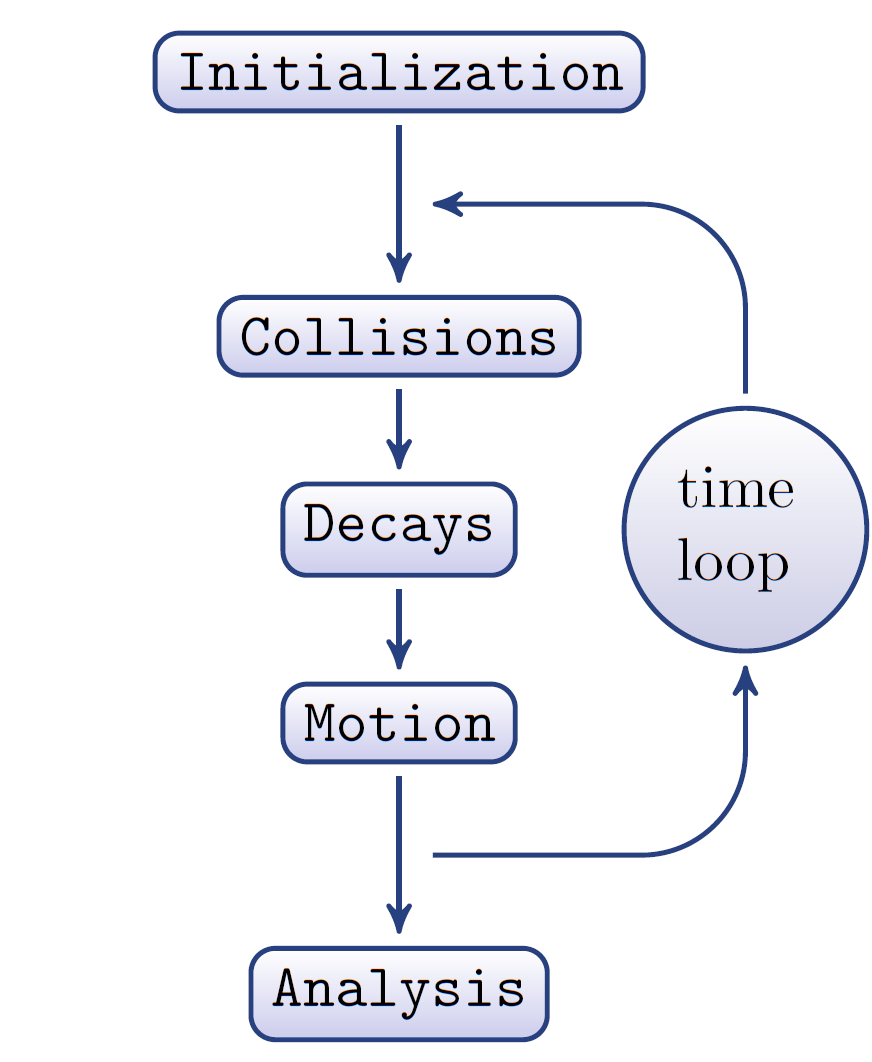}
  \end{center}
  \vskip -5mm
  \caption{(Color online) Standard algorithm for molecular dynamics calculations.\label{algorythm}}
\end{figure}

\section{Simulation program}
\vskip -2mm
In this section we discuss how the results of the previous sections are used to formulate a molecular dynamics approach to describe the expanding $q/\bar q$ plasma. The expansion of such a plasma, which is presumably created in the reaction between two heavy ions at ultrarelativisitc energies (at $\sqrt{s}$ well above 10 GeV ), is presently highly debated.


We describe the expansion in a relativistic quantum molecular dynamic approach, discussed in Sec. II, which is based on the NJL Lagrangian, discussed in Sec. III. We assume that the system remains sufficiently close to a local thermal equilibrium that we can parametrize the masses of the quarks and mesons by a local temperature and a local chemical potential. The quarks interact in two ways: First, they change the mass of fellow quarks by their contribution to the chemical potential and to the local temperature; second, they interact with their fellow quarks by elastic and inelastic scattering. This transport approach is called INTEGRAL (INTEractive Generalized Relativistic ALgorithms).

The basic structure of a molecular dynamics program is described in Fig. \ref{algorythm}. We discuss in the following each of these steps.

\subsection{Initial conditions} \label{hpm}
\vskip -3mm
Principally any phase-space distribution of partons can be taken as an initial condition for our calculations. In these first studies we assume a quite smooth initial distribution which is determined as follows. In a first step we calculate the radius of the colliding (identical) nuclei by 
\begin{equation}
  R = r_0 \ A^{1/3}
\end{equation}
with $r_0 = 1.25$. For a finite impact parameter $b$ we approximate the overlap region by an ellipse with the axes $\mathbf{x}_E$ and $\mathbf{y}_E$:
\vspace{-2mm}
\begin{equation}
  \mathbf{x}_E = R - b/2,
  \quad
  \mathbf{y}_E = \sqrt{(R - b/2)(R + b/2)}.
  \label{ellipse_size}
\end{equation}
The extension in the third dimension is assumed to be proportional to the creation time of the QGP. We take
\begin{equation}
  \mathbf{z}_E = 2 \tau_0 = 2 \text{ fm.}
\end{equation}
Knowing the atomic number of the colliding nuclei, $A$, and the impact parameter, $b$, we can construct the overlap zone in coordinate space.

\begin{figure} [b]
  \vspace{-2mm}
  \begin{center}
    \includegraphics[width=7.4cm]{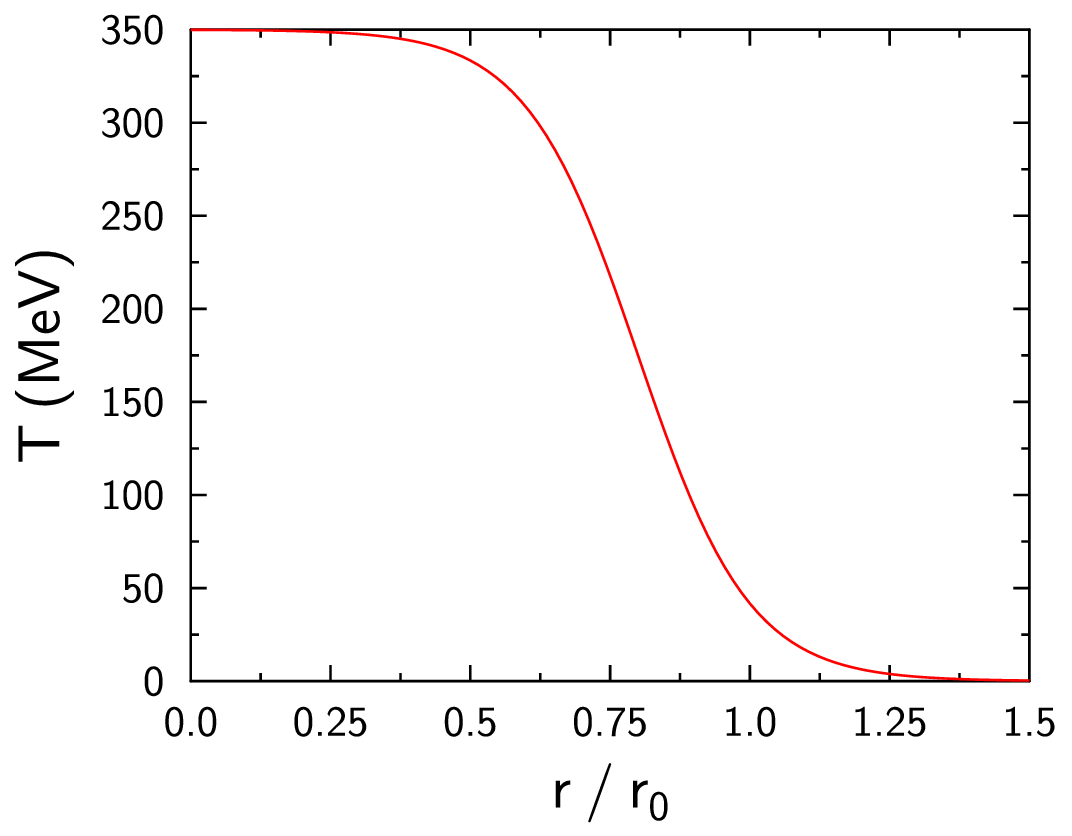}
  \end{center}
  \vskip -6mm
  \caption{(Color online) Distribution of the temperature $T$ as a function of the normalized radius $r / r_0$ for $T_0 = 350$ MeV.\label{T_radius}}
\end{figure}

\begin{figure}
  \begin{center}
    \includegraphics[width=7.5cm]{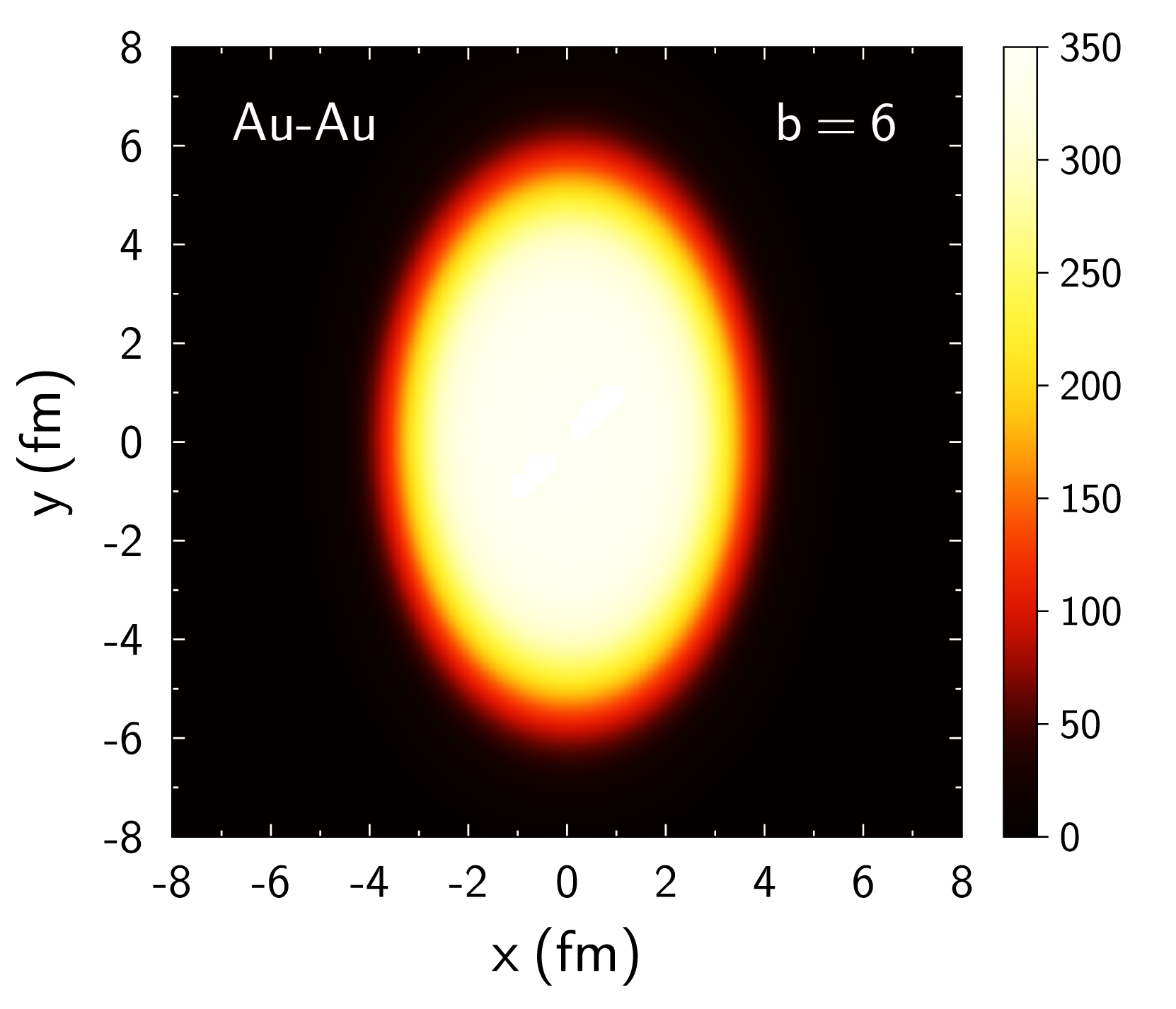}
  \end{center}
  \vskip -5mm
  \caption{(Color online) Distribution of the temperature $T$ (in MeV) in the transverse plane $x,y$.\label{T_initial}}
  \vspace{-2mm}
\end{figure}

In the present study we assume that the system is close to local equilibrium. The mass of the partons and, consequently, their energy depends then on the local $(T,\mu)$, which depend on the initial $T_0$ and $\mu_0$ for the center of the collision. The initial local temperature depends on the position of the parton. At a point $(x,y)$  with $r = \sqrt{x^2 + y^2}$ the local temperature is a function of $r/r_0$, where the vector ${\bf r_0}$ points in the direction of ${\bf r}$ and $r_0 = \sqrt{\mathbf{x}_E^2 + \mathbf{y}_E^2}$. The initial temperature is given by
\begin{equation}
  T(r) = \frac{T_0}{1 + \exp[10\ (r/r_0 - 0.8)]}.
  \label{temp}
\end{equation}
The central initial temperature $T_0$ can be parametrized as
\begin{equation}
  T_0 (\text{MeV}) = 68 \ \frac{\log \left[\sqrt{s_{NN}}(\text{GeV}) + 1\right]}{1 + \exp[1.5\ (1 - \mathbf{x}_E)]}.
\end{equation}
Figure \ref{T_radius} shows the initial temperature as a function of $r/r_0$. Figure \ref{T_initial} is a contour plot of the initial temperature. For RHIC energies this parametrization corresponds to the initial temperature of hydrodynamical calculations. By knowing the critical temperature $T_c = 165$ MeV, this equation gives us for RHIC ($\sqrt{s_{NN}} = 200$ $A$GeV) $T_0 \simeq 2.2 T_c$ and for LHC ($\sqrt{s_{NN}} = 2760$ $A$GeV) $T_0 \simeq 3.5 T_c$. The calculations we present here are calculated with $\mu_0 = \mu(r) = 0$. 

Knowing $(T(r),\mu=0)$ we can determine the mass and the initial density of quarks and antiquarks:
\begin{equation}
  \rho = \frac{N}{V} = g \int f^\pm(p) \frac{d^3 p}{(2\pi)^3(\hslash c)^3},
  \label{inirhob}
\end{equation}
where $f^\pm$ is the Fermi-Dirac distribution [eq. (\ref{Fermi0})],
\begin{equation}
  f^\pm(p,m(T)) = [ 1 + \exp ( \sqrt{p^2 + m^2} \mp \mu / T ) ]^{-1},
  \label{Fermi}
\end{equation}
$g$ is the degeneracy of the considered parton, $N$ is the quark number, and $V$ is the volume in the center-of-mass system of the reaction (an ellipse with constant thickness),
\begin{equation}
  V = \pi \mathbf{x}_E \mathbf{y}_E \mathbf{z}_E.
\end{equation}

Knowing density and volume of a slice with a given temperature we determine the local number of partons with the help of Eq. (\ref{temp}). The partons are then placed randomly in this slice. The initial momentum $\mathbf{p}$ of each parton is obtained by applying a Monte Carlo procedure which models the local Fermi-Dirac distribution with $(T,\mu)$. In the spirit of the core corona model close to the surface we assume thermalization only in the longitudinal $z$ direction and limit the transverse momentum in the outward direction by limiting the azimuthal angle $\phi$. This procedure ensures that fast partons in the corona are comovers and can hadronize easily. The spatial distribution of these partons is quite smooth. That is why we call this initial conditions model the hot pancake model (HPM).

\subsection{Transport model}
\begin{figure*}
  \begin{center}
    \includegraphics[width=7cm]{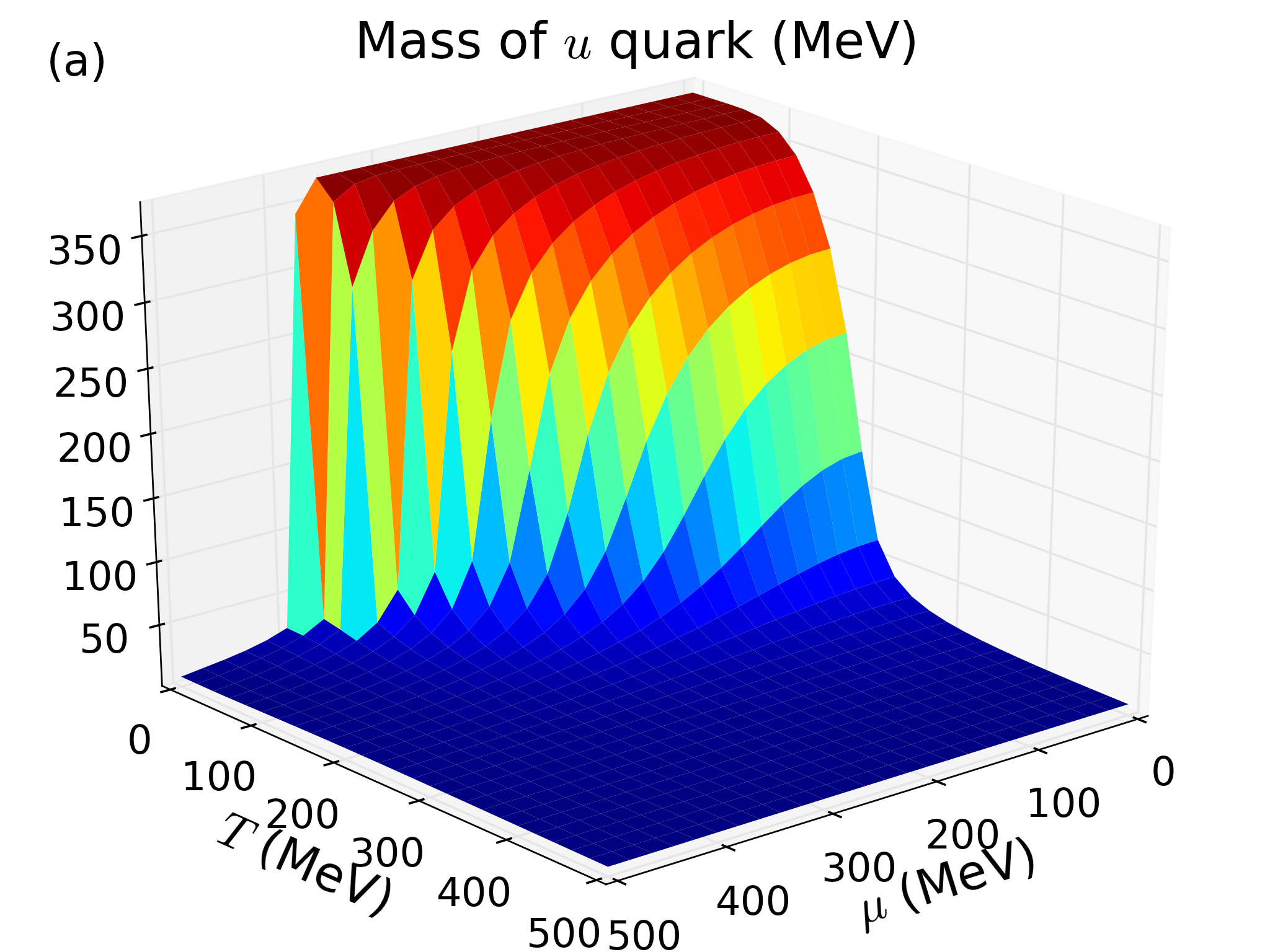}
    \includegraphics[width=7cm]{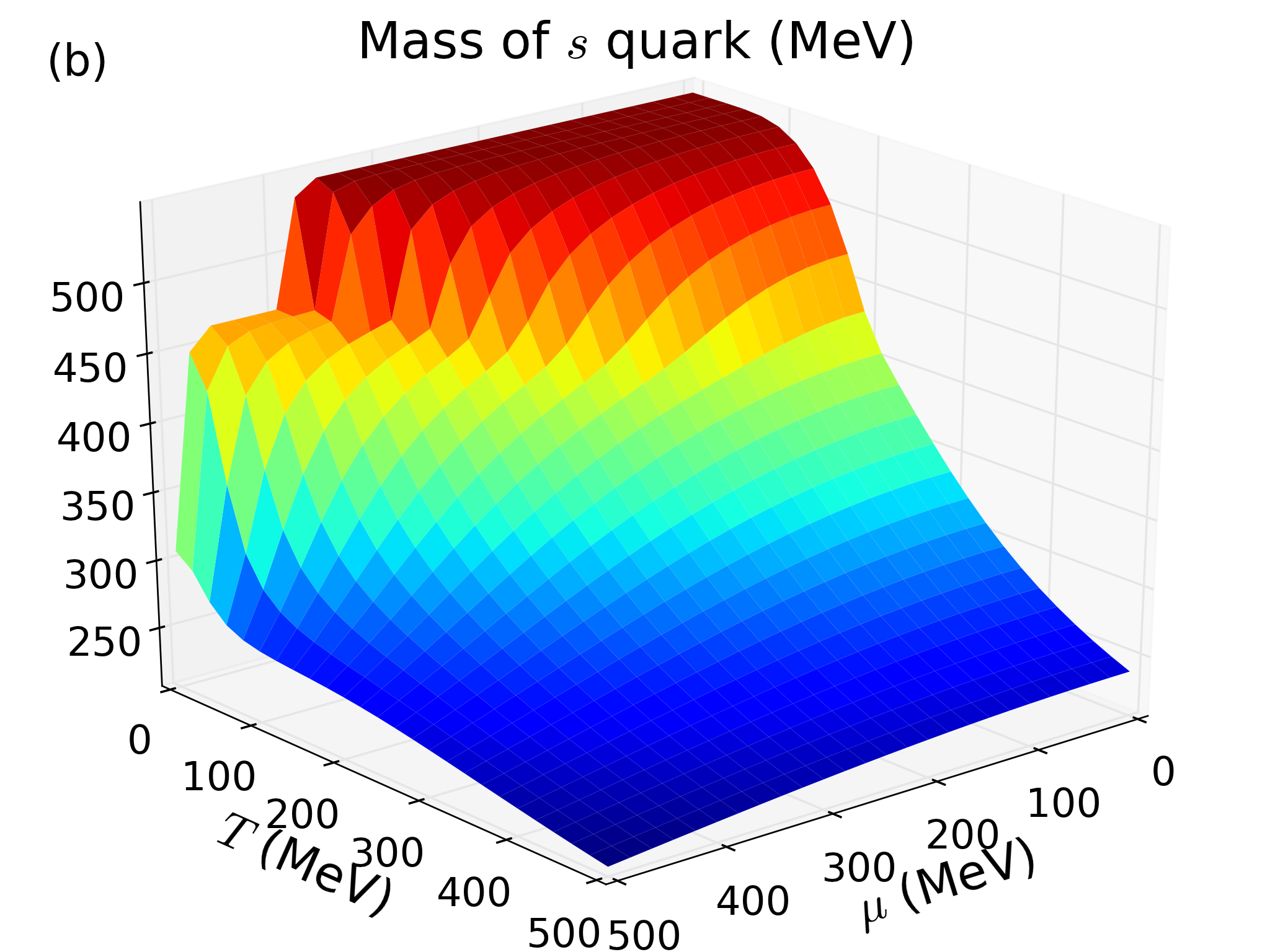}\\ \vspace{4mm}
    \includegraphics[width=7cm]{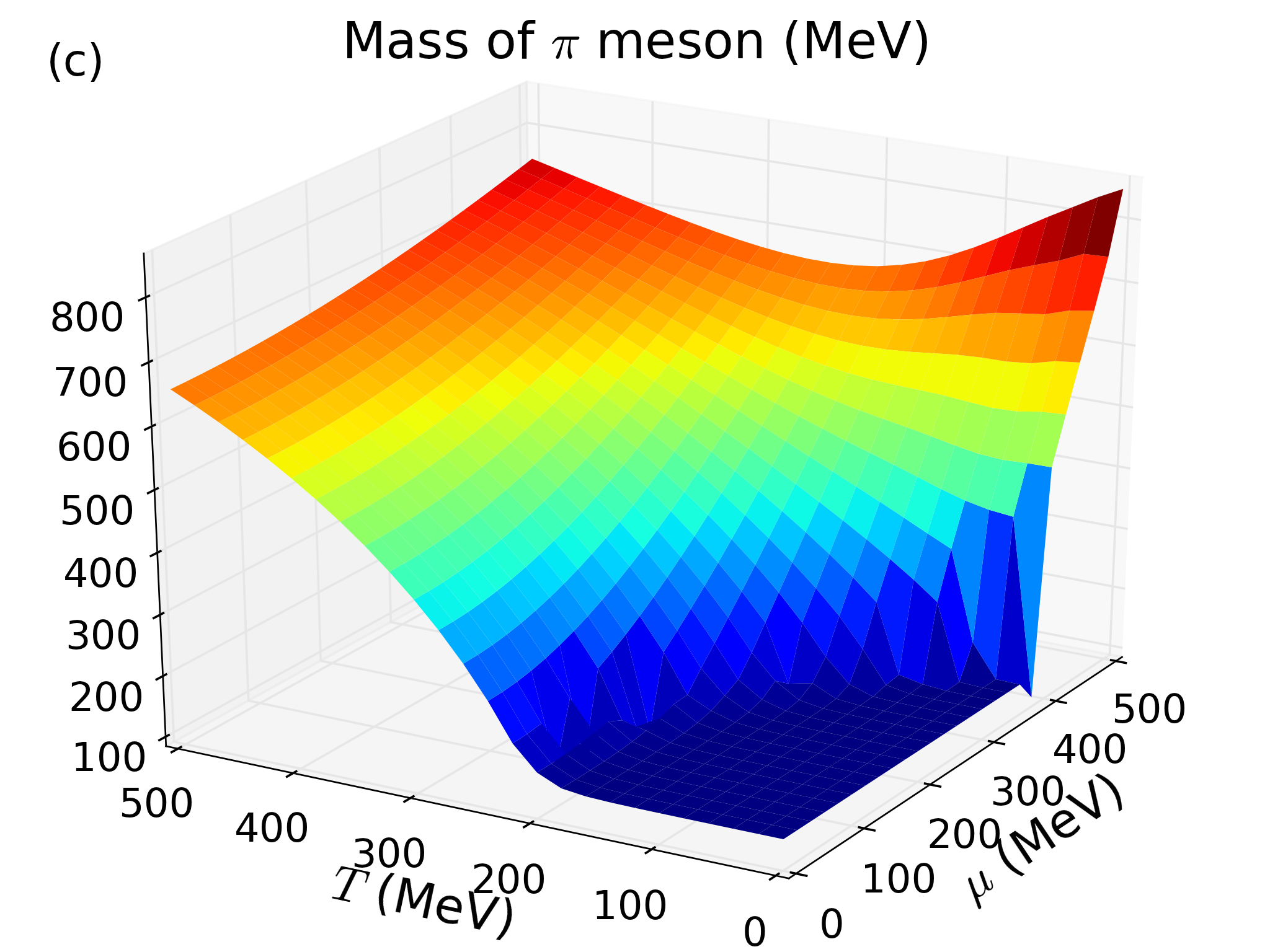}
    \includegraphics[width=7cm]{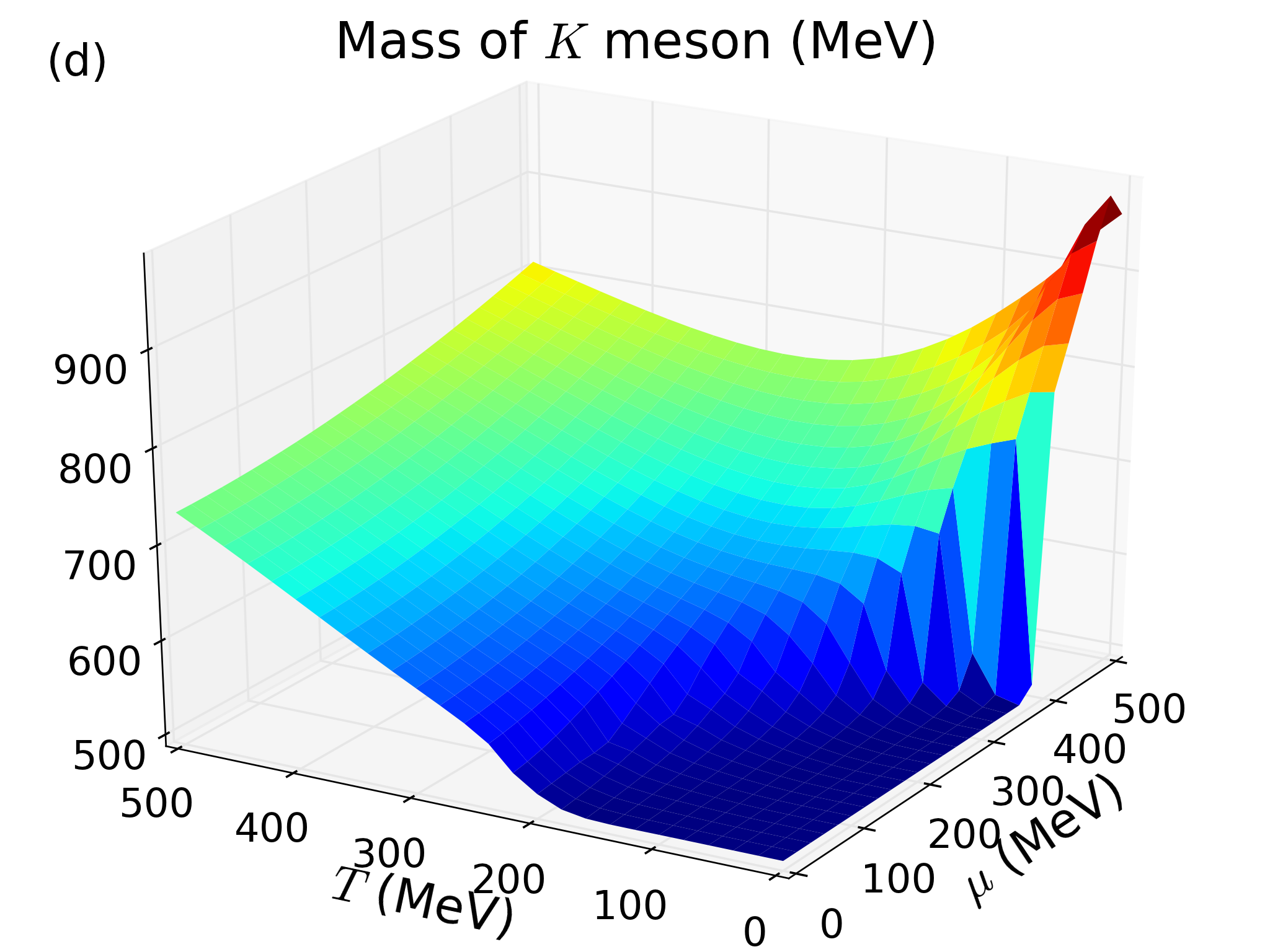}
  \end{center}
  \vskip -5mm
  \caption{(Color online) Dependence of the masses on the temperature and on the chemical potential of the environment in the NJL \label{masses_NJL} model. We display the masses of  $u$ (a) and  $s$ (b) quarks, as well as that of $\pi$ (c) and  $K$ (d) mesons.}
\end{figure*}

\vskip -3mm
The partons in the expanding system are described by their positions and their momenta. The equations of motion of the particles are given by Eqs. (\ref{finaleom})
\begin{equation}
  \frac{d q_i^\mu}{d \tau} = \frac{p_i^\mu}{E_i},
  \quad
  \frac{d p_i^\mu}{d \tau} = - \sum_{k=1}^N \frac{1}{2 E_k}
                             \frac{\partial V_k (q_T')}{\partial {q_i}_\mu}.
  \label{j1}
\end{equation}
In the NJL model the potential between the particles is a scalar interaction. This interaction acts like a mass which depends in our local equilibrium assumption on the temperature and chemical potential of the environment. Therefore we can reformulate our energy constraints, Eq. \eqref{finalconst}, by
\begin{equation}
  K_i = {p_i}^\mu {p_i}_\mu - {m^{*2}_i(T,\mu)} = 0.
\end{equation}
This modifies the equation we have to solve
\begin{equation}
  \frac{d p_i^\mu}{d \tau} =
  - \sum_{k=1}^N \frac{m^*_k}{E_k} \frac{\partial m^*_k}{\partial {q_i}_\mu},
\end{equation}
with
\begin{equation}
  \frac{\partial m^*_k}{\partial {q_i}_\mu} =
  \frac{\partial m^*_k}{\partial T_k}
  \frac{\partial T_k}{\partial {q_i}_\mu}
  +
  \frac{\partial m^*_k}{\partial \mu_k}
  \frac{\partial \mu_k}{\partial {q_i}_\mu}
\end{equation}
and with $T_k (\mu_k) $ being the local temperature (chemical potential) of the environment of particle $k$. The mass dependence of the partons as well as that of $\pi$'s and $K$'s on the chemical potential and on the temperature is displayed in Fig. \ref{masses_NJL}. Here we assume that the chemical potentials of up, down and strange quarks are identical. The masses show the expected behavior of a cross over at high $T$ and $\mu \simeq 0$.

At high $(T,\mu)$ we see the bare mass of the partons. When approaching low $(T,\mu)$ we observe a steep rise of the mass due to the scalar potential, which becomes finite. At $(T=0,\mu=0)$ the light partons have a constituent mass of around 370 MeV. The 't Hooft term connects up and down quarks with strange quarks. Therefore the dependence of the strange quark mass on temperature and chemical potential becomes more complex. We see a first steep rise of the mass when the chemical potential arrives from above the transition temperature of the light quarks $u$ and $d$ and a second rise when the genuine transition of the $s$ quark takes place.

To solve Eqs. \eqref{finaleom}, the differential equations are converted into finite-difference equations with a variable time step. Its definition will be discussed in Sec. \ref{dt}. For the solution we employ an adaptive method, depending on the time step size, with either an Euler algorithm or a Runge-Kutta algorithm of second order (RK2) or of fourth order (RK4). The cross sections and masses which have been calculated in \cite{Gastineau2002a}, \cite{Thomere2009}, and \cite{Nebauer2000} have been tabulated as a function of $(T,\mu,\sqrt{s})$ and a linear interpolation has been applied to accelerate the calculations.  
\vspace{-8mm}
\subsection{Thermodynamical medium}
\vskip -4mm
In our local equilibrium approximation the effective mass $m^*$ of the partons depends on the temperature and chemical potential of the local environment. Therefore we have to construct these two quantities from the information on the system which is available, the four-positions and four-momenta of all particles. For this we define two densities, the fermionic density $\rho_F$ and the baryonic density $\rho_B$
\begin{equation}
  \begin{aligned}
    \rho_F (T,\mu) = \frac{N_q}{V}&+ \frac{N_{\bar{q}}}{V}
    = g \int_0^\infty \frac{\mathrm{d}^3 p}{(2\pi)^3 (\hslash c)^3} \\
    \times[&2 \left( f^+(p,m_u) + f^- (p,m_{u})\right) \\
           &+ \left( f^+(p,m_s) + f^-(p,m_s)\right)],
  \end{aligned}
  \label{density_thermo1}
\end{equation}
\begin{equation}
  \begin{aligned}
    \rho_B (T,\mu) = \frac{N_q}{V}&- \frac{N_{\bar{q}}}{V}
    = g \int_0^\infty \frac{\mathrm{d}^3 p}{(2\pi)^3 (\hslash c)^3} \\
    \times[&2 \left( f^+(p,m_u) - f^- (p,m_{u})\right) \\
           &+ \left( f^+(p,m_s) - f^-(p,m_s)\right)],
  \end{aligned}
  \label{density_thermo2}
\end{equation}
\vskip 10mm
with the degeneracy factor  $g = 2 \times 3 = 6$, and $f^{\pm}$ defined in Eq. \eqref{Fermi}. Neither $\rho_F$ nor $\rho_B$ are Lorentz invariants. In order to express $T_i$ and $\mu_i$ as a function of the phase-space coordinates $(q^\mu_j, p^\mu_j)$ the following procedure is applied: we introduce a Lorentz-invariant Gaussian function $ R_{ij}(q_T')$ inspired from the Wigner density equation \eqref{wigner},
\begin{equation}
 R_{ij}(q_T') = \left( \frac{1}{L \sqrt{\pi}} \right)^3 \exp \left( \frac{{q_T'}_{ij}^2}{L^2} \right),
\end{equation}
to calculate the contribution of a neighboring parton $j$ to the density of parton $i$. For the width we take $L = 0.5$ fm, which is about the electromagnetic radius of known hadrons. This allows us to rewrite the density as
\begin{equation}
  \rho_{F i} = \sum_{j \ne i} R_{ij},
  \quad
  \rho_{B i} = \sum_{j \ne i} R_{ij} \ \textrm{Sign}(j),
\end{equation}
\vspace{-1mm}
with
\begin{equation}
  \textrm{Sign}(j) =
  \begin{cases}
   \ \ 1&\text{ for fermions},\\
      -1&\text{ for antifermions}.
  \end{cases}
  \label{def_density}
\end{equation}
For $\mu=0$ only one of these densities is necessary to determine the temperature. We use for this the Fermi density. Our approach corresponds to a Gaussian smearing of the density of a particle. These formulas apply to free quarks and antiquarks. We also have to consider the partons which are bound in hadrons. For practical reasons, especially to avoid a sudden increase of the density and hence the temperature when mesons are produced, we consider mesons like one parton.

By knowing $\rho_{Fi}$ and $\rho_{Bi}$, Eqs. \eqref{density_thermo1} and \eqref{density_thermo2} allow us to determine $T_i$ and $\mu_i$. For $\mu = 0$ and $T \gg m$ the relation between $T_i$ and $\rho_{Fi}$ is analytical (see the Appendix):
\begin{equation}
  \rho_F = \ \ell \ \frac{g}{\pi^2} \left( \frac{T}{\hslash c} \right)^3,
\end{equation}
where $\ell = 0.90154$ being a normalization factor for the Fermi integral and the degeneracy factor becomes $g = 2 \times 2 \times 2 \times 3 \times 3 = 36$. Then we find
\begin{equation}
    T_i   = (\hslash c) \left( \frac{\pi^2}{\ell\ g} \right)^{1/3}
                      \left( \sum_{j \ne i} R_{ij} \right)^{1/3}.
  \label{T_mu_def}
\end{equation}
In the general case Eqs. (\ref{density_thermo1}) and (\ref{density_thermo2}) have to be solved numerically. $T_i$ and $\mu_i$ vary from time step to time step and therefore the mass has to be also updated in each time step for each routine (collision, decay, and motion).

\begin{figure}
  \begin{center}
    \includegraphics[width=6.8cm]{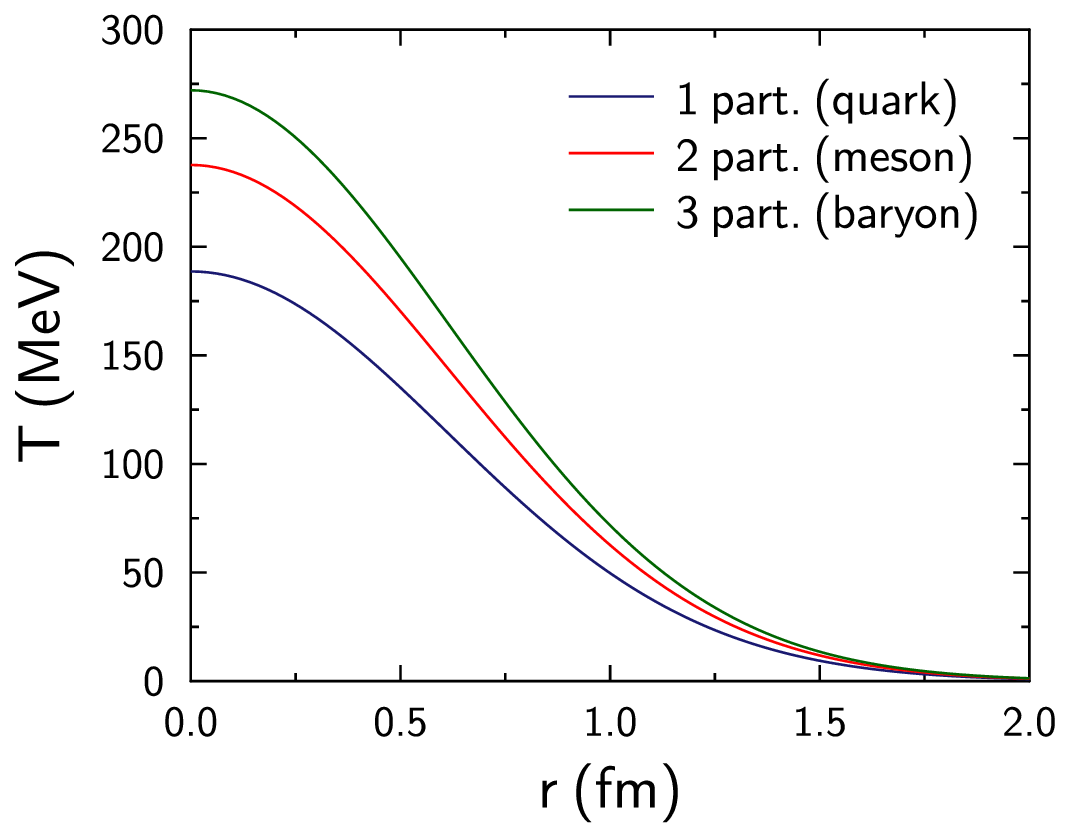}
  \end{center}
  \vskip -6mm
  \caption{(Color online) Local temperature at $r=0$ as a function of the distance $r$ and for a different number of fellow particles which all have a distance $r$ to the considered particle.\label{T_local}}
  \vspace{-5mm}
\end{figure}

How the distance between particles is related to the temperature can be demonstrated by assuming that there are one, two, or three particles which have an identical distance $r$ to the considered particle. This is shown in Fig. \ref{T_local}. The derivatives of the temperature with respect to the phase space variables, necessary to solve Eqs. \eqref{j1}, are developed in detail in the Appendix.

\subsection{Cross sections and decays}
\vskip -3mm
In addition to the potential interaction, which generates the mass of the partons, the partons interact also by collisions. Collisions are characterized by cross sections. As in all transport theories these cross sections are converted into a geometrical concept which allows us to decide which and when particles collide \cite{Aichelin1991}. If two particles come closer than $\Delta r = \sqrt{\sigma/\pi}$ a collision between the particles takes place. In the program the collision is executed at that time point at which the distance between the particles is minimal. 

In the present approach we have four types of processes :
\begin{equation*}
  q q \to q q,
\end{equation*}
\begin{equation*}
  q \bar{q} \to q \bar{q},
\end{equation*}
\begin{equation*}
  q \bar{q} \to M M \text{ (and backward process)},
\end{equation*}
and
\begin{equation*}
  M \to q \bar{q},
\end{equation*}
where quarks are characterized by $q$, antiquarks by $\bar{q}$, and mesons by $M$. These collisions increase the number of partons because for an expanding plasma $q \ \bar{q} \to M \ M$ , in which two partons are produced, is dominating over the backward reaction. Elastic collisions are primarily responsible for the thermalization of the plasma whereas the inelastic collisions are responsible for the hadronization. Both cross sections are small at temperatures well above the Mott temperature and therefore thermalization should happen only at the last stage of the expansion of the plasma shortly before the system hadronizes.

\begin{figure} [t]
 \begin{center}
    \includegraphics[width=6.7cm]{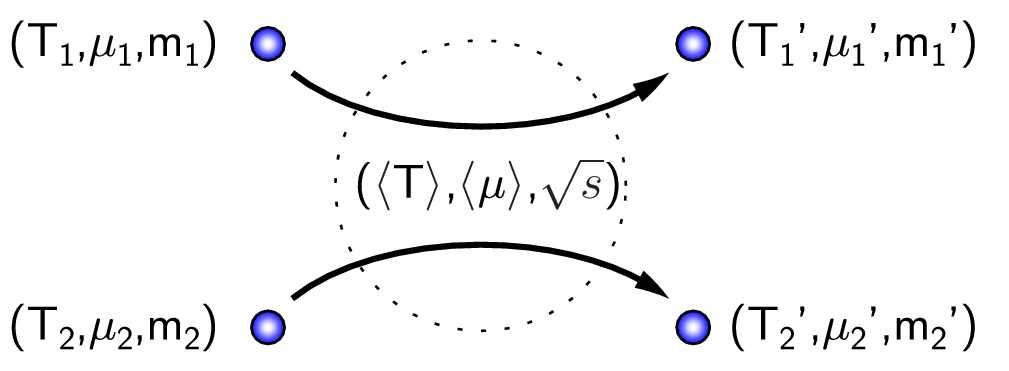}
  \end{center}
  \vskip -6mm
  \caption{(Color online) Collision between two particles in a medium.\label{particle_collision}}
  \vspace{-4mm}
\end{figure}

\begin{figure} [b]
  \vspace{-1mm}
  \begin{center}
    \includegraphics[width=4.3cm]{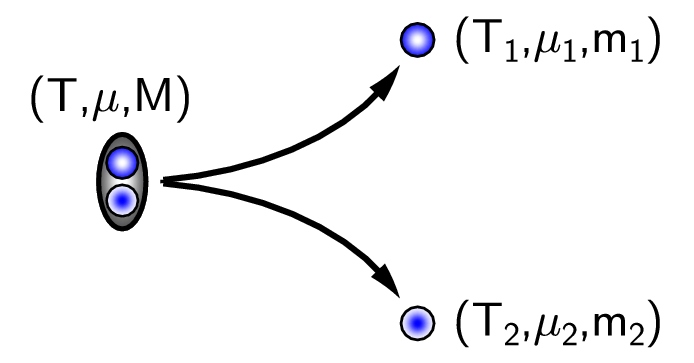}
  \end{center}
  \vskip -6mm
  \caption{(Color online) Decay of a meson in a medium.\label{meson_dissociation}}
\end{figure}

Figures \ref{particle_collision} and \ref{meson_dissociation} show a schema of a binary collision and a decay. The environment of the two particles which enter a collision may be different and therefore we do not expect them to have the same $T$ and $\mu$. In order to determine the cross section which depends on $T$ and $\mu$, we average over both particles.

In the NJL approach quarks are not confined. Nothing prevents them from expanding into the vacuum. Nevertheless, applying our cross sections to the expanding system we find that at the end of the expansion almost all partons are bound in hadrons. The reason for this is the very large cross section for hadronization close to $T_c$. Hence, when the system expands, close to $T_c$ hadron production becomes important. Hadrons formed slightly above $T_c$ live sufficiently long to survive until the system has passed $T_c$ and they become stable.     
\vspace{-5mm}
\subsection{Mean free path and time interval} \label{dt}
\begin{figure*}
 \begin{center}
    \includegraphics[width=6.4cm]{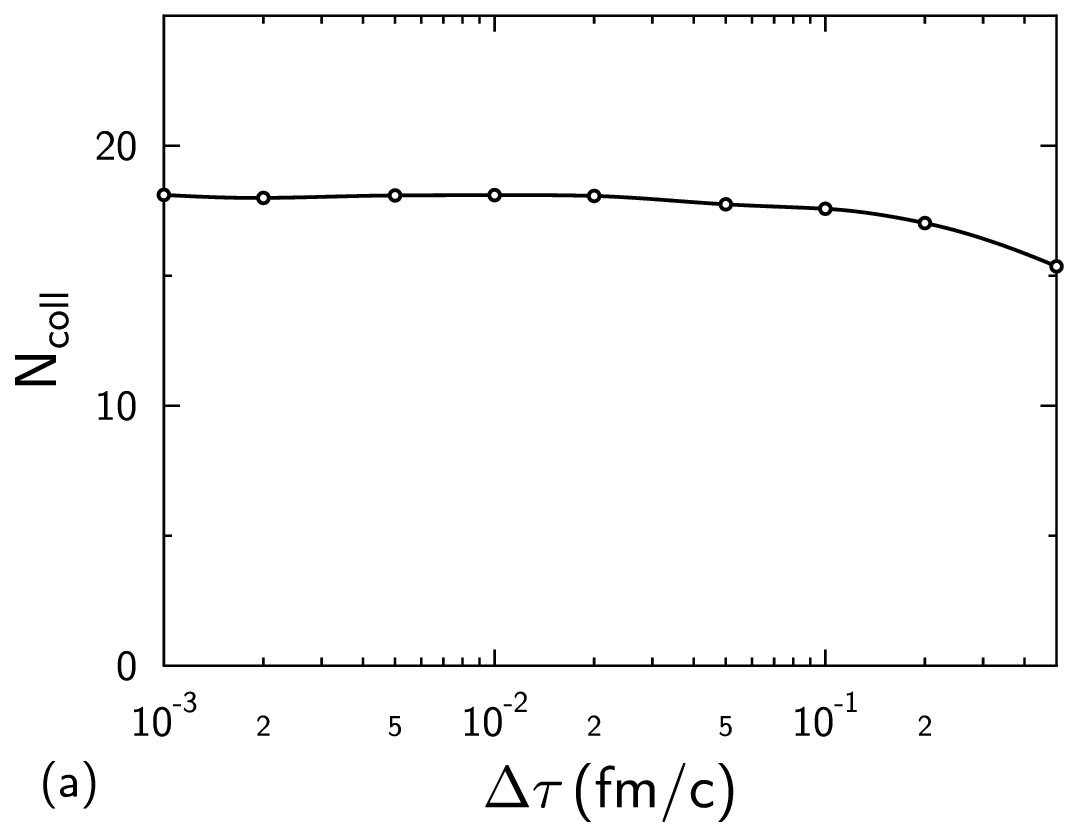}
    \includegraphics[width=6.4cm]{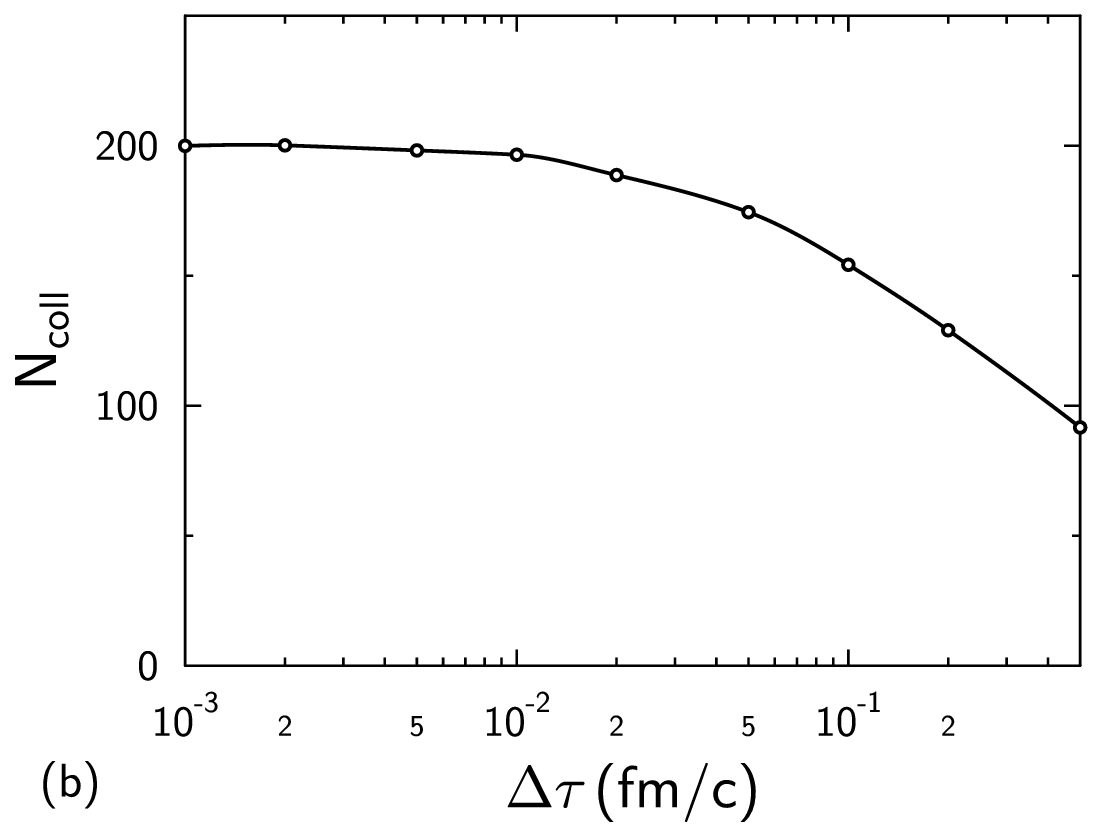}
  \end{center}
  \vskip -5mm
  \caption{Total number of collisions for the same initial condition in a box simulation as a function of the time step $\Delta\tau$  for constant cross sections of 1 mb (a) and 5 mb (b).\label{test_coll}}
\end{figure*}

\vskip -3mm
The geometrical interpretation of the cross section requires a careful study of the time step of the simulation. This can be seen by performing calculations in a box with periodic boundary conditions. As shown in Fig. \ref{test_coll} for the same initial condition (box size of: $a = 3$ fm, filled with 30 free particles, for a duration of 10 fm/$c$) the total number of collisions depends on the chosen time step. In Fig. \ref{test_coll}(a) [\ref{test_coll}(b)] we see the total number of collisions performed in the simulation program for the same initial condition as a function of the time step $\Delta \tau$ and for a total cross section of 1 (5) mb. We miss collisions if the time step is above a critical value of $\Delta \tau$.

\begin{figure*}
  \begin{center}
    \includegraphics[width=6.5cm]{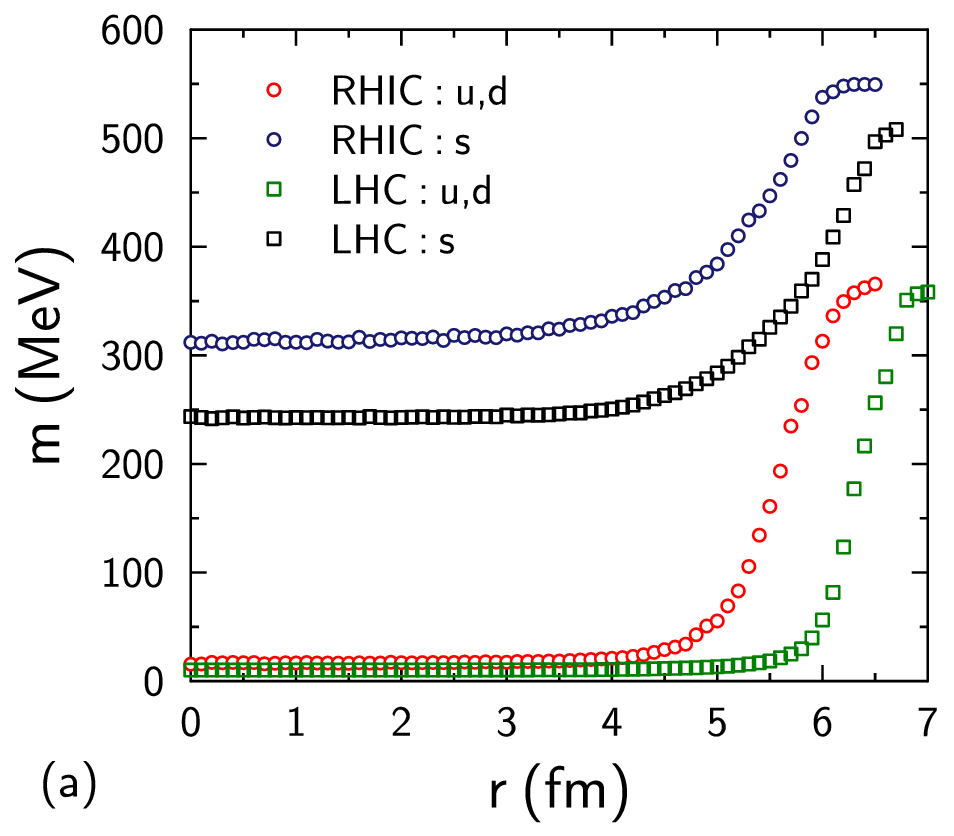}
    \includegraphics[width=6.5cm]{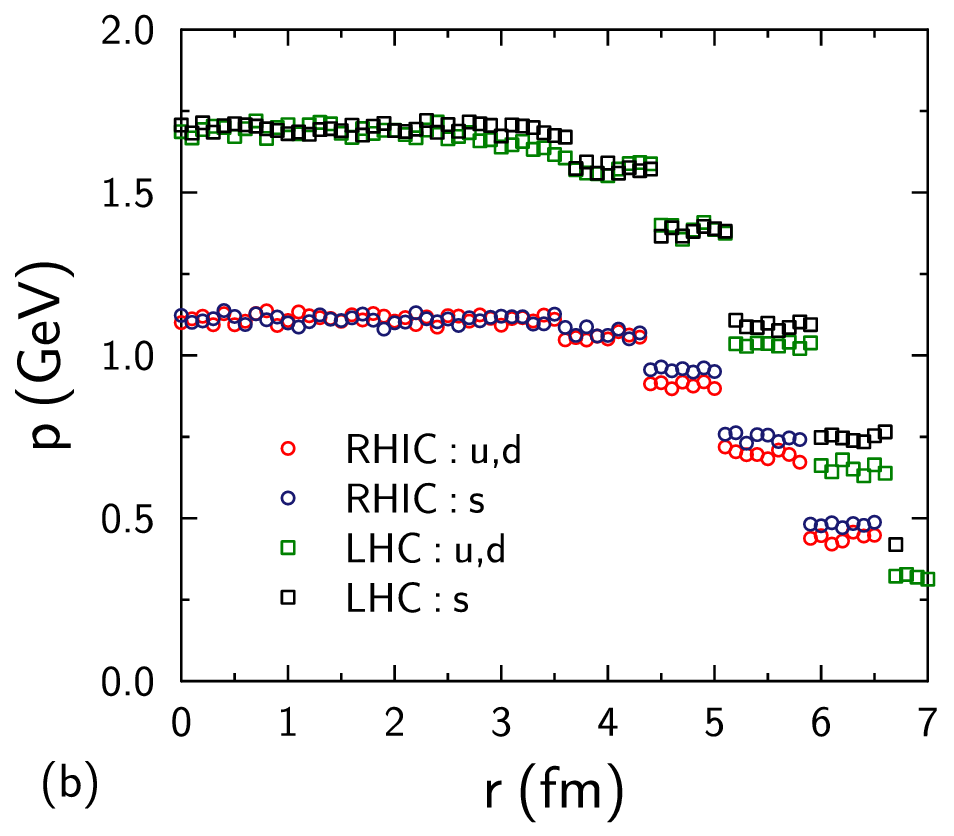} \\
    \includegraphics[width=6.5cm]{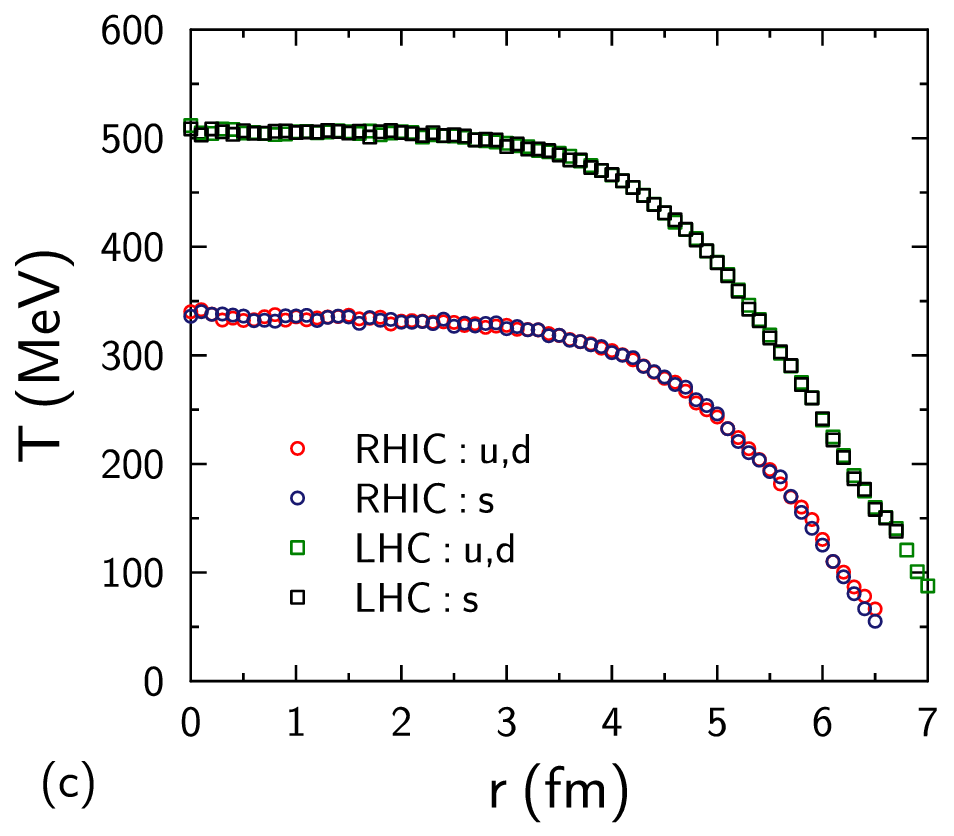}
    \includegraphics[width=6.5cm]{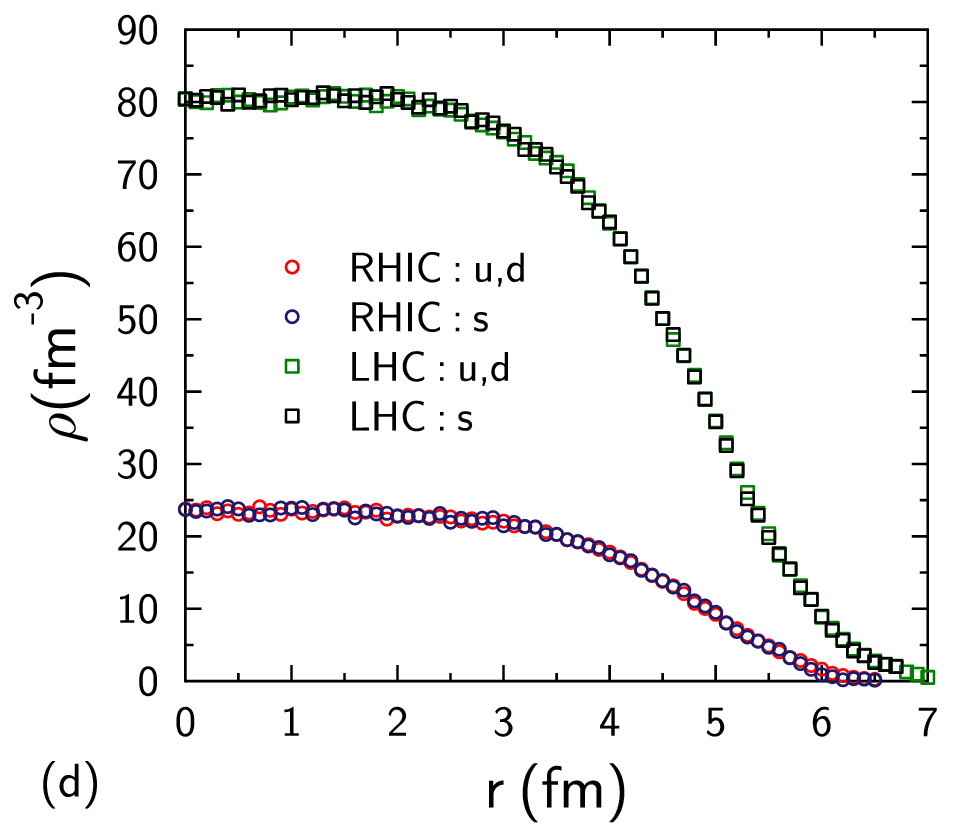}
  \end{center}
  \vskip -5mm
  \caption{(Color online) Distributions of the mass (a), momentum (b), temperature (c) and scalar density (d) for light and strange quarks as a function of the distance $r$ from the center and for central collisions.\label{distrib_r}}
  \vspace{-2mm}
\end{figure*}

The reason for this observation is that, if the mean free path is smaller than the time step, it is possible to have more than one collision per time step for the same particle, but numerically we only apply the first collision. Therefore the time step must be smaller than the mean free path. In our simulations the time between two collisions is given by the mean free path~$\ell$,
\begin{equation}
  \Delta \tau = \ell = \left( \sigma \ \rho \right)^{-1},
\end{equation}
which yields
\begin{equation*}
  \Delta \tau = 10 \text{ fm/c} \quad\text{for}\quad \sigma = 1 \text{ mb},
\end{equation*}
and
\begin{equation*}
  \Delta \tau = 2 \text{ fm/c} \quad\text{for}\quad \sigma = 5 \text{ mb},
\end{equation*}
Figure \ref{test_coll} shows that in order to have the correct number of collisions the time step has to be much smaller than $\ell$. We~need
\begin{equation*}
  \Delta \tau_{\text{opt}} = 5 \times 10^{-2} \text{ fm/c} \quad\text{for}\quad \sigma = 1 \text{ mb},
\end{equation*}
\begin{equation*}
  \Delta \tau_{\text{opt}} = 10^{-2} \text{ fm/c} \quad\text{for}\quad \sigma = 5 \text{ mb},
\end{equation*}
To be on the safe side in our simulations we use 
\begin{equation}
  \Delta \tau_{\text{opt}} = \left( 1000 \ \langle v_{\textrm{rel.}} \rangle
                                         \ \langle \sigma \rangle
                                         \ \langle \rho \rangle \right)^{-1},
  \label{mean_free_path_new}
\end{equation}
with the mean cross section $\langle \sigma \rangle$ calculated from the collisions during the previous time step, and the mean relative velocity $\langle v_{\textrm{rel.}} \rangle$ and the mean scalar density $\langle \rho \rangle$ calculated for each time step. 

%% file: section_5_results.tex
\vspace{-4mm}
\section{Results}
\vskip -3mm
\subsection{Set up of the simulations}
\vskip -4mm
The results we present here are obtained for simulations of Au-Au collisions at RHIC energies, $\sqrt{s_{NN}} = 200$~$A$GeV, or for Pb-Pb collisions at LHC energies, $\sqrt{s_{NN}} = 2760$~$A$GeV. We use the HPM (see Sec. \ref{hpm}) for the initial condition.

Figure \ref{distrib_r} displays the initial mass, momentum, temperature, and density of the light ($u$ and $d$) and heavy ($s$) quarks for RHIC and LHC initial conditions as a function of the position of the quarks measured with respect to the center of the collision. In the center the quark mass is close to the bare mass. The more the surface is approached, where the density is smaller and the temperature is lower, the more the mass increases, and close to the surface we approach the constituent quark mass. The decreasing temperature is also responsible for the decrease of the average momentum.

\begin{figure}
  \begin{center}
    \includegraphics[width=7.4cm]{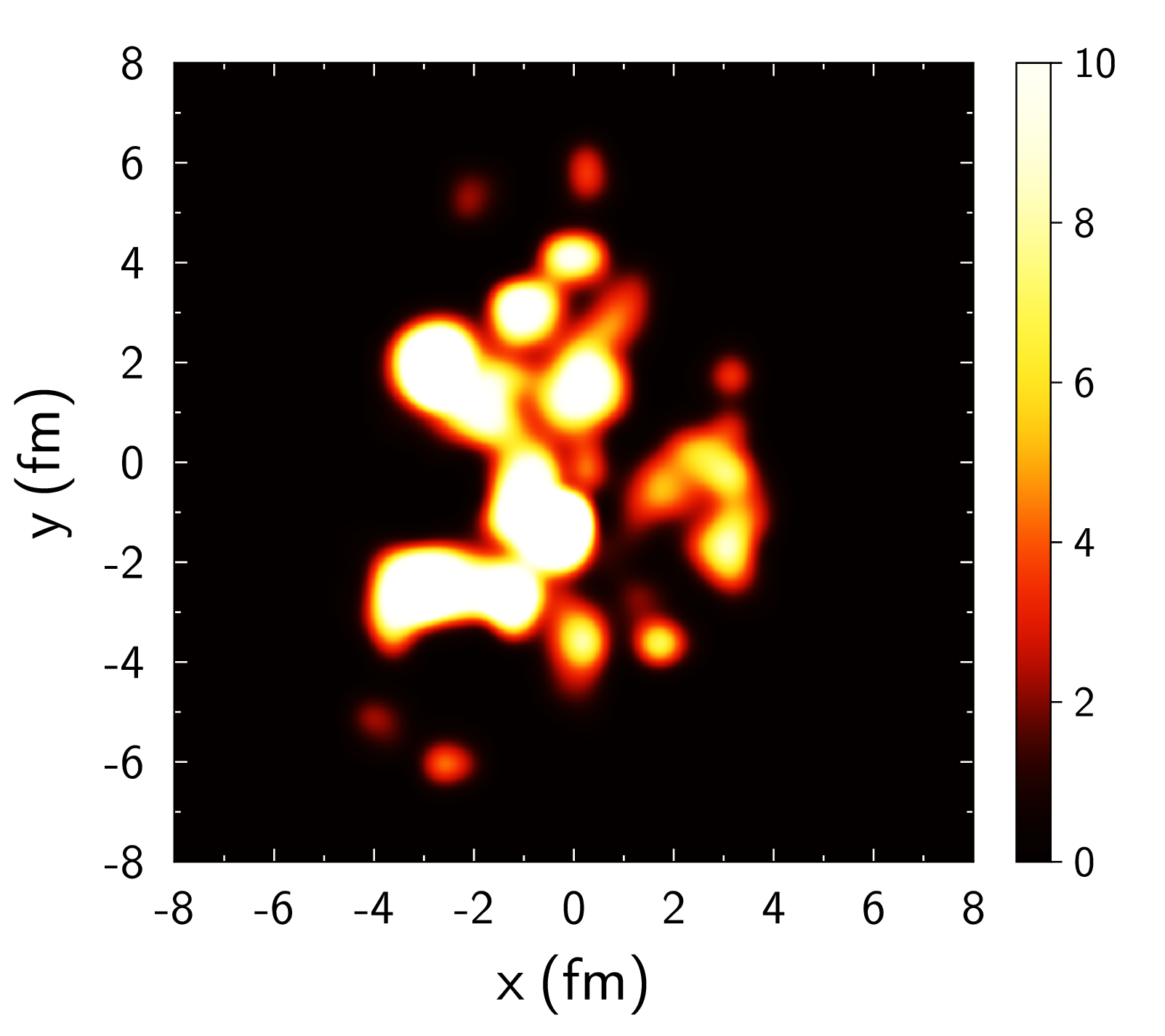}
  \end{center}
  \vskip -5mm
  \caption{(Color online) Energy density distribution (in GeV/fm$^3$) in the transverse plane $x,y$ for EPOS \cite{Werner2010} for a RHIC collision at $b = 6$ fm. Colored areas are QGP bubbles.\label{EPOS_energy_density}}
  \vspace{-1mm}
\end{figure}

We have chosen the smooth initial condition (Fig. \ref{T_initial}) to show in a simple and controlled way how the expansion takes place. For a quantitative comparison with experiments one has to include initial energy fluctuations, as shown in Fig. \ref{EPOS_energy_density} \cite{Werner2010}. Such fluctuations are visible in the final spectra of the mesons and can therefore not be neglected. We leave calculations with such more realistic initial conditions to future investigations.

\begin{figure} [b]
  \begin{center}
    \includegraphics[width=7.4cm]{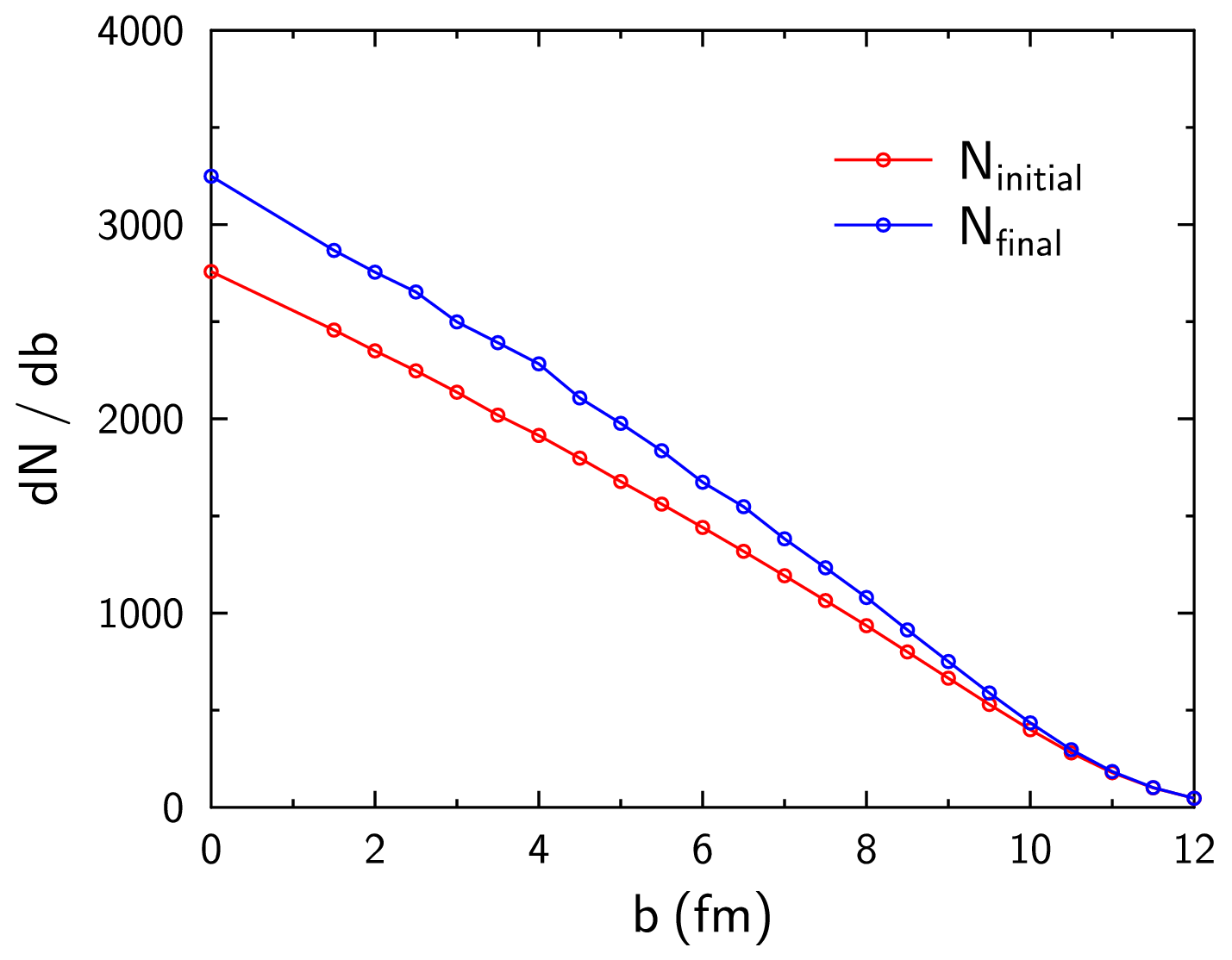}
  \end{center}
  \vskip -5mm
  \caption{(Color online) Number of initial and final particles as a function of the impact parameter $b$.\label{N_part_b}}
\end{figure}

Figure \ref{N_part_b} shows the number of initial partons and final particles (partons or hadrons) as a function of the impact parameter. (The number of partons can increase due to the decay of mesons.) The number of particles increases strongly with the centrality of the collisions and therefore also the computing time, which depends quadratically on the number of particles. In order to provide sufficient statistics the program has to be parallelized on modern computing architecture such as graphic cards.

\subsection{Check of the algorithm}
\begin{figure} [t]
  \begin{center}
    \includegraphics[width=6.9cm]{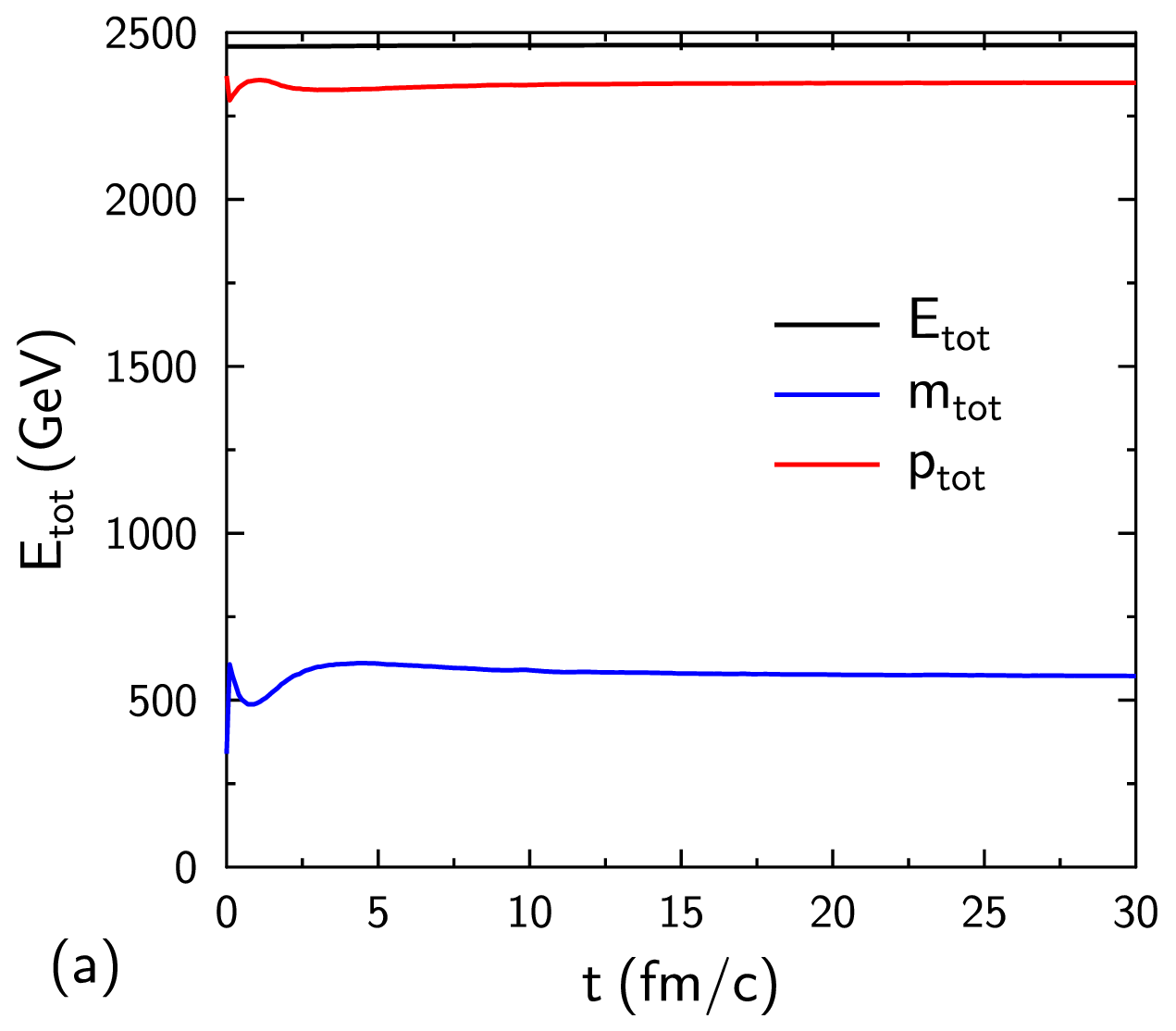}\\ [2mm]
    \includegraphics[width=6.9cm]{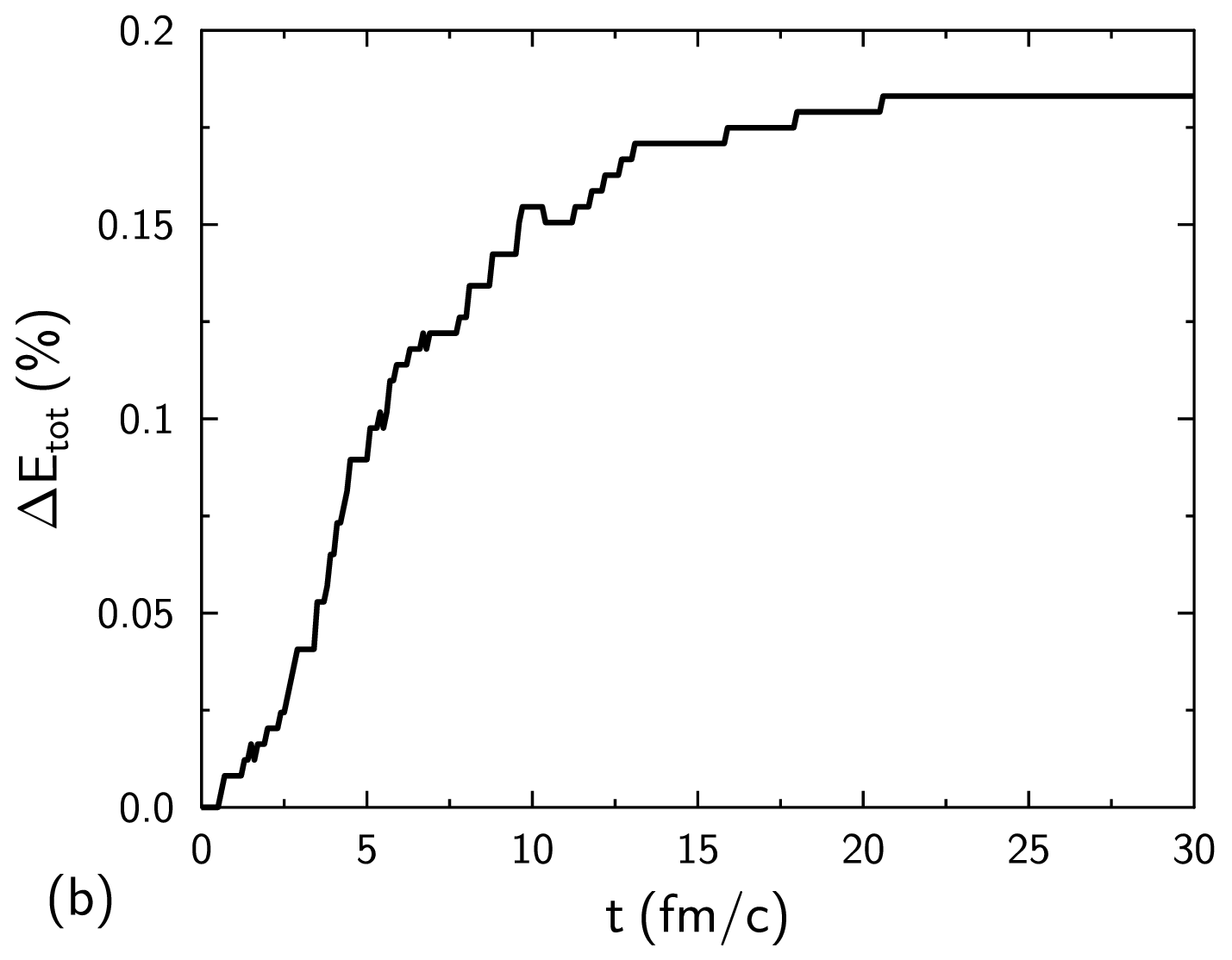}
  \end{center}
  \vskip -5mm
  \caption{(Color online) Evolution of the total energy, mass, and momentum of the system (a), and the variation of the total energy (b) for a central RHIC collision as a function of time.\label{delta_E}}
  \vspace{-3mm}
\end{figure}

\vskip -3mm
The most  important check for the consistency of the derivation and its numerical realization is for energy conservation. In a molecular dynamics calculation it has to be strictly conserved. Figure \ref{delta_E}(b) displays the variation of the total energy of the system as a function of time for a simulation of a central RHIC collision. Such a simulation contains a couple of thousand partons. We see that the energy varies by less than 0.2\%. The small variation of the total energy does not come from the solution of the differential equation (Euler or Runge-Kutta), but from the local density jump when a meson decay appears in a ``low"-density area.

\vspace{-5mm}
\subsection{First results}
\begin{figure}
  \begin{center}
    \includegraphics[width=7.2cm]{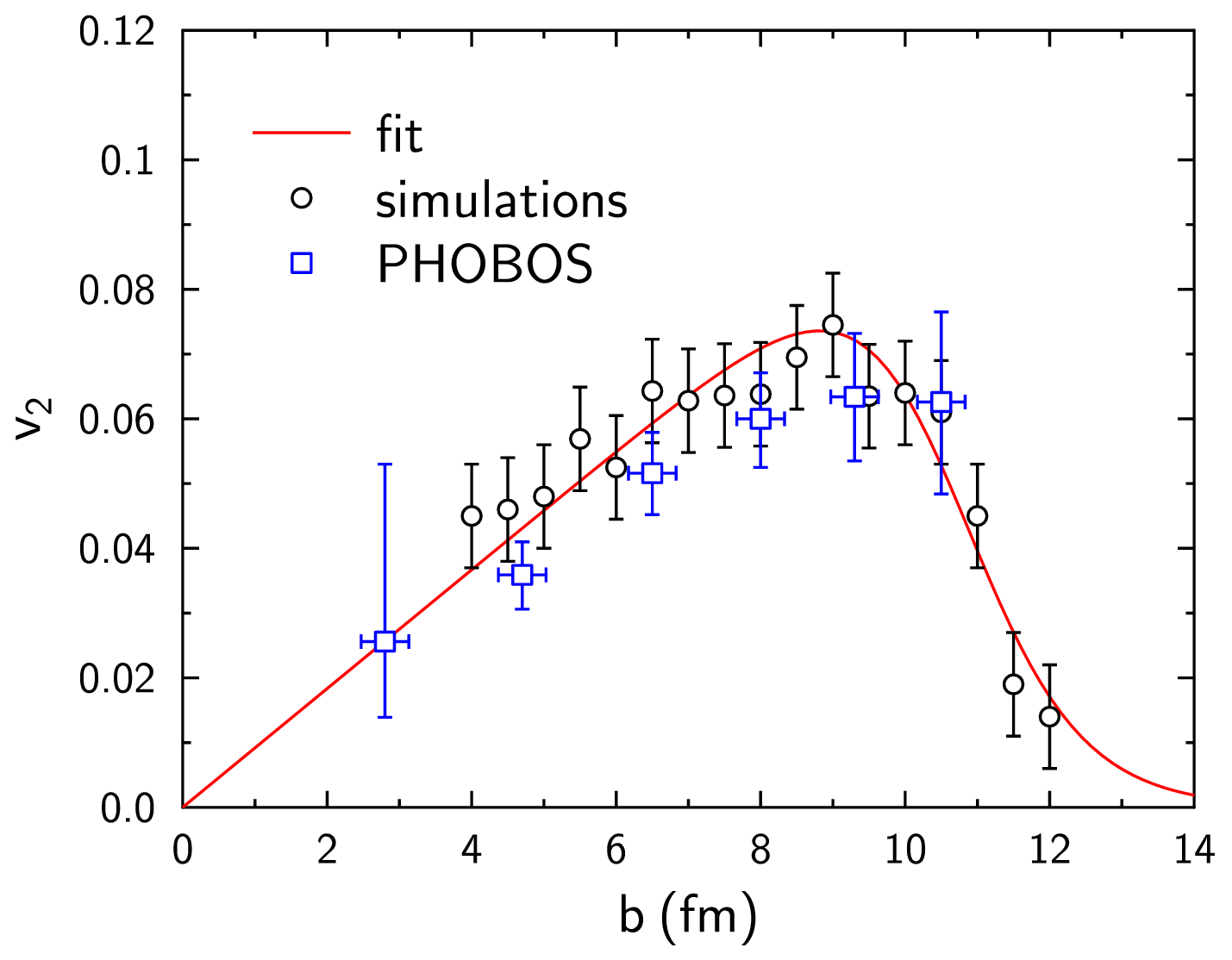}
  \end{center}
  \vskip -5mm
  \caption{(Color online) $v_2$ compared to experimental data from the PHOBOS experiment \cite{Alver2007} as a function of the impact parameter $b$.\label{v2}}
  \vspace{-2mm}
\end{figure}

\vskip -3mm
In this section we present some preliminary results which we have obtained for initial conditions adapted from RHIC and LHC heavy ion experiments. They show that basic observables are well reproduced in our approach. In Fig. \ref{v2} we display the elliptic flow, $v_2$, as a function of the impact parameter. These results are compared with the experimental data from the PHOBOS Collaboration \cite{Alver2007}. We see that the results of our approach agree quantitatively quite well with the experimental finding. In this  plot error bars come from the variations of the mean value after $N$ simulations.

\begin{figure} [b]
  \vspace{-4mm}
  \begin{center}
    \includegraphics[width=7.2cm]{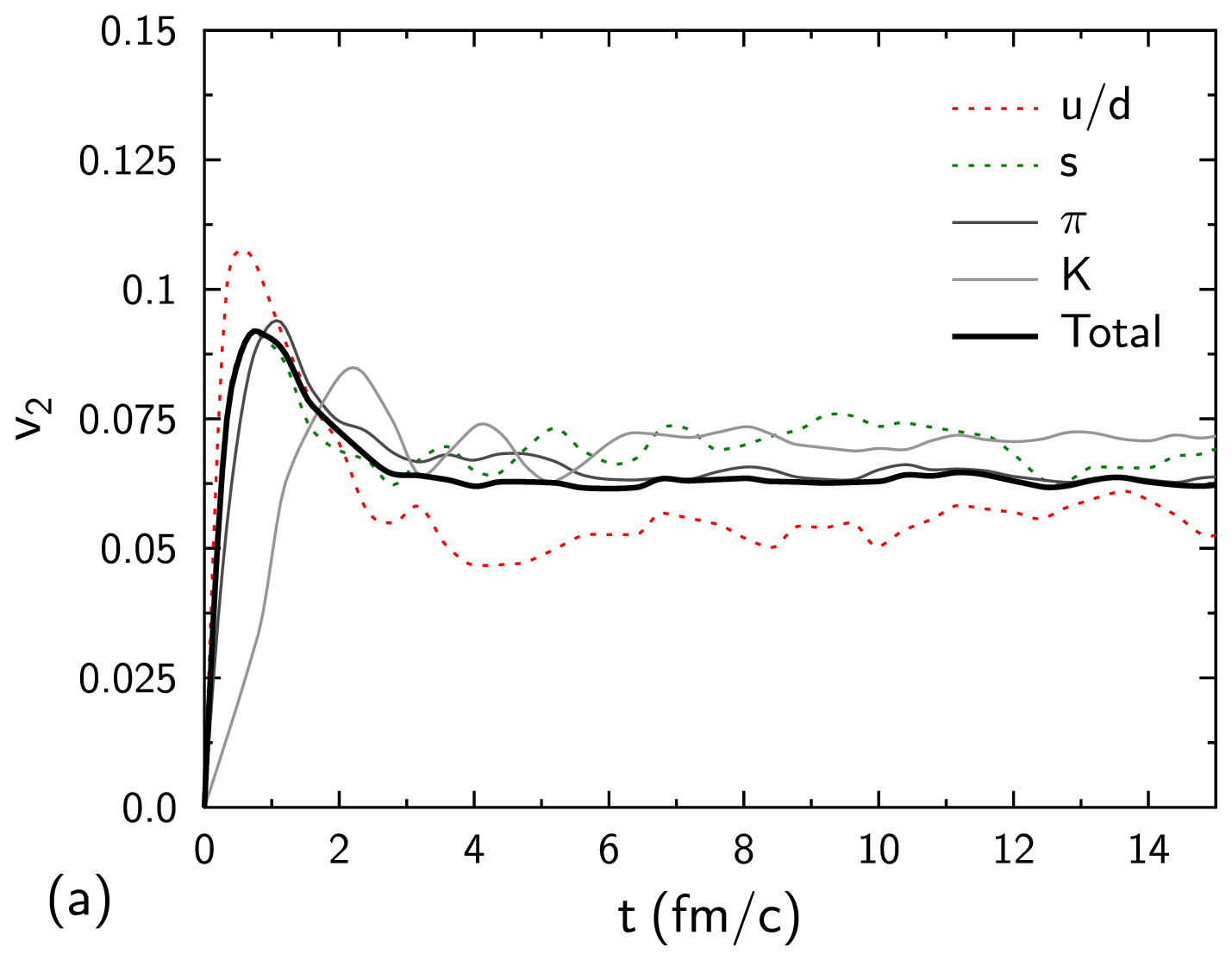}\\[3mm]
    \includegraphics[width=7.2cm]{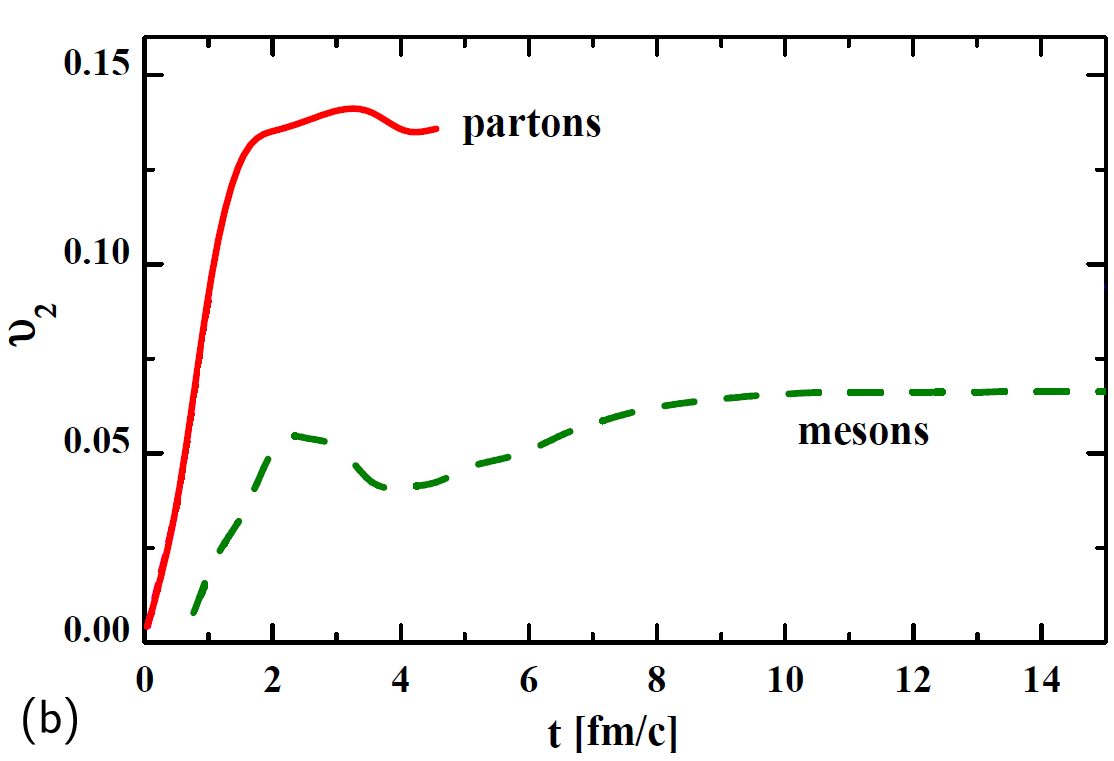}
  \end{center}
  \vskip -5mm
  \caption{(Color online) Time evolution of the elliptic flow $v_2$ for $b=6$ fm (a), and the comparison with PHSD calculations for similar initial conditions \cite{Cassing2008} (b).\label{comparison_v2_formation}}
\end{figure}

Figure \ref{comparison_v2_formation} displays how the elliptic flow develops as a function of time. In Fig. \ref{comparison_v2_formation}(a) we display our results; those of PHSD calculations \cite{Cassing2008} are shown in Fig. \ref{comparison_v2_formation}(b). By definition initially there is no elliptic flow (under the assumption of thermal equilibrium). In the two calculations the flow develops very similarly and both calculations agree also on the final value. In our case, despite the small cross section of about 4--6 mb, we observe initially many collisions due to the high density. These collisions thermalize the plasma rapidly and lead to an elliptic flow in less than 1 fm/$c$. The flow is lowered later by the change of the masses in NJL [see Fig. \ref{delta_E} (a)].

\begin{figure}
  \begin{center}
    \includegraphics[width=7.2cm]{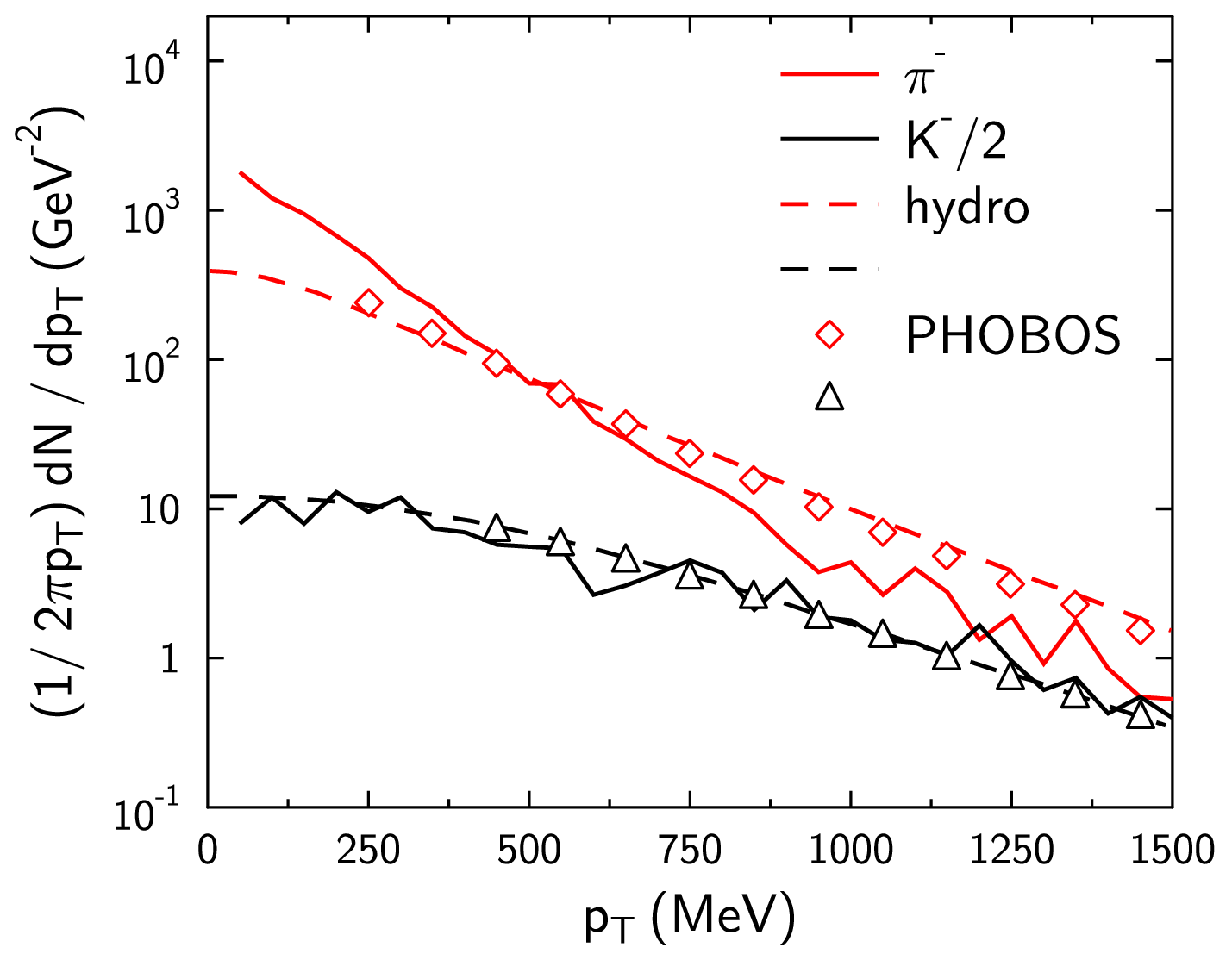}
  \end{center}
  \vskip -6mm
  \caption{(Color online) The $\mathrm d N / 2 \pi \mathbf{p}_\textrm{T} \mathrm d \mathbf{p}_\textrm{T}$ spectrum for $b = 4$ fm, and the results of a hydrodynamical calculation and of the experiment for similar conditions (centrality 0\%-5\%) \cite{Schenke2010}.\label{dN_dpt}}
  \vspace{-4mm}
\end{figure}

\begin{figure} [b]
  \vspace{-6mm}
  \begin{center}
    \includegraphics[width=6.3cm]{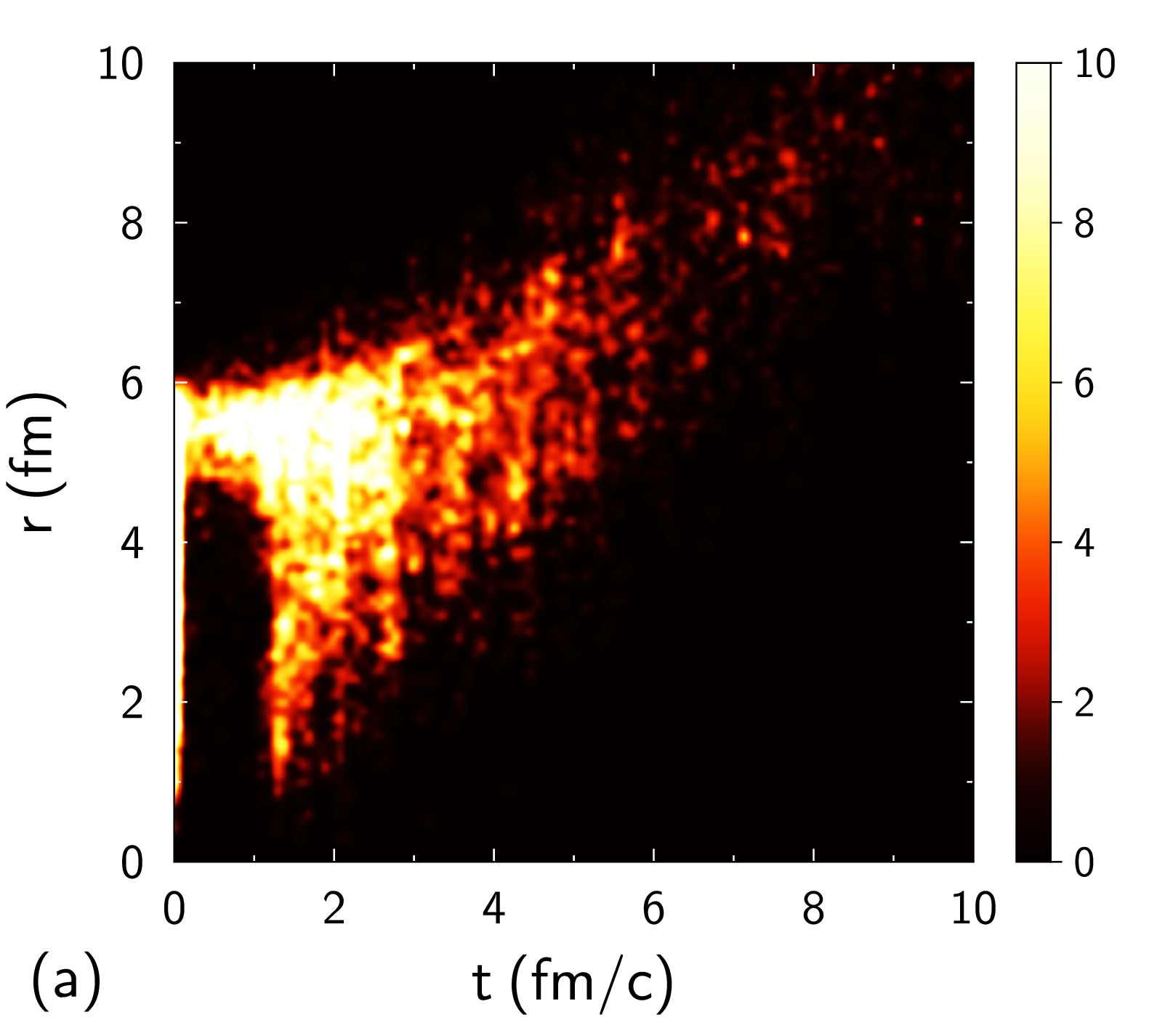}\\
    \includegraphics[width=6.3cm]{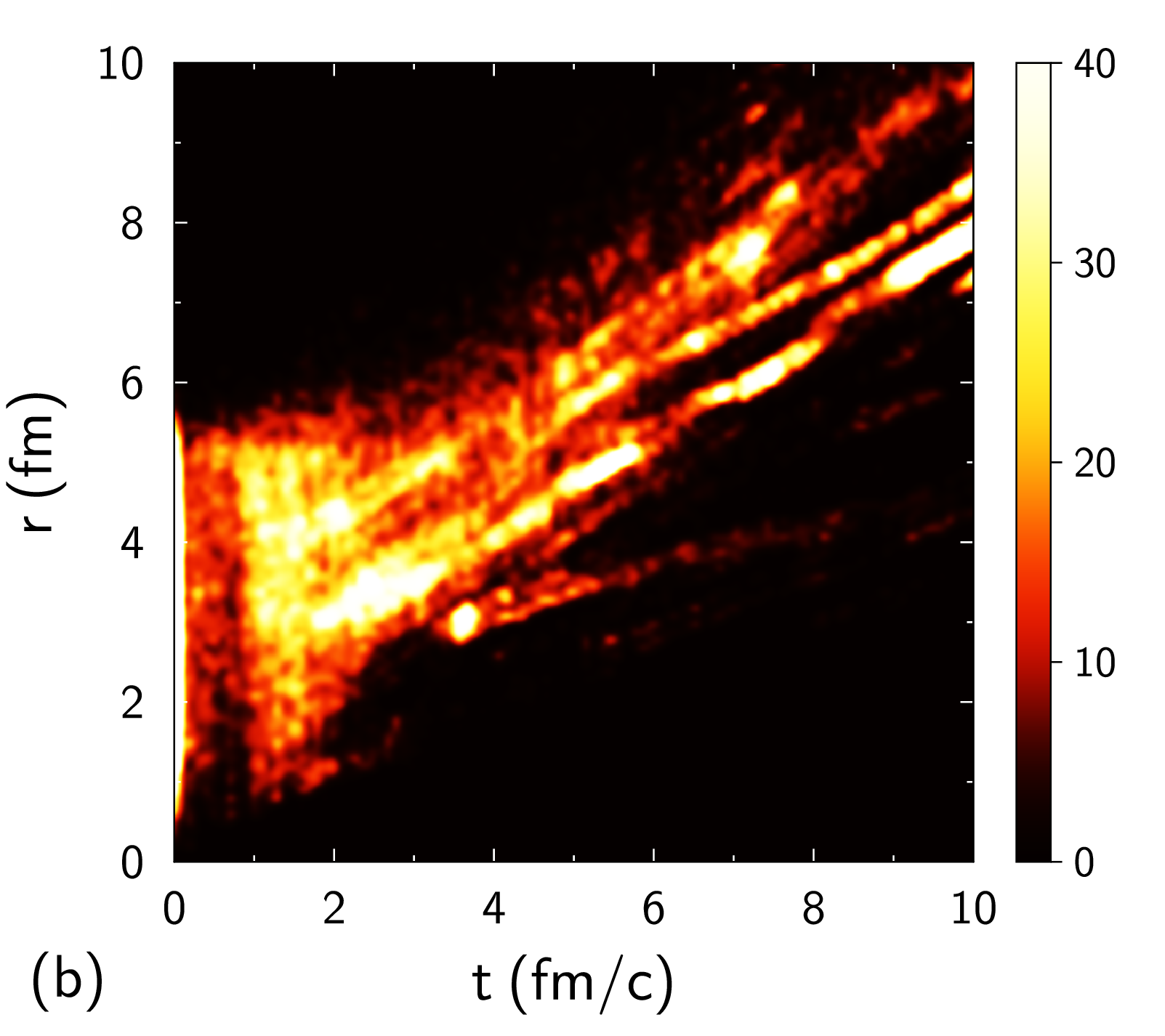}
  \end{center}
  \vskip -6mm
  \caption{(Color online) ($t,\mathbf{r}$) distributions for inelastic (a) and elastic (b) collisions at $b = 0$ fm for RHIC conditions.\label{collision_evolution}}
\end{figure}

Experimentally it has been found that the transverse momentum spectra of $\pi$ and $K$ have a different shape \cite{Schnedermann1993}. This can be seen in Fig. \ref{dN_dpt} where we compare the experimental data with results from hydrodynamical calculations \cite{Schenke2010} and our results. We observe the same difference of the slopes as seen in experiments which is usually attributed to the hydrodynamical evolution of the system.

Figure \ref{collision_evolution} displays a contour plot of the number of collisions as a function of the distance to the center of the initial ellipse $r$ and time $t$ for inelastic collisions [Fig. \ref{collision_evolution}(a)] and elastic collisions [Fig. \ref{collision_evolution}(b)].

Initially we have a very high density zone where elastic and inelastic collisions take place frequently because the mean free path is small despite the small cross section. When the system expands the density becomes lower but the cross section does not increase. Therefore we observe fewer collisions. When the system approaches the critical temperature the cross sections becomes very large; this largely compensates for the decrease of the density and there the collision rate becomes large again for elastic as well as for inelastic collisions. Here the hadrons are created which finally survive.

\begin{figure}
  \begin{center}
    \includegraphics[width=7.7cm]{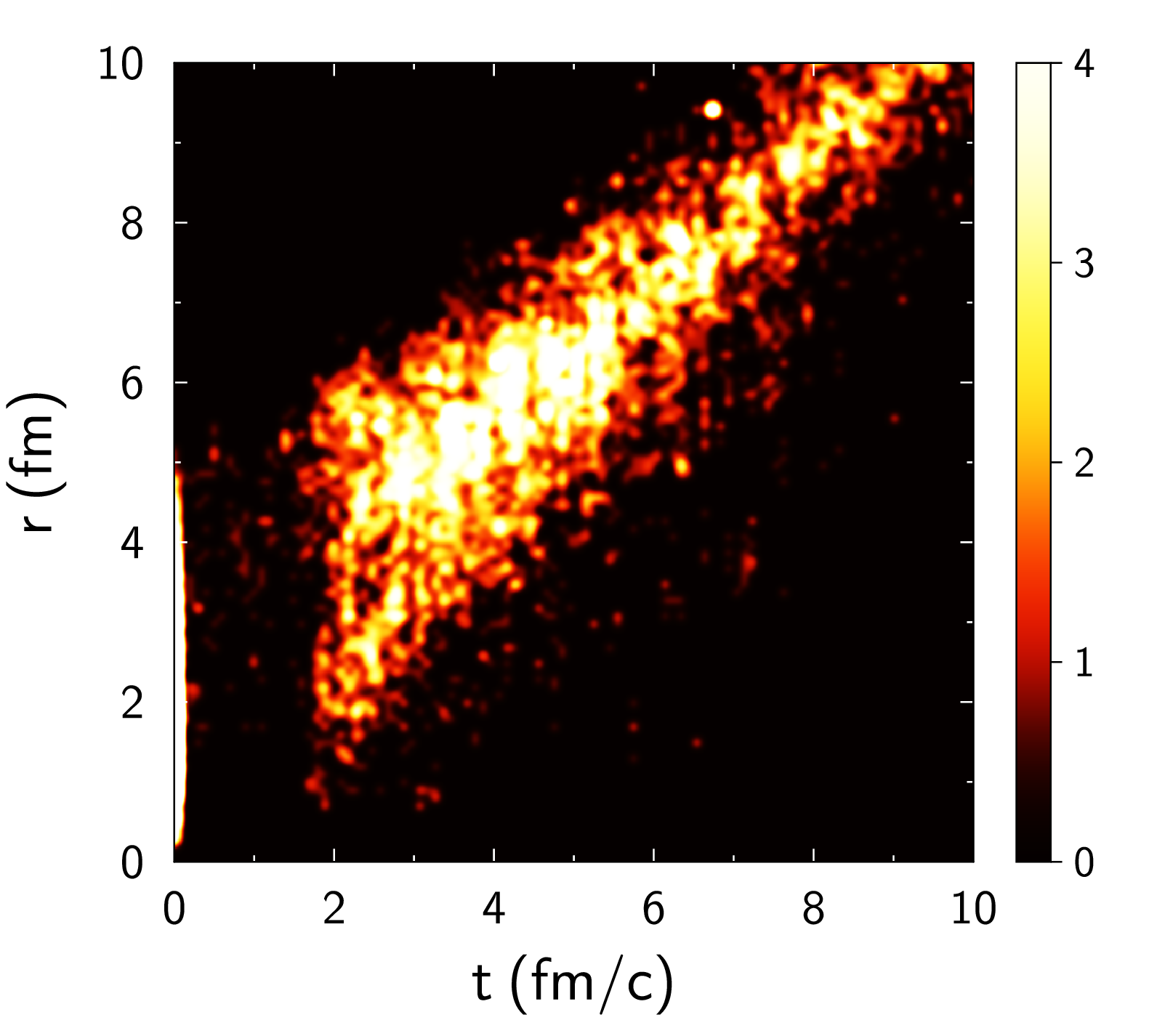}
  \end{center}
  \vskip -5mm
  \caption{(Color online) ($t,\mathbf{r}$) distributions for inelastic collisions at $b = 9$ fm for LHC conditions.\label{total_collision_evolution_bis}}
  \vspace{-2mm}
\end{figure}

For the LHC initial condition, Fig. \ref{total_collision_evolution_bis}, we see the same phenomenon but a longer lifetime of the plasma. In contradistinction to the simulation for RHIC energies the corona partons do not hadronize early but the stream of partons from the interior heats up the surface. So the system expands in the quark phase and hadronization takes place only much later over a large space-time area.

\begin{figure}
  \begin{center}
    \includegraphics[width=7.2cm]{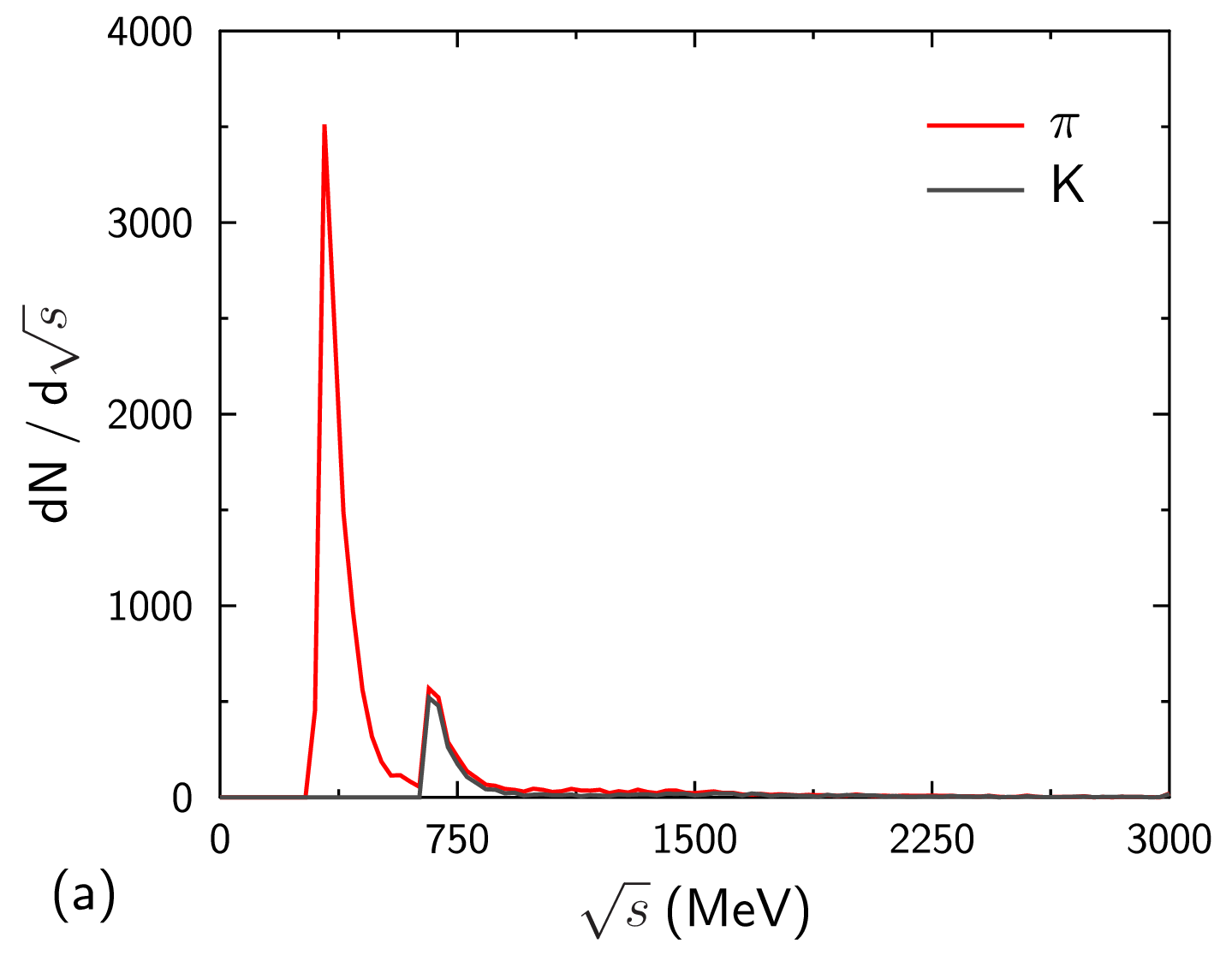}\\[3mm]
    \includegraphics[width=7.2cm]{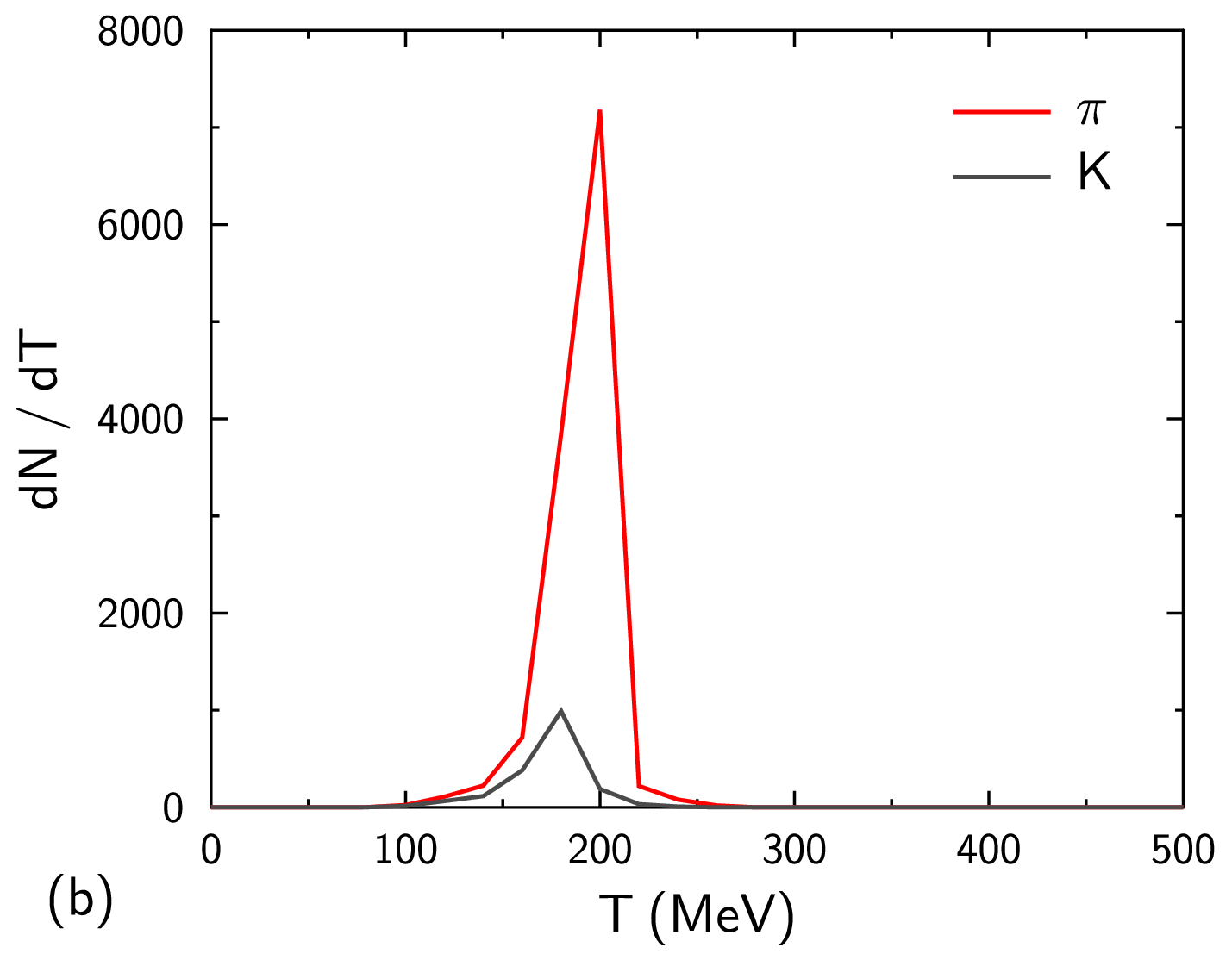}
  \end{center}
  \vskip -5mm
  \caption{(Color online) $\sqrt{s}$ distribution for inelastic collisions (a), and the distribution of the temperature at the production points, $T$,  for pions and kaons (b).\label{production_spectrum}}
  \vspace{-2mm}
\end{figure}

Figure \ref{production_spectrum} shows the distribution of $\sqrt{s}$ and of $T$ at which the final hadrons are produced. We see a broad distribution around the critical temperature and not a single freeze-out temperature as assumed in the Cooper-Frye formula \cite{Cooper1974}, which is used to created hadrons in hydrodynamical calculations \cite{Schenke2010}. The temperature at the $K$ production points is slightly lower than that for the $\pi$ production points, as expected by the NJL cross sections.

\begin{figure*}
  \begin{center}
    \includegraphics[width=6.5cm]{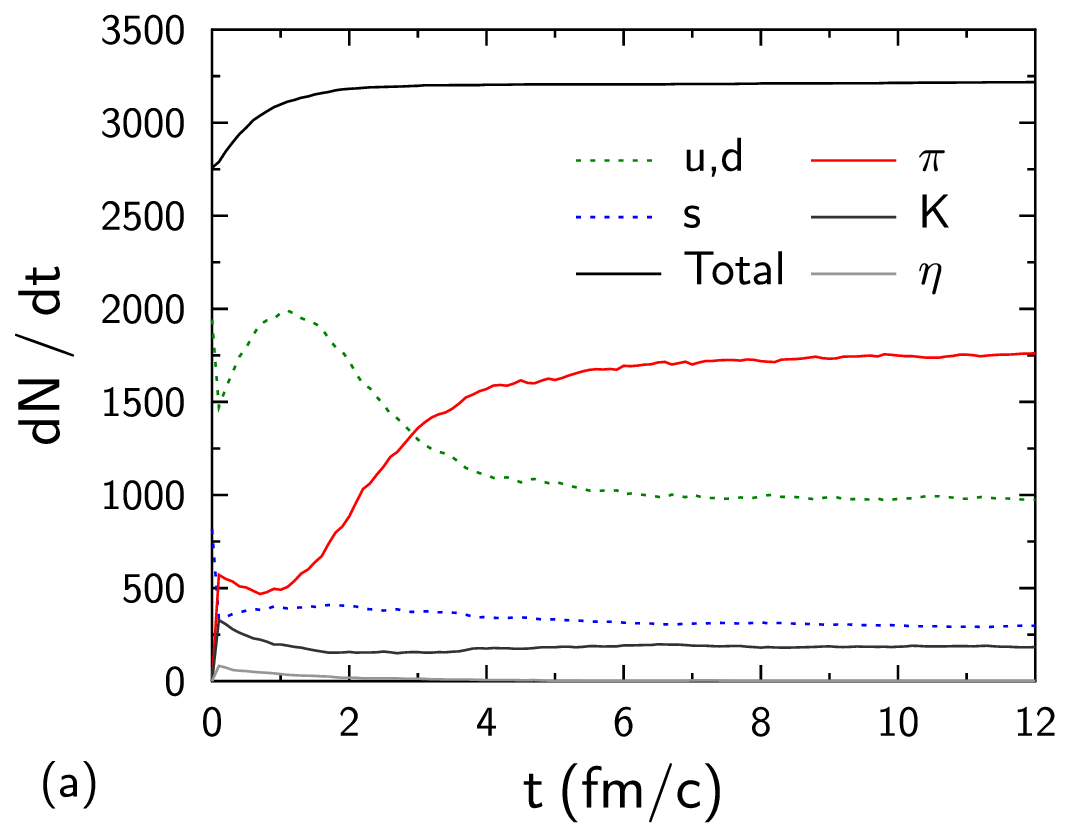} \ \ \
    \includegraphics[width=6.5cm]{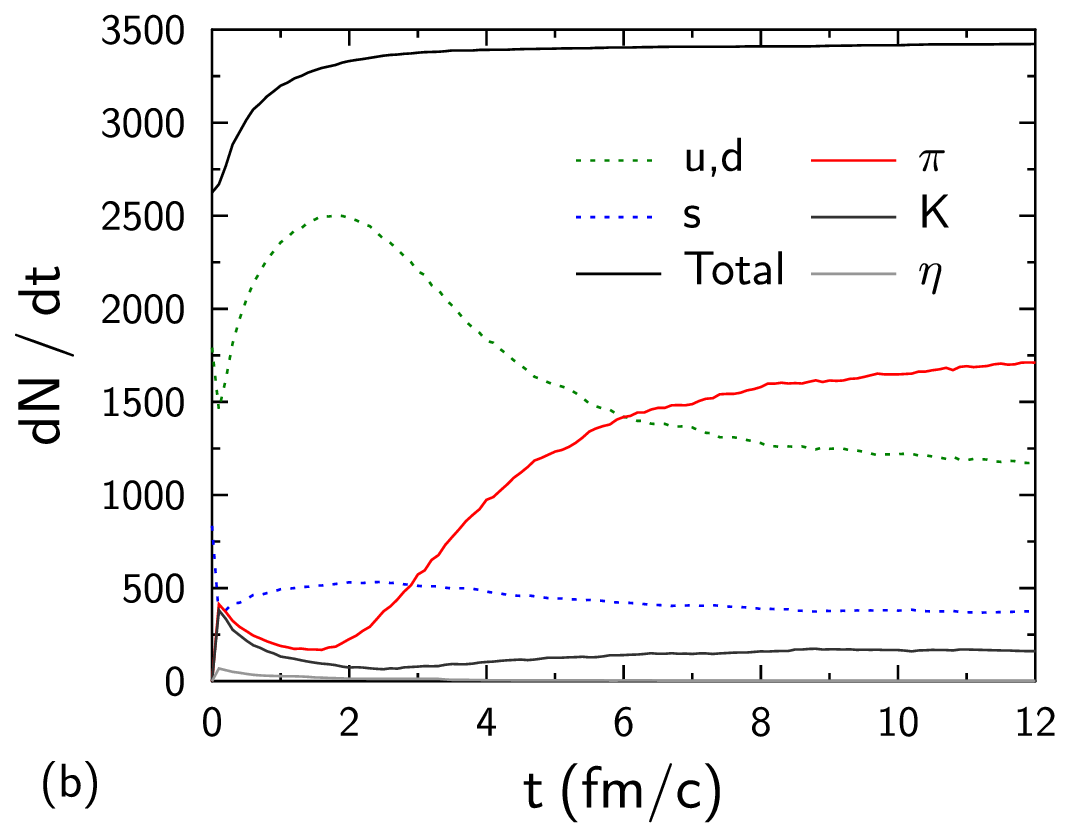}\\[3mm]
    \includegraphics[width=9cm]{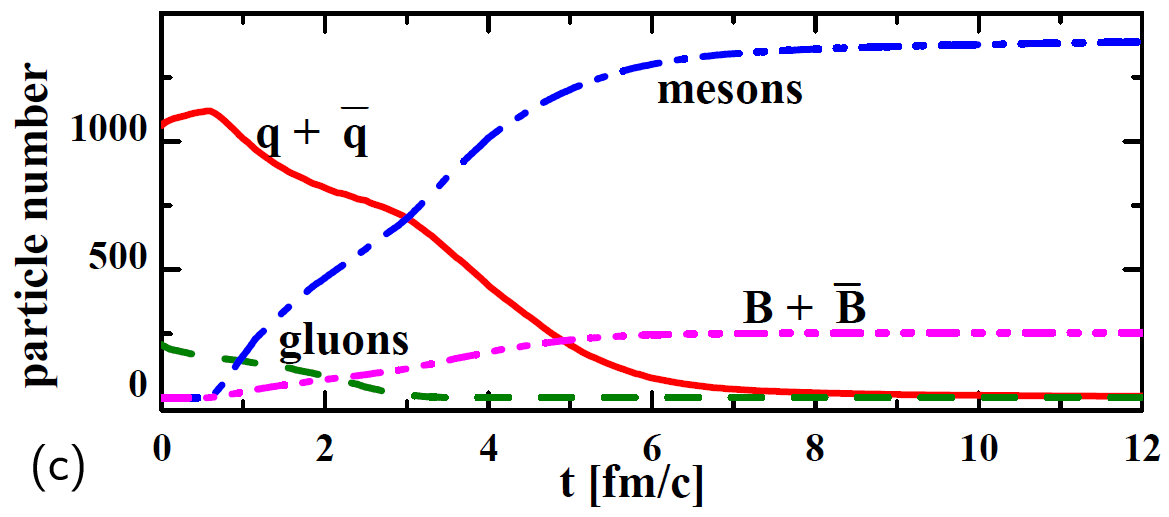}
  \end{center}
  \vskip -5mm
  \caption{(Color online) $\mathrm d N / \mathrm d t$ for our simulations at $b=0$ fm and RHIC conditions (a) and $b=9$ and LHC conditions (b) and for PHSD for $b=0$ fm and RHIC conditions (c) \cite{Cassing2008}.\label{multiplicity_time}}
  \vspace{-2mm}
\end{figure*}

Figure \ref{multiplicity_time} compares the hadronization in our approach [Fig. \ref{multiplicity_time}(a) and \ref{multiplicity_time}(b)] with the results of PHSD calculations [Fig. \ref{multiplicity_time}(c)] \cite{Cassing2008}. We observe in both calculations a hadronization time of around 5 fm/$c$ for the RHIC conditions. The difference between the particle numbers from PHSD and those in our model comes from the fact that we have different initial conditions. The hadronization time is longer for the LHC initial condition [Fig. \ref{multiplicity_time}(b)] due to the higher density of partons.

%% file: section_6_summary.tex
\section{Summary}
\vskip -3mm
We have presented in this paper a relativistic molecular dynamics approach. We have shown that for a specific choice of constraints it is possible to recover the \emph{classical} relativistic equations of motion. These constraints give us physical trajectories with causal motion and conservation of energy of a strongly interacting system. Using the Nambu--Jona-Lasinio Lagrangian to describe the potential interactions and the scattering among the partons we find that is it is possible to model the expansion of a quark-antiquark plasma. Close to the cross over the elastic as well as the hadronization cross sections increase very rapidly. The large hadronization cross section is the reason why the large majority of quarks form mesons which can finally be observed.

Our results show that a approach which does not enforce thermal equilibrium like hydrodynamics and in which the transition to the hadronic world is not sudden, as in the Cooper-Frye approach, used frequently in hydrodynamical calculations,  gives qualitative agreement with some key observables. Further studies involving an in-depth comparison with existing models will be the subject of a future publication.

The approach is in spirit close to the PHSD approach but differs completely as far as the temperature and density dependence of the mass of the quarks is concerned. Therefore it will be fruitful to compare the observables obtained in both approaches for the same initial condition.

%% file: acknowledgment.tex
\vspace{-3mm}
\section*{Acknowledgment}
\vskip -3mm
We thank E. Bratkovskaya and W. Cassing for fruitful discussions and E. Bratkovskaya also for her continuous interest. RM especially appreciates the ``HIC for FAIR'' framework of the ``LOEWE'' program for support of this work. The computational resources were provided by the LOEWE-CSC.

%% file: appendix.tex
\vspace{-3mm}
\section*{Appendix}\appendix
\vskip -3mm
\subsection{Relativistic calculations}
\vskip -2mm
\subsubsection{\bf\emph{Derivatives of transverse distances}} \label{deriv_qt}
\vskip -4mm
The calculation of the derivatives for transverse distances can be done rigorously as follow:
\begin{eqnarray}
    \frac{\partial {q_T}_{ij}^2}{\partial q_{k\nu}}&=&
    2 {q_T}_{ij \mu} \frac{\partial {q_T}_{ij}^\mu}{\partial q_{k\nu}}  \nonumber\\
                                                   &=&
    2 {q_T}_{ij \mu} \frac{\partial}{\partial q_{k\nu}}
    \big[ {q_{ij}}^{\mu} - (q_{ij \sigma} u_{ij}^{\sigma}) u_{ij}^{\mu} \big]  \nonumber\\
                                                   &=&
    2 {q_T}_{ij \mu} \bigg[ (\delta_{ik} - \delta_{jk}) \eta^{\mu\nu} \nonumber\\
    & &- \frac{\partial (q_{ij \sigma} u_{ij}^{\sigma})}{\partial q_{k\nu}} u_{ij}^{\mu}
     - (q_{ij \sigma} u_{ij}^{\sigma}) \ \frac{\partial u_{ij}^{\mu}}{\partial q_{k\nu}} \bigg],\nonumber\\
    \frac{\partial {q_T}_{ij}^2}{\partial q_{k\nu}}&=&
    2 {q_T}_{ij}^\nu (\delta_{ik} - \delta_{jk}),
\end{eqnarray}
\begin{eqnarray}
    \frac{\partial {q_T}_{ij}^2}{\partial p_{k\nu}}&=&
    2 {q_T}_{ij \mu} \frac{\partial {q_T}_{ij}^\mu}{\partial p_{k\nu}}  \nonumber\\
                                                   &=&
    2 {q_T}_{ij \mu} \frac{\partial}{\partial p_{k\nu}}
    \big[ {q_{ij}}^{\mu} - (q_{ij \sigma} u_{ij}^{\sigma}) u_{ij}^{\mu} \big]  \nonumber\\
                                                   &=&
    2 {q_T}_{ij \mu} \bigg[ 0 - \frac{\partial (q_{ij \sigma} u_{ij}^{\sigma})}{\partial p_{k\nu}}
    u_{ij}^{\mu} - (q_{ij \sigma} u_{ij}^{\sigma}) \
    \frac{\partial u_{ij}^{\mu}}{\partial p_{k\nu}} \bigg],\nonumber\\
    \frac{\partial {q_T}_{ij}^2}{\partial p_{k\nu}}&=&
    - 2 {q_T}_{ij}^\nu (\delta_{ik} + \delta_{jk})
    \frac{(q_{ij \sigma} u_{ij}^{\sigma})}{\sqrt{p_{ij}^2}} \stackrel{\text{cms}}{=} 0,
\end{eqnarray}
and the same kind of derivatives can be found for $q_T'$:
\begin{eqnarray}
    \frac{\partial {q_T'}_{ij}^2}{\partial q_{k\nu}}&=&
    2 {q_T'}_{ij \mu} \frac{\partial {q_T'}_{ij}^\mu}{\partial q_{k\nu}}  \nonumber\\
                                                    &=&
    2 {q_T'}_{ij \mu} \frac{\partial}{\partial q_{k\nu}}
    \big[ {q_{ij}}^{\mu} - (q_{ij \sigma} U^{\sigma}) U^{\mu} \big]  \nonumber\\
                                                    &=&
    2 {q_T'}_{ij \mu} \bigg[ (\delta_{ik} - \delta_{jk}) \eta^{\mu\nu} \nonumber\\
    & &- \frac{\partial (q_{ij \sigma} U^{\sigma})}{\partial q_{k\nu}}
    U^{\mu} - (q_{ij \sigma} U^{\sigma}) \
    \frac{\partial U^{\mu}}{\partial q_{k\nu}} \bigg],\nonumber\\
    \frac{\partial {q_T'}_{ij}^2}{\partial q_{k\nu}}&=&
    2 {q_T'}_{ij}^\nu (\delta_{ik} - \delta_{jk}),
\end{eqnarray}
\begin{eqnarray}
    \frac{\partial {q_T'}_{ij}^2}{\partial p_{k\nu}}&=&
    2 {q_T'}_{ij \mu} \frac{\partial {q_T'}_{ij}^\mu}{\partial p_{k\nu}}  \nonumber\\
                                                    &=&
    2 {q_T'}_{ij \mu} \frac{\partial}{\partial p_{k\nu}}
    \big[ {q_{ij}}^{\mu} - (q_{ij \sigma} U^{\sigma}) U^{\mu} \big]  \nonumber
\end{eqnarray}
\begin{eqnarray}
                                                    &=&
    2 {q_T'}_{ij \mu} \bigg[ 0 - \frac{\partial (q_{ij \sigma} U^{\sigma})}{\partial p_{k\nu}}
    U^{\mu} - (q_{ij \sigma} U^{\sigma}) \
    \frac{\partial U^{\mu}}{\partial p_{k\nu}} \bigg],\nonumber\\
    \frac{\partial {q_T'}_{ij}^2}{\partial p_{k\nu}}&=&
    - 2 {q_T'}_{ij}^\nu \frac{(q_{ij \sigma} U^{\sigma})}{\sqrt{P^2}} \stackrel{\text{lab}}{=} 0.
\end{eqnarray}

\subsubsection{\bf\emph{Matrix of constraints}} \label{matconst}
\vskip -3mm
We present the full calculation of the matrix of constraints and the full expression of the equations of motion for the case in which the KT \cite{Todorov1982} condition is not fulfilled. We start with the calculation of the derivative of the first constraints:
\begin{equation}
  \frac{\partial K_i}{\partial q_k^\mu} =
  \frac{\partial V_i}{\partial q_k^\mu},
  \quad
  \frac{\partial K_i}{\partial p_k^\mu} = 2 p_{i \mu} \delta_{ik} +
  \frac{\partial V_i}{\partial p_k^\mu}.
\end{equation}
Then for the time constraint we have
\begin{equation}
  \begin{aligned}
    &\frac{\partial \chi_i}{\partial q_k^\mu} = \sum_{j \ne i} (\delta_{ik} - \delta_{jk}) \frac{U_\mu}{N}, \\
    &\frac{\partial \chi_i}{\partial p_k^\mu} = \frac{1}{N \sqrt{P^2}} \sum_{j \ne i} q_{ij}^\nu \Theta_{\nu\mu},
  \end{aligned}
\end{equation}
and
\begin{equation}
  \frac{\partial \chi_N}{\partial q_k^\mu} = \frac{U_\mu}{N},
  \quad
  \frac{\partial \chi_N}{\partial p_k^\mu} = \frac{1}{N \sqrt{P^2}} \sum_j q_j^\nu \Theta_{\nu\mu}.
\end{equation}
We notice that, except for $\partial \chi_i / \partial q_k^\mu$, the derivatives of $\chi$ do not depend on $k$. For the full matrix of constraints we find
\begin{widetext}
\begin{equation}
  \begin{aligned}
  \{ K_i, \chi_j \}&= \sum_k \left[ \left( \frac{\partial V_i}{\partial q_k^\mu} \right)
                             \left( \frac{1}{N\sqrt{P^2}} \sum_{l \ne j} {q_T'}_{jl \mu} \right)
                           - \left( 2 p_i^\mu \delta_{ik}
                                  + \frac{\partial V_i}{\partial p_k^\mu} \right)
                             \left( \sum_{l \ne j} (\delta_{jk} - \delta_{lk})
                                    \frac{U_\mu}{N} \right) \right]\\
              &= \underbrace{\left( \sum_k \frac{\partial V_i}{\partial q_k^\mu} \right)}_{=0}
                             \left( \frac{1}{N\sqrt{P^2}} \sum_{l \ne j} {q_T'}_{jl \mu} \right)
                           - \left(        N \frac{\partial V_i}{\partial p_j^\mu}
                                    - \sum_l \frac{\partial V_i}{\partial p_l^\mu} \right)
                                    \frac{U_\mu}{N}
                           - \left( \frac{2 p_i^\mu U_\mu}{N}
                                    \sum_{l \ne j} (\delta_{ji} - \delta_{li}) \right),\\
  \{ K_i, \chi_j \}     &= - \left(        N \frac{\partial V_i}{\partial p_j^\mu}
                                    - \sum_l \frac{\partial V_i}{\partial p_l^\mu} \right)
                                    \frac{U_\mu}{N}
                           - \left( \frac{2 p_i^\mu U_\mu}{N}
                                    \sum_{l \ne j} (\delta_{ji} - \delta_{li}) \right).
  \end{aligned}
\end{equation}
\begin{equation}
  \begin{aligned}
  \{ K_i, \chi_N \}&= \sum_k \left[ \left( \frac{\partial V_i}{\partial q_k^\mu} \right)
                             \left( \frac{1}{N\sqrt{P^2}} \sum_l {q_T'}_{l \mu} \right)
                           - \left( 2 p_i^\mu \delta_{ik}
                                  + \frac{\partial V_i}{\partial p_k^\mu} \right)
                             \left( \frac{U_\mu}{N} \right) \right]\\
              &= \underbrace{\left( \sum_k \frac{\partial V_i}{\partial q_k^\mu} \right)}_{=0}
                             \left( \frac{1}{N\sqrt{P^2}}  \sum_l {q_T'}_{l \mu} \right)
                           - \left( \sum_k \frac{\partial V_i}{\partial p_k^\mu} \right)
                                    \frac{U_\mu}{N}
                           - \left( \frac{2 p_i^\mu U_\mu}{N} \right),\\
  \{ K_i, \chi_N \}     &= - \left(  \sum_k \frac{\partial V_i}{\partial p_k^\mu} \right)
                                    \frac{U_\mu}{N}
                           - \left( \frac{2 p_i^\mu U_\mu}{N} \right).
  \end{aligned}
\end{equation}
Using Eq. \eqref{propqt}, we can write
\begin{equation}
  \{ \chi_i , \chi_j \} = \sum_k
  \left( \sum_{l \ne i} (\delta_{ik} - \delta_{lk}) \frac{U^\mu}{N} \frac{1}{N\sqrt{P^2}} \sum_{m \ne j} {q_T'}_{jm \mu}
  - \frac{1}{N\sqrt{P^2}} \sum_{l \ne i} {q_T'}_{il}^\mu \sum_{l \ne j} (\delta_{jk} - \delta_{lk}) \frac{U_\mu}{N} \right) = 0,
\end{equation}
\begin{equation}
  \{ \chi_i , \chi_N \} = \sum_k
  \left( \sum_{l \ne i} (\delta_{ik} - \delta_{lk}) \frac{U^\mu}{N} \frac{1}{N \sqrt{P^2}} \sum_m {q_T'}_{m \mu}
  - \frac{1}{N\sqrt{P^2}} \sum_{l \ne i} {q_T'}_{il}^\mu \frac{U_\mu}{N} \right) = 0,
\end{equation}
\begin{equation}
  \{ \chi_N , \chi_N \} = 0.
\end{equation}

We can summarize these results by presenting the complete matrix of constraint :
\begin{equation}
  C_{ij}^{-1} = \{ \phi_i , \phi_j \} =
  \begin{pmatrix}
    \{    K_i , K_j \} & \{    K_i , \chi_j \} \\
    \{ \chi_i , K_j \} & \{ \chi_i , \chi_j \}
  \end{pmatrix}
\end{equation}
with
\begin{equation}
  A_{ij}^{-1} = \{    K_i    , K_j \} \ne 0,
  \quad
  B_{ij}^{-1} = \{ \chi_i , \chi_j \} = 0,
  \quad
  S_{ij}^{-1} = \{ \chi_i ,    K_j \} \ne 0.
\end{equation}
\vskip 10mm
That gives us the following relativistic factor [Eq. (\ref{lambda})]:
\begin{equation}
  \lambda_k = C_{k 2N} \qquad \ 1 < k < 2N .
\end{equation}
The final expression for the equations of motion is :
\begin{equation}
  \begin{aligned}
    &\frac{\mathrm d q_i^\mu}{\mathrm d \tau}
     = 2 \lambda_i p_i^\mu + \sum_{k=1}^N \lambda_k \frac{\partial V_k (q_T')}{\partial {p_i}_\mu}
       + \frac{1}{N \sqrt{P^2}}
     \left[   \left( \sum_{j \ne i} {q_T'}_{ij \mu} \right) \sum_{k=N+1}^{2N-1} \lambda_k 
            + \left( \sum_j         {q_T'}_{j  \mu} \right)                     \lambda_{2N}  \right],\\
    &\frac{\mathrm d p_i^\mu}{\mathrm d \tau}
     = - \sum_{k=1}^N \lambda_k \frac{\partial V_k (q_T')}{\partial {q_i}_\mu}
       - \frac{U_\mu}{N}
     \left[\sum_{k=N+1}^{2N-1} \lambda_k \left( \sum_{j \ne i} (\delta_{ik}-\delta_{jk}) \right)
           +                   \lambda_{2N} \right].
  \end{aligned}
\end{equation}
We can write these equations in a simplified form using $q_T'$ and the global reference frame ($U^\mu = (1,0,0,0)$):
\begin{equation}
    \frac{\mathrm d q_i^\mu}{\mathrm d \tau} = \frac{p_i^\mu}{E_i} + \frac{1}{N \sqrt{P^2}}
     \left[   \left( \sum_{j \ne i} {q_T'}_{ij \mu} \right) \sum_{k=N+1}^{2N-1} \lambda_k 
            + \left( \sum_j         {q_T'}_{j  \mu} \right)                     \lambda_{2N}  \right],
    \quad
    \frac{\mathrm d p_i^\mu}{\mathrm d \tau} = - \sum_{k=1}^N \frac{1}{2 E_k}
                                                \frac{\partial V_k}{\partial {q_i}_\mu}.
\end{equation}
(Notice that we only use the 3-vector part of these equations.) The second term of $\mathrm d q_i^\mu / \mathrm d \tau$ is embarrassing. For $N = 2$ particles this term is vanishing. Unfortunately for a large number of particles ($2 < N$) it does not disappear. To avoid this the KT condition must be fulfilled.
\end{widetext}

\subsection{Thermodynamical densities} \label{thermdens}
\vskip -2mm
Equation \eqref{density_thermo1} can be calculated analytically for $\mu \to 0$ and $m \ll T$ (including a factor of 2 in $g$):
\begin{equation}
  \begin{split}
  \rho_F =& \frac{4 \pi}{(2\pi)^3 (\hslash c)^3} g \int_0^\infty \left( f^+ + f^- \right) p^2 \mathrm{d} p \\
         =& \frac{4 \pi}{(2\pi)^3 (\hslash c)^3} g \ m^2 T \ \ell \ K_2\left( \frac{m}{T} \right)\\
	     =& \frac{4 \pi}{(2\pi)^3 (\hslash c)^3} g \ m^2 T \ \ell \ \frac{\Gamma(2)}{2} \ \left( \frac{2}{m/T} \right)^2, \\
  \rho_F =& \frac{\ell \ g}{\pi^2} \left( \frac{T}{\hslash c} \right)^3 .
  \end{split}
\end{equation}
Then we find:
\begin{equation}
  T_i = (\hslash c) \left( \frac{\pi^2}{\ell \ g} \right)^{1/3} \rho_F^{1 / 3} .
\end{equation}
For the baryonic density, Eq. \eqref{density_thermo2}, we have:
\begin{equation}
  \begin{split}
  \rho_B =& \frac{4 \pi}{(2\pi)^3 (\hslash c)^3} g \int_0^\infty \left( f^+ - f^- \right) p^2 \mathrm{d} p \\
         =& \frac{4 \pi}{(2\pi)^3 (\hslash c)^3} g \ \frac{\pi^2}{3} \ T^3 \left( \frac{\mu}{T} + \left( \frac{\mu}{T} \right)^3 \frac{1}{\pi^2} \right) \\
	 =& \frac{4 \pi}{(2\pi)^3 (\hslash c)^3} g  \ \frac{\pi^2}{3} \bigg( \underbrace{T^2 \mu}_{\textrm{low order}} + \frac{\mu^3}{\pi^2} \bigg), \\
  \rho_B \approx& \frac{g}{6 \pi^2} \left( \frac{\mu}{\hslash c} \right)^3 ,
  \end{split}
\end{equation}
and finally
\begin{equation}
  \mu_i = (\hslash c) \left( \frac{6 \pi^2}{g} \right)^{1/3} \rho_B^{1 / 3} .
\end{equation}
The degeneracy factor for the spin, the parity, the color, and the flavor is $g = 2 \times 2 \times 3 \times 3 = 36$.

\subsection{Derivation of potential} \label{derivpot}
\vskip -2mm
The forces coming from the derivative of the potential are:
\begin{equation}
  \begin{aligned}
  &\frac{\partial V_i}{\partial q_{k\nu}} =
  \frac{\partial m_i^2}{\partial q_{k\nu}} =
  2 m_i \frac{\partial m_i}{\partial q_{k\nu}} =
  2 m_i \frac{\partial m_i}{\partial T_i} \frac{\partial T_i}{\partial q_{k\nu}},\\
  &\frac{\partial V_i}{\partial p_{k\nu}} =
  \frac{\partial m_i^2}{\partial p_{k\nu}} =
  2 m_i \frac{\partial m_i}{\partial p_{k\nu}} =
  2 m_i \frac{\partial m_i}{\partial T_i} \frac{\partial T_i}{\partial p_{k\nu}}.
  \end{aligned}
\end{equation}
The local temperature is defined as:
\begin{equation}
  T_i = (\hslash c) \left( \frac{\pi^2}{\ell\ g} \right)^{1/3} \rho_F^{1 / 3}
      = \Pi \left( \sum_{l \ne i} R_{il} \right)^{1 / 3}
\end{equation}
with $R_{ij} = \exp({q_T'}_{ij}^2 / L^2)$, where $L$ is a weighting factor, and
\begin{equation}
  \Pi = (\hslash c) \left( \frac{\pi^2}{\ell\ g} \right)^{1/3} \simeq 190 \text{ MeV}.
\end{equation}
Its derivatives are:
\begin{equation}
  \begin{aligned}
    &\frac{\partial T_i}{\partial q_{k\nu}} = \Pi \frac{1}{3}
    \frac{1}{\left( \sum_{l \ne i} R_{il} \right)^{2 / 3}} \sum_{l \ne i} R_{il}
    \frac{1}{L^2} \frac{\partial {q_T}(')_{ij}^2}{\partial q_{k\nu}}, \\
    &\frac{\partial T_i}{\partial p_{k\nu}} = \Pi \frac{1}{3}
    \frac{1}{\left( \sum_{l \ne i} R_{il} \right)^{2 / 3}} \sum_{l \ne i} R_{il}
    \frac{1}{L^2} \frac{\partial {q_T}(')_{ij}^2}{\partial p_{k\nu}},
  \end{aligned}
\end{equation}
which can be rewritten as
\begin{equation}
  \begin{aligned}
    &\frac{\partial T_i}{\partial q_{k\nu}} = \frac{2 \Pi^3}{3 L^2 T_i^2}
    \sum_{l \ne i} R_{il} {q_T'}_{il}^\nu \left( \delta_{ik} - \delta_{lk} \right), \\
    &\frac{\partial T_i}{\partial p_{k\nu}} =   -   \frac{2 \Pi^3}{3 L^2 T_i^2}
    \sum_{l \ne i} R_{il} {q_T'}_{il}^\nu
    \frac{(q_{il \sigma} U^{\sigma})}{\sqrt{P^2}} \stackrel{\text{lab}}{=} 0.
  \end{aligned}
\end{equation}
\clearpage